\documentclass[12pt]{article}

\usepackage{filecontents}

\usepackage{xr-hyper}
\usepackage{amsthm}
\usepackage{fullpage}
\usepackage{footnote}
\usepackage{tabu}
\usepackage{booktabs}
\usepackage[title]{appendix}
\providecommand{\keywords}[1]
{
  \small	
  \textbf{\textit{Keywords---}} #1
}
\usepackage{etoolbox}

\makeatletter
\patchcmd{\@makecaption}
  {\parbox}
  {\advance\@tempdima-\fontdimen2} 
  {}{}
\makeatother 

\newcommand{\specialcell}[2][c]{%
  \begin{tabular}[#1]{@{}c@{}}#2\end{tabular}}

\usepackage{titling} 

\usepackage{amsmath,bbm,amssymb, amsfonts}
\usepackage{mathtools}
\usepackage{graphicx,psfrag,epsf}
\usepackage{enumitem}
\usepackage{natbib}
\usepackage{url} 
\usepackage{color}
\usepackage{algorithmicx}
\usepackage{booktabs}
\usepackage{siunitx}
\usepackage{array}
\usepackage{rotating}

\usepackage{adjustbox}

\usepackage{algorithm}
\usepackage{algpseudocode}

\usepackage{tikz}

\usepackage{url}

\usepackage{multicol}
\usepackage{xcolor}
\usepackage{algorithm}

\makeatletter
\def\breve{\mathpalette\wide@breve}
\def\wide@breve#1#2{\sbox\z@{$#1#2$}%
	\mathop{\vbox{\m@th\ialign{##\crcr
				\kern0.08em\brevefill#1{0.8\wd\z@}\crcr\noalign{\nointerlineskip}%
				$\hss#1#2\hss$\crcr}}}\limits}
\def\brevefill#1#2{$\m@th\sbox\tw@{$#1($}%
	\hss\resizebox{#2}{\wd\tw@}{\rotatebox[origin=c]{90}{\upshape(}}\hss$}
\makeatletter

\newcommand{\overbar}[1]{\mkern 1.5mu\overline{\mkern-1.5mu#1\mkern-1.5mu}\mkern 1.5mu}

\DeclareMathOperator*{\argmin}{argmin} 

\newtheorem{remark}{Remark}

\def\indep{{\perp \!\!\! \perp}}

\def\gobblestop#1#2{#1}
\def\killstop{%
	\aftergroup\gobblestop
}

\def\bt{{\bf t}}

\def\bw{{\bf w}}
\def\bx{{\bf x}}

\def\bX{{\bf X}}

\def\thick#1{\hbox{\rlap{$#1$}\kern0.25pt\rlap{$#1$}\kern0.25pt$#1$}}

\def\btheta{\boldsymbol{\theta}}

\def\smbalpha{\boldsymbol{{\scriptstyle{\alpha}}}}

\def\smbalpha{\widehat{\smbalpha}}

\def\hbar{\bar{ h}}

\def\Dsc{{\cal D}}

\def\transpose{{\sf \scriptscriptstyle{T}}}

\def\trans{^{\transpose}}

\def\mybox#1{\vskip1mm \begin{center}
        \hspace{.0\textwidth}\vbox{\hrule\hbox{\vrule\kern6pt
\parbox{.9\textwidth}{\kern6pt#1\vskip6pt}\kern6pt\vrule}\hrule}
        \end{center} \vskip-5mm}
\def\lboxit#1{\vbox{\hrule\hbox{\vrule\kern6pt
      \vbox{\kern6pt#1\vskip6pt}\kern6pt\vrule}\hrule}}

\def\thickboxit#1{\vbox{{\hrule height 1mm}\hbox{{\vrule width 1mm}\kern6pt
          \vbox{\kern6pt#1\kern6pt}\kern6pt{\vrule width 1mm}}
               {\hrule height 1mm}}}

\def\bbE{\bb_{\scriptscriptstyle E}}

\def\bbP{\mathbb{P}}

\def\calA{\mathcal{A}}

\def\calD{\mathcal{D}}
\def\calE{\mathcal{E}}

\def\calH{\mathcal{H}}

\def\calN{\mathcal{N}}

\def\calV{\mathcal{V}}

\def\calX{\mathcal{X}}
\def\calY{\mathcal{Y}}

\def\bbE{\mathbb{E}}

\def\bbP{\mathbb{P}}

\def\bbR{\mathbb{R}}

\def\fat#1{\hbox{\rlap{$#1$}\kern0.25pt\rlap{$#1$}\kern0.25pt$#1$}}

\def\calD{\Dsc}

\newcolumntype{R}{@{\extracolsep{0.5cm}}r@{\extracolsep{0pt}}}%
\newcolumntype{E}{@{\extracolsep{0.25cm}}c@{\extracolsep{0pt}}}%

\makeatletter
\newcommand{\distas}[1]{\mathbin{\overset{#1}{\kern\z@\sim}}}%
\makeatother

\usepackage{xr-hyper}
\usepackage[colorlinks, linkcolor=blue, anchorcolor=blue, citecolor=blue]{hyperref}

\newtheorem{thm}{Theorem}[section]

\newtheorem{lem}[thm]{Lemma}

\usepackage[left=0.9in,right=0.9in,top=0.95in,bottom=0.95in]{geometry}

\usepackage{xr}
\newcommand{\blind}{1}

\begin{document}

\if1\blind
{
	\title{\bf Independence weights for causal inference with continuous treatments}
	\author{Jared D. Huling$^{1}$\thanks{corresponding authors: huling@umn.edu}, 
		Noah Greifer$^{2}$,
		Guanhua Chen$^{3}$\thanks{corresponding authors: gchen25@wisc.edu} \\
		\\
		$^{1}$Division of Biostatistics, University of Minnesota \\ [8pt]
		$^{2}$Institute for Quantitative Social Science, \\ Harvard University \\ [8pt]
		$^{3}$Department of Biostatistics and Medical Informatics, \\University of Wisconsin-Madison \\ [8pt]
	}
} \fi

\if0\blind
{
	\bigskip
	\bigskip
	\bigskip
	\begin{center}
		{\Large \bf placeholder }
	\end{center}
	\medskip
} \fi

\date{}

\maketitle

\begin{abstract}
Studying causal effects of continuous treatments is important for gaining a deeper understanding of many interventions, policies, or medications, yet researchers are often left with observational studies for doing so. In the observational setting, confounding is a barrier to the estimation of causal effects. Weighting approaches seek to control for confounding by reweighting samples so that confounders are comparable across different treatment values. Yet, for continuous treatments, weighting methods are highly sensitive to model misspecification. In this paper we elucidate the key property that makes weights effective in estimating causal quantities involving continuous treatments. We show that to eliminate confounding, weights should make treatment and confounders independent on the weighted scale. We develop a measure that characterizes the degree to which a set of weights induces such independence. Further, we propose a new model-free method for weight estimation by optimizing our measure. We study the theoretical properties of our measure and our weights, and prove that our weights can explicitly mitigate treatment-confounder dependence. The empirical effectiveness of our approach is demonstrated in a suite of challenging numerical experiments, where we find that our weights are quite robust and work well under a broad range of settings.
 \\[3em]
\end{abstract}%

\def\spacingset#1{\renewcommand{\baselinestretch}%
{#1}\small\normalsize} \spacingset{1.5}

\newpage
\spacingset{1.725} 
\setlength{\abovedisplayskip}{7pt}%
\setlength{\belowdisplayskip}{7pt}%
\setlength{\abovedisplayshortskip}{5pt}%
\setlength{\belowdisplayshortskip}{5pt}%


\section{Introduction}
\label{sec:intro}

Confounding is a major barrier to studying causal effects of treatments or exposures from observational data. Considerable work has focused on the development of approaches for studying causal effects of binary or otherwise discrete-valued treatments from observational data. 
With continuous treatments, however, the choices of methods for confounding control are far more limited, and clear guidance that can help practitioners choose among the available methods is lacking. 
A common approach to reduce confounding by observed variables is using the propensity score, which was initially proposed for binary treatments \cite{rosenbaum1983central} and has been generalized to the setting of continuous treatments \citep{hirano2004propensity, imai2004causal, zhu2015boosting, kennedy2017non, galvao2015uniformly}. With binary treatments, the causal effect of interest is often the average treatment effect, which can be estimated as a difference in the weighted averages of the treatment group outcomes, where the weights are proportional to the inverse of the propensity score \citep{robins1994estimation, robins2000marginal}; this method is known as inverse probability weighting (IPW).
With continuous treatments, the interest is often in estimation of the causal dose-response functionals \citep{robins2000marginal,van2003unified} such as the causal average dose-response function (ADRF), which can be estimated using a weighted regression of the outcome on the treatment, where the weights are proportional to the inverse of the conditional density of the treatment given the covariates, the generalized propensity score (GPS).

In the binary treatment setting, IPW estimators can be unstable due to extreme weights and susceptible to model misspecification \citep{kang2007demystifying, fan2021optimal}. These issues carry over to IPW estimators for continuous treatments and are substantially more challenging to address. A key reason is that IPW estimation via the GPS requires inverse weighting by a conditional {density} estimate, not just a conditional probability. Even if the conditional mean of the treatment given covariates is correctly specified, GPS weights can fail to perform well if the \textit{distribution} of the conditional density is misspecified \citep{naimi2014constructing}. The difficulty of correctly specifying a conditional distribution is exacerbated with increased dimension of the pre-treatment covariates to be controlled for, i.e., the confounders.
The tailored use of flexible machine learning estimation approaches such as that proposed in \citet{zhu2015boosting} can in many cases yield substantial improvements, however, as shown in our simulations and data analysis, they are still susceptible to the issues of GPS weighting and can perform poorly and/or yield large weights in practice. The ReGPS approach of \citet{colangelo2020double} directly estimates the inverse GPS without the need for correct specification of a model for the GPS.

As misspecification of a conditional density model can be difficult to diagnose and assess, several works have focused on directly estimating weights to reduce the correlation magnitude between marginal moments of (pre-treatment) covariates and the treatment. Approaches along this line of work include the generalized covariate balancing propensity score (CBPS) approach of \citet{fong2018covariate} which is an extension of the CBPS approach for discrete treatments \citep{imai2014covariate}, covariate association eliminating weights \citep{yiu2018covariate}, and entropy balancing weights \citep{tubbicke2020entropy, vegetabile2020nonparametric}. Because these approaches focus on estimation of weights directly as opposed to estimating a conditional density explicitly and inverting it, they tend to be more effective empirically than direct modeling of the GPS. 
While intuitively appealing, these approaches require careful choices of which moments of \textit{both} the covariates and the treatment to ``decorrelate''. Yet, there is no guidance on specifying the right moments necessary to mitigate bias in estimation of the ADRF due to confounding. Missing important moments can leave substantial residual dependence between the covariates and treatment, see e.g. Figure \ref{fig:mech_power_pao2} from our analysis of electronic health record data. Our simulations show that the choice of moments is indeed critically important in practice and that numerical instability can arise when too many moments are used. The tension between including enough moments to reduce bias and the instability of weights as more moments are included can make these methods challenging to use in practice. 
 The general setup of \citet{ai2021unified} relies on sieve/series estimators; little finite-sample guidance is provided. Kernel Optimal Orthogonality Weighting (KOW) \citep{kallus2019kernel} and a generalization of KOW proposed in \citet{martinet2020balancing} are kernel-based nonparametric extensions of direct weights estimation ideas. They focus on estimating weights to decorrelate over function spaces of treatment and covariates such that there is no need to choose models or which moments to decorrelate. Yet, careful tuning is still required when flexible kernels are used, and, on the other hand, when inflexible kernels are used, there is no guarantee that the resulting weights fully mitigate confounding.  Thus, the user is often left with a difficult choice of which kernel to use and unclear guidance on how to choose the kernel's tuning parameters. Further, no theoretical justification of the approach of \citet{kallus2019kernel} is provided. The theoretical results of \citet{martinet2020balancing} are limited to the convergence of the weighted distribution functions and do not explore properties involving estimation of the ADRF. More extensive discussion of the existing literature can be found in the Section \ref{supsec:connection_other_methods} of the Supplementary Material. 

Our work aims to achieve several goals. First, we provide clarity on the role of weights in estimation of the ADRF. To do so, we provide a general decomposition of the error of a weighted nonparametric estimator of the ADRF and demonstrate that, under broad conditions, the ideal weights should induce complete \textit{independence} between the treatment variable and pre-treatment covariates to guarantee mitigation of confounding bias when estimating the ADRF. While already intuitively understood in the literature, our decomposition precisely quantifies the impact of this dependence on the estimation error in finite samples. We also show that dependence plays a key role in other estimands, such as the causal quantile dose-response function. 
Second, we develop a measure based on energy statistics \citep{szekely2007measuring,szekely2013energy} that allows one to assess how well a set of weights is able to induce independence, where smaller values of our measure indicate the weights yield less treatment-covariates dependence and a value of zero indicates complete independence between the treatment and covariates in the weighted data. \citet{huling2020energy} developed a weighted energy distance to mitigate distributional imbalance of covariates in a discrete-treatment setting. In their setting, the energy distance is used to measure distributional imbalance of covariates between different treatment groups, whereas in the setting of this work, a modified distance covariance is used to measure statistical \textit{dependence} between treatment and confounders. Distributional imbalance plays an important role in confounding control with discrete treatments, but this concept does not naturally generalize to continuous treatments. However, removing statistical dependence between discrete treatments and confounders implies distributional balance has been achieved. As such, our work involves a more general notion of confounding control that can in principle be applied to discrete treatments.

Finally, we propose a new approach for estimating weights, which we call the distance covariance optimal weights (DCOWs), by optimizing our measure. The proposed weights directly aim to mitigate dependence between the treatment and confounders; our error decomposition illustrates that the DCOWs reduce finite sample dependence and thus source of error due to confounding in a weighted nonparametric estimate of the ADRF. In other words, the DCOWs aim to create a pseudo population where treatment and confounders are statistically independent. Our weight construction approach does not require modeling a conditional density, careful tuning of hyperparameters, or choosing which moments of covariates and treatment to decorrelate, making it readily accessible and easy to use for practitioners with varying degrees of statistical sophistication.

We provide some theoretical results for our proposal, showing that our weights indeed reduce dependence between treatment and covariates and fully induce independence asymptotically. Further, we show that with a small penalty on the variability of the weights, our proposal results in the same convergence rate as a nonparametric regression estimate of the ADRF in a scenario with no confounding. Although adding a penalty to reduce weight variability involves the inclusion of a tuning parameter, our proposed approach rarely results in weights with large variability even without penalization. Careful tuning of the parameter that controls weight variability is rarely necessary, as evidenced by our simulation studies, which investigate a wide variety of scenarios with strong and complex confounding and scenarios with moderately high-dimensional confounding, in all of which we fix the tuning parameter to its default value.

Our proposed weights can be used beyond simple weighted nonparametric estimators of the ADRF. \citet{kennedy2017non} and \citet{ diaz2013targeted} extended the idea of doubly-robust estimation to continuous treatments, allowing for estimates that combine outcome regression models \citep{imbens2004nonparametric,hill2011bayesian} and the conditional density models. This allows for relaxed dependence on the correctness of the regression and conditional density models. However, doubly-robust estimators are not immune to highly variable weights and their finite sample performance can suffer if the conditional density model is misspecified.
We show that our weights can enhance doubly-robust estimators. 
Pairing a reasonable outcome regression model with our weights in a doubly-robust fashion can be effective in estimating the ADRF.

The remainder of this paper is organized as follows. We investigate the role of dependence between the treatment and covariates in the estimation of the ADRF in Section \ref{sec:setup} and we develop a criterion that assesses how much dependence is mitigated by a set of weights, propose a new weight estimation strategy, and provide some corresponding theory in Section \ref{sec:methods}. We demonstrate the effectiveness of our approach in finite samples with a suite of challenging simulation studies in Section \ref{sec:simulations} and illustrate the use of our approach in a real-world study of electronic health record data in Section \ref{sec:data}. We conclude the paper with some discussion.



\section{Confounding, Weighting, and Dependence}
\label{sec:setup}

\subsection{Setup, notation, and assumptions}

The observable quantities we consider consist of the random triplet $(\bX,A,Y)$, where $\bX\in\calX\subseteq\bbR^p$ is a vector of pre-treatment covariates, $A\in\calA\subseteq\bbR$ is a continuous-valued treatment variable indicating the assigned dose for a unit, and $Y\in\calY\subseteq\bbR$ is an outcome of interest. The variate $(\bX,A,Y)$ has a joint distribution $F_{\bX,A,Y}$ with respect to a dominating measure. We denote the marginal density of the treatment and covariates as $f_A(a)$ and $f_\bX(\bx)$, respectively,  the conditional density of the treatment given $\bX$ as $f_{A|\bX}(a|\bx)$, and their joint density as $f_{\bX,A}(a,\bx)$. Similarly, corresponding distribution functions are denoted $F_A(a) = \bbP(A\leq a)$, $F_\bX(\bx) = \bbP(\bX\leq \bx)$, $F_{A|\bX}(a|\bx) = \bbP(A\leq a \:|\: \bX = \bx)$, and $F_{\bX,A}(\bx,a) = \bbP(\bX\leq\bx,A\leq a)$. Our observed data consists of $n$ i.i.d. samples $(\bX_i,A_i,Y_i)_{i=1}^n$ from $(\bX,A,Y)$. Note that we drop the subscripts on the density and cumulative distribution functions when there is no ambiguity. 

We work under the potential outcomes framework, wherein the potential outcome function $Y(a)$ for $a\in\calA$ is the outcome that would be observed if $A$ were set to the value $a$. The causal quantity of  interest in this paper is the mean potential outcome function, also called the causal \textit{average dose-response function} (ADRF), which is 
    $\mu(a) \equiv \bbE\left[ Y(a) \right], \text{ for } a\in\calA$.

The causal ADRF $\mu(a)$ can be identified, or expressed in terms of observational data, under standard causal assumptions. These  assumptions, which we employ throughout this paper,  are 1) \textit{consistency}, which posits that $A=a$ implies $Y=Y(a)$, 2) \textit{positivity}, which states that all values of the treatment are possible across the covariate space in the sense that $f(a|\bX=\bx) \geq \nu >0$  for all $\bx\in\calX$ for some constant $\nu$, and 3) \textit{ignorability} of the assignment mechanism: $Y(a) \indep A \:|\: \bX$ for all $a\in\calA$, where $\indep$ denotes (conditional) independence. Under assumptions 1)-3), the dose-response function is identified as 
    $\mu(a) = \bbE_\bX\left(\bbE\left[ Y \:|\: \bX,A=a\right]\right) = \bbE_\bX\left[\mu(\bX,a)\right]$.
Estimation of the dose-response function via regression-based estimation of the mean function $\mu(\bX,a) \equiv \bbE\left[ Y \:|\: \bX,A=a\right]$, however, can be highly challenging. Misspecification of the regression function can result in poor estimation of $\mu(a)$, and nonparametric estimation of the regression function is also difficult, especially when $\bX$ is not low dimensional. Instead, this paper focuses on weighting-based estimators of $\mu(a)$. A benefit of weighting estimators is that the dose-response function can be flexibly estimated by univariate (weighted) nonparametric regression, whereas with regression-based estimation, the need to incorporate covariates in a regression may make flexible estimation of the ADRF more difficult due to the additional dimensions. A conceptual benefit of weighting methods over regression-based methods is that they provide a clear separation between design and analysis phases of a study. This separation is critical when substantial or iterative model-building is required to control for confounding.

\subsection{Ignorability, independence, and the GPS}

To reliably estimate the causal dose response function using observational data, sources of structural bias should be mitigated, among which the bias due to confounding is the most common. Consider the setting where the covariate vector $\bX$ contains all confounders in studying the causal relationship between $A$ and $Y$. Blocking the backdoor path $A \leftarrow \bX \rightarrow Y$ mitigates confounding. One way of blocking the backdoor path is by removing the arrow between $A$ and $\bX$ (i.e., making $A$ independent of $\bX$). In a randomized trial, the independence between $A$ and $\bX$ holds due to randomization. This motivates us to create a pseudo-population mimicking the one we would observe under such a trial by reweighting the subjects in an observational study such that $A$ is independent of $\bX$ in the pseudo-population.
Weighting by the generalized propensity score $f_{A|\bX}(A\:|\:\bX)$ (GPS) \citep{hirano2004propensity} achieves such and extends the pioneering work of \cite{rosenbaum1983central} for binary treatments to continuous treatments.

For continuous treatments, stabilized GPS weights are computed as $f_{A}(A)/f_{A|\bX}(A\:|\:\bX)$ \citep{robins2000marginal}, which naturally arise in estimating equations for the dose-response function via semiparametric theory \citep{kennedy2017non}. Estimating the weights requires correct specification not only of the mean of the conditional density of the treatment, but also of its higher order properties such as \textit{shape}. When any of these is misspecified, bias can result in estimates of the ADRF \citep{naimi2014constructing, zhu2015boosting}, indicating a particularly sensitive reliance on correct modeling of the conditional distribution of the treatment given the covariates. Furthermore, similar to standard propensity score weights, GPS modeling can yield extreme weights, leading to unstable estimation. Weight trimming/capping may alleviate the problem of large weights but can be seen as ad hoc and may change the estimand \citep{crump2009dealing}: the estimated ADRF will correspond to the population represented by the newly weighted sample rather than to the original target population.

The stabilized GPS weights $f_{A}(A)/f_{A|\bX}(A\:|\:\bX)$ have several key properties that have motivated work to improve upon the GPS weights. Namely, when weighting by the GPS weights in the population sense, they 1) result in independence of $\bX$ and $A$, 2) preserve the marginal distributions of $\bX$ and $A$, and 3) have mean 1. These properties are listed more explicitly in Section \ref{sec:appendix_aa} of the Supplementary Material.
Instead of indirectly estimating the weights by estimating the GPS, a nascent line of work has involved methods which directly estimate weights designed to satisfy the above three properties via balancing criteria. For example, work has focused on estimation of weights that induce zero marginal correlation between treatment and covariates  \citep{fong2018covariate, yiu2018covariate, vegetabile2020nonparametric}. Although they are more robust than GPS weights, these approaches rely on both the correct choice of moments of the covariates and the choice of moments of the treatment variable to decorrelate. Yet, there is little guidance in deciding which set of moments in both covariates and treatment to focus on, as these choices depend on the form of the true potential outcome-generating model.

The aforementioned three properties of GPS weights are intuitively appealing, but it is not immediately clear which of or in what manner these properties are important in mitigating bias in a weighted nonparametric estimator of the ADRF. 
To justify which properties are crucial in weights estimation, we derive the relationship between the properties of weighted dose-response function estimators using generic weights and the systematic source of error of the weighted estimator for the ADRF. We demonstrate that the ability of a set of weights to induce independence between $\bX$ and $A$ is critical for reducing the bias of a weighted estimator.

\subsection{A general error decomposition for weighted nonparametric  estimators of the ADRF}
\label{sec:error_decomp}

In this section we aim to provide an explicit mechanistic connection between dependence and bias in weighted estimates of the ADRF.
Although it is understood that using weights constructed from a well-estimated conditional density is consistent \citep{kennedy2017non}, it is unclear what role weights play more generally. It is not immediately clear what the connection is between the balancing criteria that aim to ``decorrelate'' moments of covariates and moments of the treatment variable and the systematic bias in estimating the ADRF. In the following, we investigate the precise source of the systematic bias of an estimator and illuminate the role of the weights in influencing the bias. We focus on weighted Nadaraya-Watson estimators of the ADRF for clarity of presentation, though the key message applies to weighted local polynomial regression and other weighted nonparametric regression.

The response can be expressed as $Y_i = \mu(\bX_i, A_i) + \varepsilon_i$, where $\varepsilon_i \equiv Y_i(A_i) - \mu(\bX_i, A_i)$. By construction, $\varepsilon_i$ have mean zero but are not necessarily identically distributed.
Given any set of weights $\bw = (w_1, \dots, w_n)$ and a kernel $K_h(A_i - a_0) = K(\frac{A_i - a_0}{h})/h$ centered at $A=a_0$ with bandwidth $h>0$, the weighted Nadaraya-Watson (NW) estimator of the ADRF at $A=a_0$ is 
\begin{equation}
    \widehat{\mu}_{NW}^{\bw}(a_0)
=  \frac{\sum_{i=1}^n Y_i w_i K_h(A_i - a_0)}{ \sum_{i=1}^n K_h(A_i - a_0)}. \label{eqn:weighted_nw_estimator}
\end{equation}
This class of estimators of the causal ADRF is motivated by the identification results of \citet{colangelo2020double}, who showed under certain causal conditions and assumptions regarding the kernel $K_h$ that $\mu(a_0) = \lim_{h\to 0}\bbE\left[YK_h(A-a_0)/f_{A|\bX}(a_0\mid\bX) \right]$, which implies the use of the inverse of $f_{A|\bX}(A\mid\bX)$ as weights, since $\mu(a_0) = \lim_{h\to 0}\bbE\left[YK_h(A-a_0)w^*(\bX,A) \right]/\bbE\left[K_h(A-a_0)\right] = \lim_{h\to 0}\bbE\left[YK_h(A-a_0)w^*(\bX,a_0) \right]/f_A(a_0)$, where $w^*(\bx,a) = f_A(a)/f_{A|\bX}(a\mid\bX=\bx)$.
Given \textit{any} weights $\bw$, the error of \eqref{eqn:weighted_nw_estimator} at $A=a_0$ can be decomposed as 
\begingroup
\allowdisplaybreaks
\begin{align}
\widehat{\mu}_{NW}^{\bw}(a_0) - \mu(a_0)  = {} & \int_{\calX} \int_\calA \mu(\bx, a_0) \mathrm{d} \left[ F^n_{\bX,A,\bw} - F^n_{\bX}F^n_{A} \right](\bx,a)  \label{eqn:error_decomp_indep_first} \\
& + \int_{\calX} \mu(\bx,a_0)\mathrm{d} \left[ F^n_{\bX} - F_\bX \right](\bx)   + \left(\frac{f_A(a_0)}{\widehat{f}^n_{A,h}(a_0)} - 1\right) \int_{\calX} \mu(\bx, a_0)\mathrm{d}F^n_{\bX}(\bx)  \nonumber \\
    & + \left(\frac{f_A(a_0)}{\widehat{f}^n_{A,h}(a_0)} - 1\right) \int_{\calX} \int_\calA \mu(\bx, a_0) \mathrm{d} \left[ F^n_{\bX,A,\bw} - F^n_{\bX}F^n_{A} \right](\bx,a)  \nonumber \\
    &+{\widehat{f}^n{}}^{-1}_{A,h}(a_0)\int_\calX\int_\calA \left[\mu(\bx,a)K_h(a-a_0) - \mu(\bx,a_0)f_A(a_0)\right]\mathrm{d}F^n_{\bX,A,\bw}(\bx,a)  \nonumber \\
    & + \frac{1}{n}\sum_{i=1}^n \varepsilon_i w_i {\widehat{f}^n{}}^{-1}_{A,h}(a_0) {K_h(A_i - a_0)}, \label{eqn:error_decomp_wts_error_first}
\end{align}
\endgroup
where $\widehat{f}^n_{A,h}(a_0) = \int_\calA K_h(a-a_0)\mathrm{d}F^n_{A}(a)$ is a kernel density estimate of $f_A(a_0)$, $F^n_{\bX}(\bx) = \\ n^{-1}\sum_{i=1}^nI(\bX_i\leq \bx)$ is the empirical cumulative distribution function (CDF) of $\{\bX_i\}_{i=1}^n$, $F^n_{A}(a) = n^{-1}\sum_{i=1}^nI(A_i\leq a)$ is the empirical CDF of $\{A_i\}_{i=1}^n$, and $F^n_{\bX,A,\bw}(\bx,a) = n^{-1}\sum_{i=1}^nw_iI(\bX_i\leq \bx, A_i\leq a)$ is the weighted empirical CDF of $\{\bX_i, A_i\}_{i=1}^n$ using weights $\bw$.
 We provide a derivation of this decomposition in Section \ref{sec:appendix_b} of the Supplementary Material. The second term on the right is due to sampling variability only and has mean zero and converges at rate $n^{-1/2}$ if the sample is representative of the super-population. The expectations of the third, fourth, and fifth terms on the right above go to 0 when $h\to 0$ regardless of the weights $\bw$. The last term has mean zero regardless of both the weights and the bandwidth, though its variability is impacted by the weights. We note, however, that the third through fifth terms are not guaranteed to converge to zero without additional conditions on the variability of the weights.

On the other hand, the first term \eqref{eqn:error_decomp_indep_first} on the right above is the source of systematic bias of the weighted estimator unrelated to kernel smoothing. In other words, taking limits of the bandwidth of the kernel to 0 and sample size to infinity does not make \eqref{eqn:error_decomp_indep_first} vanish. This term also provides insight into why using the weights we propose later performs well in finite sample settings when used in treatment effect estimators, as targeting this term can help decrease the magnitude of the systematic component of the bias of an estimator. If a given set of weights induces finite-sample independence of $\bX$ and $A$ in the sense that $F^n_{\bX,A,\bw}(\bx,a) = F^n_{\bX}(\bx)F^n_{A}(a)$ for all $\bx\in\calX$, $a\in\calA$, then the source of bias \eqref{eqn:error_decomp_indep_first} of  $\hat \mu^{\bw}(a_0)$ will be zero. The mean-squared error of the estimator will, however, depend primarily on both the bias term \eqref{eqn:error_decomp_indep_first} and the variance of \eqref{eqn:error_decomp_wts_error_first}. Mitigating the variance of \eqref{eqn:error_decomp_wts_error_first} merely amounts to controlling the squares of the weights; however, providing a measure that can characterize \eqref{eqn:error_decomp_indep_first} is non-trivial and none exists in the literature. The term \eqref{eqn:error_decomp_indep_first} is bounded by the distance between $F^n_{\bX,A,\bw}(\bx,a)$ and $F^n_{\bX}(\bx)F^n_{A}(a)$ provided that $\mu(\bx, a_0)$ is bounded.
Without modeling the response function, constructing a measure that bounds \eqref{eqn:error_decomp_indep_first} is critical for assessing a set of weights.

\begin{remark}\label{remark1}
The role of weights in their ability to induce independence between treatment and covariates is not unique to estimation of the ADRF and applies to a wide variety of estimands.
Consider estimation of the causal dose-response quantile function $q_{Y(a_0)}(\alpha) = \inf\{y: F_{Y(a_0)}(y) \leq \alpha\}$, where $F_{Y(a_0)}(y) = \bbP(Y(a_0) \leq y) = \bbE_{\bX}\left\{ \bbP(Y\leq y\vert \bX, A=a_0) \right\} = \bbE_{\bX}\left\{ F_{Y|\bX,A}(y\vert \bX,A=a_0) \right\}$. By replacing $Y_i$ with $I(Y_i\leq y)$ in \eqref{eqn:weighted_nw_estimator}, we can show that the estimation error of $F_{Y(a_0)}(y)$ also depends on how well weights mitigate dependence between $A$ and $\bX$. More details are included in Section \ref{sec:appendix_b} of the Supplementary Material.
\end{remark}

In practice, the estimator $\widehat{\mu}_{NW}^{\bw}(a_0)$ in \eqref{eqn:weighted_nw_estimator} may be unstable, as the weights only appear in the numerator, so the estimated ADRF may lie outside the range of the observed values of the response. Instead, a more stable estimator is the following weighted average of the responses
\begin{equation}
    \widehat{\mu}_{NWs}^{\bw}(a_0)
=  \frac{\sum_{i=1}^n Y_i w_i K_h(A_i - a_0)}{ \sum_{i=1}^n w_iK_h(A_i - a_0)}, \label{eqn:weighted_nw_estimator_stabilized}
\end{equation}
which can be viewed as the minimizer of a weighted least squares problem where the $i$th weight is $w_iK_h(A_i - a_0)$.  The estimator \eqref{eqn:weighted_nw_estimator_stabilized} is also a valid estimator of the ADRF as long as the denominator divided by $\sum_{i=1}^nw_i$ is a consistent estimator of $f_A(a_0)$; in this case the key source of systematic bias still depends on the term \eqref{eqn:error_decomp_indep_first}.


\section{Measuring and Controlling Weighted Dependence with Energy Statistics}
\label{sec:methods}

\subsection{A criterion to evaluate the quality of a set of weights}

Having established the relationship between the dependence of $A$ and $\bX$ and the error in a weighted nonparametric estimate of the ADRF, we now construct a criterion that can assess how well a given set of weights induces independence on the weighted scale, i.e., we aim to characterize and bound the distance between $F^n_{\bX,A,\bw}$ and $F^n_{\bX}F^n_{A}$. We do so by building on the ideas of distance covariance \citep{szekely2007measuring}, which is a measure of dependence between two random vectors of arbitrary but dimensions. The population distance covariance is zero if and only if the vectors are independent. Hence, a \textit{weighted} distance covariance will be a useful component for our measure. 
Let $\bw = (w_1, \dots, w_n)$ and define the weighted distance covariance to be
\begin{equation}\label{eqn:weighted_distance}
    \calV^2_{n,\bw}(\bX,A) = \frac{1}{n^2}\sum_{k,\ell=1}^nw_kw_\ell C_{k\ell}D_{k\ell}, 
\end{equation}
where 
\vspace{-15pt}
\begin{alignat*}{2}
    c_{k\ell} & = \lVert \bX_k - \bX_\ell \rVert_2, \quad \overbar{c}_{k\cdot}  &&= \frac{1}{n}\sum_{\ell=1}^nc_{k\ell}, \quad \overbar{c}_{\cdot\ell}  = \frac{1}{n}\sum_{k=1}^nc_{k\ell}, \\
    \overbar{c}_{\cdot\cdot} & = \frac{1}{n^2}\sum_{k,\ell=1}^nc_{k\ell}, \quad C_{k\ell} &&= c_{k\ell} - \overbar{c}_{k\cdot} - \overbar{c}_{\cdot\ell} + \overbar{c}_{\cdot\cdot},
\end{alignat*}
for $k,\ell =1, \dots, n$. Similarly define $d_{k\ell} = \lvert A_k - A_\ell \rvert$, $\overbar{d}_{k\cdot}=\frac{1}{n}\sum_{\ell=1}^nd_{k\ell}$, $\overbar{d}_{\cdot\ell} = \frac{1}{n}\sum_{k=1}^nd_{k\ell}$, and $D_{k\ell} = d_{k\ell} - \overbar{d}_{k\cdot} - \overbar{d}_{\cdot\ell} + \overbar{d}_{\cdot\cdot}$. 
The quantity \eqref{eqn:weighted_distance} simplifies to the original distance covariance when weights are all $1$. Since the original distance covariance is always non-zero, \eqref{eqn:weighted_distance} is also always non-zero if the weights are positive.
We now provide further insight and motivation of the form of $\calV^2_{n,\bw}(\bX,A)$ and its interpretation in terms of weighted distributions.

Letting $i=\sqrt{-1}$, we define the (weighted) empirical characteristic functions as $\varphi^n_{\bX,A,\bw}(\bt,s) =$ $\frac{1}{n}\sum_{j=1}^nw_j\exp\{i\mathbf{t}\trans {\bX}_j +isA_j \}$, $\varphi^n_{\bX,\bw}(\bt) =$ $\frac{1}{n}\sum_{j=1}^nw_j\exp\{i\mathbf{t}\trans {\bX}_j \}$, $\varphi^n_{A,\bw}(s) = \frac{1}{n}\sum_{j=1}^nw_j\exp\{isA_j\}$, and empirical characteristic functions $\varphi^n_{\bX,A}(\bt,s)$, $\varphi^n_{\bX}(\bt)$ and $\varphi^n_{A}(s)$ are defined accordingly.

\begin{thm}\label{thm:weighted_distance_duality}
Let $\bw = (w_1,\dots, w_n)$ be a vector of weights such that $\sum_{i=1}^nw_i = n$ and $w_i\geq 0$ for all $i=1,\dots,n$. Then $\calV^2_{n,\bw}(\bX,A) \geq 0$ and 
\begin{align}  
\calV^2_{n,\bw}(\bX,A)= {} & \int_{\bbR^{p+1}}\vert \varphi^n_{\bX, A, \bw}(\bt, s) - \varphi^n_{\bX,\bw}(\bt)\varphi^n_{A,\bw}(s) \nonumber \\ 
& \;\;\quad\quad + (\varphi^n_{\bX,\bw}(\bt) -\varphi^n_{\bX}(\bt))(\varphi^n_{A,\bw}(s) - \varphi^n_{A}(s)) \vert^2 \omega(\bt, s) \mathrm{d}\bt \; \mathrm{d}s \label{eqn:extrac} 
\end{align}
where $\omega(\bt, s) = (c_pc_1\lVert\bt\rVert_2^{1+p}| s|^2)^{-1}$ with $c_d = \frac{\pi^{(1+d)/2}}{\Gamma((1+d)/2)}$, and $\Gamma(\cdot)$ is the complete gamma function. 
\end{thm}

Based on \eqref{eqn:extrac}, it is clear that if $F^n_{\bX,A,\bw} = F^n_{\bX}F^n_{A}$, then $\calV^2_{n,\bw}(\bX,A)=0$. However, the converse is not necessarily true. Yet, if the weights preserve the marginal distribution of treatment and covariates, i.e. $F^n_{\bX,\bw} = F^n_{\bX}$ and $F^n_{A,\bw} = F^n_{A}$, then $\calV^2_{n,\bw}(\bX,A)=0$ implies that $F^n_{\bX,A,\bw} = F^n_{\bX}F^n_{A}$. In other words, if one can add additional terms to \eqref{eqn:weighted_distance} that also measure the distance between $F^n_{\bX,\bw}$ and $F^n_{\bX}$ along with that between $F^n_{A,\bw}$ and $F^n_{A}$, then \eqref{eqn:weighted_distance} can be utilized to construct a measure that determines the distance between $F^n_{\bX,A,\bw}$ and  $F^n_{\bX}F^n_{A}$, i.e., a measure for the weighted dependence between $\bX$ and $A$. We leverage the weighted energy distance proposed in \citet{huling2020energy} to help construct such a measure.

Applied to our setting, the weighted energy distance between $F^n_{\bX,\bw}$ and $F^n_\bX$ is
\begin{align*}
\quad \calE(F^n_{\bX,\bw}, F^n_{\bX}) \equiv {} & \frac{2}{n^2}\sum_{i = 1}^n \sum_{j = 1}^nw_i\lVert {\bX}_i  
-  {\bX}_j \rVert_2 - \frac{1}{n^2}\sum_{i = 1}^{n}\sum_{j = 1}^{n}w_i w_j \lVert  {\bX}_i - {\bX}_j\rVert_2 - \frac{1}{n^2}\sum_{i = 1}^{n}\sum_{j = 1}^{n}\lVert  {\bX}_i - {\bX}_j \rVert_2.
\end{align*}
Due to Proposition 1 of \citet{huling2020energy}, it can be shown that $\calE(F^n_{\bX,\bw}, F^n_{\bX}) = \int_{\bbR^p}\vert \varphi^{n}_\bX(\bt) - \varphi^n_{\bX,\bw}(\bt) \vert^2 \omega(\mathbf{t}) \mathrm{d}\mathbf{t}$, where $\omega(\mathbf{t}) = 1/(C_p\lVert\bt\rVert_2|^{1+p})$, $C_p = \pi^{(1+p)/2} / \Gamma((1+p)/2)$ is a constant. The weighted energy distance $\calE(F^n_{A,\bw}, F^n_{A})$ between $F^n_{A,\bw}$ and $F^n_A$ can be similarly defined.

Our proposed measure of the level of independence induced by a set of weights is defined as 
\begin{equation}\label{eqn:bal_criterion}
\calD(\bw) = \calV^2_{n,\bw}(\bX,A) + \calE(F^n_{\bX,\bw}, F^n_{\bX}) + \calE(F^n_{A,\bw}, F^n_{A}).
\end{equation}
The following result demonstrates that $\calD(\bw)$ indeed achieves its stated goal.
\begin{thm}\label{thm:weighted_distance_criterion_duality}
Let $\bw = (w_1, \dots, w_n)$ be such that $w_i>0$ and $\sum_{i=1}^nw_i = n$. Then $\calD(\bw) \ge 0$ with equality to zero if and only if $\varphi^n_{\bX, A, \bw}(\bt, s) = \varphi^n_{\bX}(\bt)\varphi^n_A(s)$, $\varphi^n_{\bX,\bw}(\bt) = \varphi^n_{\bX}(\bt)$, and $\varphi^n_{A,\bw}(s) = \varphi^n_{A}(s)$ for all $(\bt, s)\in \bbR^{p+1}$. Further,  $\int|\varphi^n_{\bX,A,\bw}(\bt,s) - \varphi^n_{\bX}(\bt)\varphi^n_{A}(s)|^2 \omega(\bt,s)\mathrm{d}\bt \mathrm{d}s \leq 3 \calD(\bw)$. 
\end{thm}
\noindent Thus, smaller values of $\calD(\bw)$ indicate smaller potential for dependence between $\bX$ and $A$ after weighting and better preservation of the marginal distributions, and larger values indicate the opposite. $\calD(\bw)=0$ implies the weights induce complete independence between $\bX$ and $A$ and that the marginal distributions of $\bX$ and $A$ are exactly preserved.

We also have the following result, which shows how the proposed distance acts as a bound on integration errors over a class of functions. 
\begin{lem}\label{thm:lem1}
Let $\calH$ be the native space induced by the radial kernel $\Phi(\cdot,\cdot) = -\| \cdot \|_2\times -| \cdot |$ on $\calX\times\calA$ equipped with inner product $\langle \cdot, \cdot \rangle_\calH$ and norm $\|g\|_\calH = \sqrt{\langle g, g \rangle_{\mathcal{H}}}$ for any $g(\cdot,\cdot) \in\calH = \calH_{\calX} \otimes \calH_{\calA}$, where $\calH_{\calX}, \calH_{\calA}$ are defined in Theorem 4 of \citet{mak2018support}. Then, for any weights $\bw$ satisfying $\sum_{i=1}^nw_i = n, w_i \ge 0$, we have
\begin{equation}\label{eqn:distance_bound}
\left[ \int_\calX\int_\calA g(\bx,a) \mathrm{d} \left[ F^n_{\bX, A, \bw} - F^n_\bX F^n_A  \right](\bx,a) \right]^2 \leq C_g \mathcal{D}(\bw),
\end{equation}
where $C_g=3\|g\|^2_\calH \geq 0$ is a constant depending on only $g$.
\end{lem}
For any $a\in \calA$ if $\mu(\cdot,a)\in\calH_{\calX}$, we can see that $\calD(\bw)$, modulo a constant, acts as a bound on the systematic bias term \eqref{eqn:error_decomp_indep_first} as long as $\mu(\cdot,a)$ is sufficiently smooth. Thus, if $\mu(\bx,a)$ is contained in $\calH$, then we can expect weights with smaller $\calD(\bw)$ to lead to smaller systematic bias. The space $\calH$ is a reasonably broad class of functions as it contains the Sobolev space of functions with square-integrable functions with $r < \lceil (p+1)/2\rceil$-th differentials \citep{mak2018support, huling2020energy}. 
Our goal for the next section is to define weights that are optimal in terms of our criterion. The weights that minimize $\calD(\bw)$ will result in mitigation of the dependence of $\bX$ and $A$ induced by nonrandom selection into treatment.

In contrast to the measure \eqref{eqn:bal_criterion} comprised of the distance covariance term \eqref{eqn:extrac}, it is natural to wonder whether it would instead be more appropriate to simply define a distance as $\int|\varphi^n_{\bX,A,\bw}(\bt,s) - \varphi^n_{\bX}(\bt)\varphi^n_{A}(s)|^2 \omega(\bt,s) \mathrm{d}\bt \mathrm{d}s$ and construct a relationship between this distance and Euclidean norms computable from data. However, we have found that the resulting quantity can be empirically problematic and unreliable.  Further, we have found that \textit{optimization} of such a quantity cannot be achieved reliably by existing algorithms and is thus not suitable for the proactive construction of weights. Our proposed quantity, while more complicated, does not exhibit any of these issues in the sense that it reliably measures dependence, and, as we will demonstrate, it is straightforward to optimize with existing quadratic programming software.

\subsection{A new proposal: distance covariance optimal independence weights}

We define the \textit{distance covariance optimal weights} (DCOWs) to be
\begin{align} 
\bw^d_n \in {} &  \argmin_{\bw = (w_1, \dots, w_{n})} \calD(\bw) \text{ such that }  \sum_{i=1}^nw_i = n, \text { and } w_i \geq 0 \text{ for } i=1,\dots,n. \label{eqn:dcows}
\end{align}

The name reflects that the weights are constructed as the optimizers of our distance-covariance-based criterion.
Due to Theorem \ref{thm:weighted_distance_criterion_duality}, the DCOWs $\bw^d_n$ are designed to minimize dependence between $\bX$ and $A$ on the weighted scale while keeping the weighted marginal distributions of $\bX$ and $A$ close to those of the unweighted data. The constraint $\sum_{i=1}^nw_i = n$  ensures that $F^n_{\bX,A,{\bw^d_n}}$ is a valid distribution function. Since $\calD(\bw)$ tracks with the dependence induced by a set of weights, the DCOWs can be thought of as \textit{optimal independence weights} (i.e., optimal with respect to achieving independence between the treatment and covariates).

Although, as we will show later, the DCOWs result in consistent weighted dose-response estimators, they may not guarantee optimal convergence rates without additional constraints. Instead, a small change to our criterion to include penalization of the squares of the weights provides control of the variability of the weights without sacrificing their bias-reduction property. This additional penalty is akin to focusing more directly on mean squared error instead of bias and can be interpreted as discouraging the effective sample size (ESS) after weighting from being too small. In particular, the ESS is typically approximated as $(\sum_iw_i)^2/\sum_iw_i^2$ \citep{kish1965survey}, which, due to our constraints, is $n^2/\sum_iw_i^2$ and is precisely the inverse of our proposed penalty. Further, combining a ``balance'' criterion with a means of weight variability mitigation is in line with the recommendations of \citet{chattopadhyay2020balancing}.

We now define the \textit{penalized distance covariance optimal weights} (PDCOWs) to be
\begin{align} 
\bw^{pd}_n \in {} &  \argmin_{\bw = (w_1, \dots, w_{n})} \calD(\bw)+\lambda\frac{1}{n^2}\sum_{i=1}^nw_i^2 \nonumber \\ & \text{ such that }  \sum_{i=1}^nw_i = n, w_i \geq 0 \text{ for } 0<\lambda<\infty, i=1,\dots,n. \label{eqn:pdcows}
\end{align}
Here, the tuning parameter $\lambda$ is any positive constant and can be chosen by the user to achieve a desired ESS. A lemma provided in Section \ref{supsec:more_theory} of the Supplementary Material similar to Lemma \ref{thm:lem1} shows that the penalized version of our criterion acts as a bound on the term in the left-hand side of \eqref{eqn:distance_bound} plus the squares of the weights, which is more akin to a bound on the root mean squared error than bias as in Lemma \ref{thm:lem1}.
Although having a non-zero, positive value of $\lambda$ is necessary for the convergence rate guarantee of Theorem \ref{thm:thm3}, in practice we have found that minimal or even no penalization at all works well because the unpenalized weights, the DCOWs, tend to be quite stable. In all analyses described later, we use only the DCOWs with no weight penalization. 

Both the DCOWs and PDCOWs can be used in a wide variety of estimators for various causal estimands involving continuous treatments, not just Nadaraya-Watson-based estimators and not just estimators of the ADRF.
The weights can be used either in a simple weighted nonparametric estimator of the dose-response function or to supplement any doubly-robust estimator of such. For the former, our weights provide a fully nonparametric and empirically robust estimation approach that requires only mild moment conditions on the covariates and treatment for estimation consistency. For the latter, such an estimator using our weights is guaranteed to be consistent regardless of the correctness of the outcome model, while it still enjoys efficiency gains if the outcome model is well-specified.

\begin{remark}\label{remark3}
The optimization problems \eqref{eqn:dcows} and \eqref{eqn:pdcows} can be formulated as quadratic programming problems with linear constraints, making them straightforward to implement with commercial and open-source solvers such as OSQP \citep{stellato2020osqp}. If in practice additional emphasis on correlations of particular moments of covariates and the treatment is of importance, our framework can accommodate that by adding additional linear constraints. The details are deferred to Sections \ref{sec:computation} and \ref{sec:extensions} of the Supplementary Material.
\end{remark}
 
In a later section, we provide more formal statements on the consistency of dose-response function estimators that use our proposed distance covariance optimal weights.

\subsection{Asymptotic properties}
\label{sec:theory}

We first show that the distance covariance optimal weights do induce complete independence asymptotically. Throughout this section we define $w^*(\bx,a) = f_A(a)/f_{A|\bX}(a|\bX=\bx)$ to be the ``true'' normalized density weights. 
\begin{thm}\label{thm:weighted_distance_convergence}
Let $\bw^d_n$ be the distance covariance optimal weights defined in \eqref{eqn:dcows}. Then if $\bbE\lVert \bX \rVert_2 < \infty$ and $\bbE\lvert A \rvert < \infty$ we have
\begin{equation}\label{eqn:ecdf_convergence}
    \lim_{n \to \infty} F^n_{\bX, A, \bw^{d}}(\bx, a) = F_{\bX}(\bx)F_{A}(a)
\end{equation}
almost surely for every continuity point $(\bx, a)\in \bbR^{p+1}$ and further that $\lim_{n \to \infty} F^n_{\bX,\bw^{d}}(\bx) = F_{\bX}(\bx)$ for every continuity point $\bx\in \bbR^{p}$ and $\lim_{n \to \infty} F^n_{A, \bw^{d}}(\bx, a) = F_{A}(a)$ for every continuity point $a\in \bbR$. If, additionally $\bbE {w^*}^2(\bX,A) < \infty$ holds, then the same result holds for $\bw^{pd}$.
\end{thm}
Theorem \ref{thm:weighted_distance_convergence} is in some sense the most important property of the DCOWs, as it demonstrates the feasibility of using these weights not just in estimation of the ADRF, but also in the estimation of many causal estimands that require independence. In particular, if the source of confounding bias has the form $\int_{\calX} \int_\calA g(\bx, a) \mathrm{d} \left[ F^n_{\bX,A,\bw} - F^n_{\bX}F^n_{A} \right](\bx,a)$ for some function $g\in \calH$, then the use of our weights can be justified due to Lemma \ref{thm:lem1} and Theorem \ref{thm:weighted_distance_convergence}.

We now show that for the particular task of estimating the ADRF, using the DCOWs in a weighted Nadaraya-Watson estimator results in consistent estimation of the ADRF. 

\begin{thm}\label{thm:thm3}
Assume that the kernel $K(\cdot)$ is symmetric, second order, i.e. it meets the conditions that  $\int uK(u)\mathrm{d}u=0$, $\int K(u)\mathrm{d}u=1$, and $0 < \int u^2K(u)\mathrm{d}u < \infty$, and is bounded differentiable. Further, assume that the moment conditions required in Theorem \ref{thm:weighted_distance_convergence} hold and that $\mu(\bx,a_0)$ and is bounded and continuous on $\calX\times\calA$ and has second order derivatives, $f_A(a_0)$ is bounded and has second order derivatives, $1/f_A(a_0)$ is uniformly bounded. When $h \to 0$, $nh \to \infty$, then for $\bw=\bw^d \text{ and } \bw=\bw^{pd}$
\begin{equation}
\lim_{ n \to\infty} \widehat{\mu}_{NW}^{\bw}(a_0) = \lim_{ n \to\infty} \widehat{\mu}_{NWs}^{\bw}(a_0) = \mu(a_0) 
\end{equation}
in probability for all continuity points $a_0 \in\calA$. 
\end{thm}
Thus, both the DCOWs and PDCOWs result in consistent estimation of the causal ADRF using either the stabilized or unstabilized estimator. It can also be  shown (Supplementary Material Section F.3) that $\widehat{\mu}_{NW}^{\bw,DR}(a_0)$ is still consistent even if $\widehat{\mu}(\bx,a_0)$ is inconsistent for $\mu(\bx,a_0)$ as long as $\widehat{\mu}(\bx,a_0)$ converges to any finite function uniformly almost surely. 

\begin{remark}\label{remark2}
Our distance metric has some relationship with maximum mean discrepancy based distances and the kernel-based independence test via the results in \citet{sejdinovic2013equivalence}, where our distance induces a particular kernel $\Phi(\cdot,\cdot)$, defined in our Lemma \ref{thm:lem1}. However, despite this connection, our Theorem \ref{thm:thm3} does not require the response function $\mu(\bx,a_0)$ to be in the native space induced by $\Phi(\cdot,\cdot)$. Thus, while our method has some connection with kernel-based distances, our weights result in consistent estimation of the ADRF without correct specification of the kernel $\Phi(\cdot,\cdot)$.
\end{remark}

The following shows the convergence rate of the penalized distance covariance optimal weights under additional mild moment conditions on the covariates, treatment, and $w^*(\bX,A)$. This theorem builds on a key lemma on the squares of the weights presented in Section \ref{supsec:more_theory} of the Supplementary Material.
\begin{thm}\label{thm:thm4}
Assume the conditions required in Theorem \ref{thm:thm3} hold, that the moment conditions (A1) and (A2) listed in the Appendix hold, and that $\bbE {w^*}^2(\bX,A) < \infty$. Additionally assume $\bbE[ \varepsilon^2\:|\:\bX = \bx, A = a_0] < c$ for some $c$ uniformly over $\bx\in\calX$. Then
\begin{equation}
\widehat{\mu}_{NW}^{\bw^{pd}}(a_0) - \mu(a_0) = O_p(1/\sqrt{nh}+h^2).
\end{equation}
\end{thm}

This convergence rate is the standard rate for unweighted Nadaraya-Watson estimators of a univariate regression function; thus, the convergence rate of the weighted estimator based on our weights is unaffected by the nonparametric nature of the estimation of $\bw^{pd}$.

\subsection{Augmented estimation with independence weights}
\label{sec:drest}

Another class of estimators for causal ADRFs are doubly-robust/augmented estimators such as in \citet{kennedy2017non} and \citet{colangelo2020double}, which combine sample weights and an outcome model $\widehat{\mu}(\bx,a_0)$, ideally a consistent estimator of ${\mu}(\bx,a_0)$. Though the term ``doubly-robust'' is reflective of the property that if either the weights or outcome model is correctly specified, the estimator will be consistent, a more consequential property of doubly-robust estimators is that the error rates of each model are multiplied. 

Although in the previous section we showed that the use of the DCOWs alone results in the ideal convergence rate for the ADRF, DCOWs can be enhanced by using doubly-robust/augmented estimators, or conversely, that the use of DCOWs can significantly enhance doubly-robust estimators. The DCOWs assure the analyst that the estimator will converge at the right rate regardless of whether the outcome model is correctly specified, but allow for using  an outcome model to provide an opportunity to fine-tune performance by reducing residual variance, resulting in an estimator that works well empirically. 

Here, for simplicity of presentation, we focus on the following Nadaraya-Watson-based augmented estimator based on any estimator $\hat{\mu}(\bx,a_0)$ of $\mu(\bx,a_0)$ as
\begin{equation*}
    \widehat{\mu}_{NW}^{\bw,DR}(a_0) = \frac{1}{n}\sum_{i=1}^n\hat{\mu}(\bX_i,a_0) + \frac{\sum_{i=1}^n (Y_i - \hat{\mu}(\bX_i,a_0)) w_i K_h(A_i - a_0)}{ \sum_{i=1}^n K_h(A_i - a_0)}.
\end{equation*}
In Section \ref{sec:appendix_b} of the Supplementary Material, we derive a decomposition of the error $\widehat{\mu}_{NW}^{\bw,DR}(a_0) - \mu(a_0)$ and show that the systematic bias term not related to smoothing is 
\begin{equation}
    \int_{\calX}  \left\{\mu(\bx, a_0) - \widehat{\mu}(\bx, a_0)\right\} \int_\calA \mathrm{d} \left[ F^n_{\bX,A,\bw} - F^n_{\bX}F^n_{A} \right](\bx,a). \label{eqn:doubly_robust_bias}
\end{equation}
Lemma \ref{thm:lem1} implies that \eqref{eqn:doubly_robust_bias} is less than or equal to $3\|\mu(\bx,a_0)-\widehat{\mu}(\bx,a_0)\|_\calH\calD(\bw)$ provided that $\mu(\cdot,a_0)-\widehat{\mu}(\cdot,a_0) \in \calH$.
With the DCOWs, we provide a set of weights $\bw^d$ that makes $\calD(\bw)$ as small as possible, though it may not be exactly zero for a finite sample.

We now formalize the above claims and provide asymptotic results for the augmented estimator $\widehat{\mu}_{NW}^{\bw,DR}(a_0)$ using a slightly modified version of our PDCOWs; this modification is motivated by a technical condition and in practice has little or no impact on the weights. The modified penalized distance covariance optimal weights are defined as
\begin{align} 
\widetilde{\bw}^{pd}_n \in {} &  \argmin_{\bw = (w_1, \dots, w_{n})} \calD(\bw)+\lambda\frac{1}{n^2}\sum_{i=1}^nw_i^2 \nonumber \\ & \text{ such that }  \sum_{i=1}^nw_i = n, Bn^{1/3} \geq w_i \geq 0 \text{ for } 0<\lambda,B<\infty, i=1,\dots,n. \label{eqn:pdcows_mod}
\end{align}
Here, $B$ is a pre-specified positive constant. We have found that in practice the maximum weight rarely, if ever, comes near $Bn^{1/3}$ with $B=1$ even without the constraint on the max weight. Thus, this additional constraint does little to change the empirical behavior of the PDCOWs. Further, we show in Section Section \ref{supsec:more_theory} of the Supplementary Material and in the following that the key asymptotic results of the PDCOWs (e.g. Theorem \ref{thm:weighted_distance_convergence}) also hold for $\widetilde{\bw}^{pd}_n$. 

We next show that the modified weights paired with an augmented estimator based on a correctly-specified outcome model result in asymptotic normality of the resulting causal ADRF. 

\begin{thm}\label{thm:asymptotic_normality}
Let $\widetilde{\bw}^{pd}_n$ be the distance covariance optimal weights defined in \eqref{eqn:pdcows_mod}. Let $K(\cdot)$ be a kernel with conditions listed in the statement of Theorem \ref{thm:thm3}. Assume the moment conditions (A1) and (A2) listed in the Appendix hold, that $\bbE\lVert \bX \rVert_2 < \infty$ and $\bbE\lvert A \rvert < \infty$, that $\bbE {w^*}^2(\bX,A) < \infty$, that $1/f_A(a_0)$ is uniformly bounded, $f_A(a_0)$ is bounded and has second order derivatives, and that $\mu(\bx,a_0)$ is bounded and continuous on $\calX\times\calA$ and has second order derivatives. Further, assume that $\bbE|\varepsilon_i|^3<\infty$ and $\bbE[\varepsilon^2_i] = \sigma ^ 2 < \infty$ for all $i$. Assume that the outcome regression model satisfies $\mu(\cdot,a_0)-\hat{\mu}(\cdot,a_0) \in \calH_{\calX}$ for each $n$, $||\mu-\hat{\mu}||_\calH = O_p(1)$,  and $\int_{\calX} (\mu(\bx, a_0) - \hat{\mu}(\bx, a_0))^2 \mathrm{d}F_{\bX}(\bx) = o_p(1)$. Then 
\begin{equation}\label{eqn:asympt_distr}
    \frac{\sqrt{nh}f_A(a_0)}{\sigma \sqrt{\frac{1}{n}\sum_{i=1}^nw_i^2K^2_h(A_i-a_0)}}\left(\hat \mu_{NW}^{\bw,DR}(a_0) - \mu(a_0) - h^2\kappa_2B(a_0)\right) \xrightarrow{d} \calN(0,1)
\end{equation}
as $h\rightarrow 0$, $nh\rightarrow\infty$, and $nh^5=O_p(1)$  for $\bw = \widetilde{\bw}^{pd}_n$, where $\kappa_2 = \int u^2K(u)\mathrm{d}u$, $B(a_0) \equiv \int_\calX B(\bx,a_0)\mathrm{d}F_\bX(\bx)$, and $B(\bx,a_0) \equiv \frac{\partial^2}{\partial a_0^2}\mu(\bx,a_0)/2 + \frac{\partial}{\partial a_0}\mu(\bx,a_0)\frac{\partial}{\partial a_0}f_A(a_0) / f_A(a_0).$
\end{thm}
The conditions required regarding $\hat{\mu}(\cdot,a_0)$ and $\varepsilon_i$ are analogous to those required in Theorem 3 of \citet{wong2017kernel}. We note that normalization by the squares of the weights in \eqref{eqn:asympt_distr} is necessary as our results do not rely on a proof of $\widetilde{\bw}^{pd}_n$ or their squares to converge to anything in particular. We note, however, that the expression in \eqref{eqn:asympt_distr} can be simplified if it is possible to show that $\frac{1}{n}\sum_{i=1}^n\left(\widetilde{w}_i^{pd} - w^*(\bX_i, A_i)\right)^2$ converges to 0 in probability. In particular, it would simplify to a form similar to the asymptotic distribution of the augmented ADRF estimator in \citet{colangelo2020double}. We provide an informal investigation into the convergence of the PDCOWs to the true GPS weights in Section \ref{supsec:wt_convergence_sim} of the Supplementary Material.




\section{Numerical experiments}
\label{sec:simulations}

We evaluate our proposed methodology using two tracks of simulation experiments. The first track utilizes existing data to conduct simulation studies. In this approach, we fix the confounding structure of a complex dataset and simulate outcomes under a wide range of outcome models. In the second track of simulation experiments, we generate synthetic data under the data-generating setup of \cite{vegetabile2020nonparametric}, which amounts to a markedly different data-generating process from the first track of simulation experiments.

\subsection{Comparator methods}

We use the following methods to estimate weights. We use a na\"{i}ve method which uses weights equal to identity (unweighted). We use stabilized GPS weights computed four ways: a linear regression model for estimating $\bbE(A|\bX)$ (i.e., the conditional mean of dose given covariates) and a normal conditional density (``GPS normal''); a gamma regression model for estimating $\bbE(A|\bX)$ and a gamma conditional density (``GPS gamma''); a generalized boosted model for estimating $\bbE(A|\bX)$, where the number of trees was chosen to minimize the weighted average absolute correlation between the treatment and covariates as in \citet{zhu2015boosting}, and a normal conditional density (``GBM''); and a Bayesian Additive Regression Trees model for estimating $\bbE(A|\bX)$ and a normal conditional density (``BART''). For methods that estimate weights by directly inducing a lack of correlation between moments of the treatment and covariates, we use the covariate balancing generalized propensity score of \citet{fong2018covariate} and the entropy balancing approach of \citet{tubbicke2020entropy} and \citet{vegetabile2020nonparametric}. Among this class of methods, we only consider these two as other approaches to estimating weights that decorrelate pre-specified moments behave largely similarly to each other \citep{tubbicke2020entropy, vegetabile2020nonparametric}. We use the exactly-identified version of the covariate balancing generalized propensity score (``CBPS''). We use versions of the entropy balancing approach that decorrelate either  all first order moments (``Entropy (1)''), all first order moments and squared terms in continuous covariates (``Entropy (2)''), or all first order moments, pairwise interactions, and squared terms in continuous covariates (``Entropy (int)''). Any resulting weights greater than 500 when standardized to sum to $n$ are truncated at 500. We use our proposed DCOWs (``DCOW'') and the proposed DCOWS where we further induce exact decorrelation of first order moments (``DCOW (dm)'') as discussed in Section \ref{sec:extensions} of the Supplementary Material, both using the dimension adjustment described also in that section. For each method, the weights are used in a weighted local linear regression used to estimate the ADRF. For the GPS (normal), GPS (gamma), DCOW, and DCOW (dm) methods, we also use the doubly-robust estimator of \citet{kennedy2017non} with a an outcome model that is linear in the covariates with additive first order terms; the methods are labeled as ``GPS (normal,DR)'', ``GPS (gamma,DR)'', ``DCOW (DR)'', and ``DCOW (dm,DR)'', respectively.

\subsection{Simulation using National Medical Expenditure Survey Data}

The National Medical Expenditure Survey (NMES) relates medical expenditures with degree of smoking among U.S. citizens \citep{johnson_disease_2003}. The NMES dataset contains information on 9708 individuals. The outcome variable is the total medical expenditures in dollars and the treatment $A \in [0.05, 216]$ is the amount of smoking in pack years. We limit the data to those with $A \leq 80$ (i.e. $\calA=[0.05, 80]$), leaving 9368 units. We limit to those with such values, as the number of patients who smoke more than $80$ pack years is exceptionally rare. Two of the covariates are continuous and the remaining are categorical, with an overall dimension equal to $18$ after converting categorical variables to dummy variables. In our simulation, we leave the treatment level (pack years) and covariates intact and simulate outcomes for each unit from the following model:
    $Y = {m}(\bX; \btheta_1) + f(A) (1 + \delta(\bX; \btheta_2))) + \varepsilon,$
where ${m}(\bX; \btheta_1)$ are main effects with parameters $\btheta_1$ generated as described in Section \ref{supsec:main_sim} of the Supplementary Material, ${\delta}(\bX; \btheta_2)$ are mean 0 interaction effects, $\varepsilon$ are i.i.d. $N(0,4)$ random variables, and the treatment effect curve $f(A) = A/4 + \frac{2}{(A/100 + 1/2)^3} - (A-40)^2/100$. The main effect function involves interactions and up to squared terms in the continuous covariates. For the constant treatment effect setting, $\delta(\bX; \btheta_2) = 0$, and for the heterogeneous treatment effect setting it is a nonzero but mean 0 function involving linear and interaction effects in covariates. Notably, the parameters in $\btheta_1$ and $\btheta_2$ also make some covariates have no contribution to the main effects and interactions, respectively.
We generate 100 different draws of the coefficients ($\btheta_1$ and $\btheta_2$) in the outcome model above, allowing for the simulation study to explore a wide variety of outcome models. For each of these 100 outcome model draws, we replicate the simulation experiment 1000 times. 
For each replication, a random subsample of size $n<9368$ is drawn without replacement from the 9368 units, and outcomes for these units are generated from the outcome model. The simulation process is repeated for each of the sample sizes $n = 100, 200, 400, 800, 1600, 3200$.

We also consider a simulation setting with the heterogeneous treatment effect where 50 additional variables are added to the covariate vector so that the overall dimension is 68. The 50 additional variables are generated so that they are correlated with both the response and treatment while preserving the original values of the response and treatment; details are provided in Section \ref{supsec:main_sim} of the Supplementary Material.

All methods are evaluated with a measure of the mean absolute bias (MAB) and integrated root mean squared error (IRMSE), both of which are used for evaluation of estimates of the ADRF in \citet{kennedy2017non}. MAB and IRMSE are defined as
\begin{align*}
    \text{MAB} = \int_\calA \left\lvert \frac{1}{S}\sum_{s=1}^S \widehat{\mu}_s(a) - \mu(a) \right\rvert \widehat{f}_A(a)\mathrm{d}a,
    \text{IRMSE} = \int_\calA \left[ \frac{1}{S}\sum_{s=1}^S \left\{ \widehat{\mu}_s(a) - \mu(a)\right\}^2 \right]^{1/2} \widehat{f}_A(a)\mathrm{d}a,
\end{align*}
where $\widehat{f}_A(a)$ is a kernel density estimate of the marginal density of the treatment variable, $s$ indexes the simulation replications, and $\calA$ is a trimmed version of the support of the treatment variable that excludes pack years greater than 80. We calculate the MAB and IRMSE statistics for each of the 100 different outcome model settings and then average them over the 100 settings. 

The MAB and IRMSE results for the constant treatment effect setting are displayed in Table \ref{tab:nmes_const_tx_simulation_10000reps} and the results for the heterogeneous treatment effect setting are displayed in 
Section \ref{supsec:main_sim} of the Supplementary Material as they track closely with the former setting. 
Results for any method are not shown if no numerical solution is found in more than 75\% of the replications.
The standard DCOWs performed the best in terms of both MAB and IRMSE across all sample sizes among all non-doubly-robust estimators, only being outperformed by the doubly-robust estimator that uses DCOWs as weights. DCOWs with exact first order moment decorrelation performed similarly to, but slightly worse than standard DCOWs, for both the non-doubly-robust estimator and the doubly-robust estimator, though the performance was much worse for $n=100$, as the exact constraints may have been too stringent for the sample size. 
The estimators using standard GPS weights, both doubly-robust and non-doubly-robust, performed poorly in terms of both MAB and IRMSE for small to moderate sample sizes, though the doubly-robust estimator with gamma regression-based weights performed well in terms of MAB and IRMSE for larger sample sizes. The machine learning approaches to GPS estimation (GBM and BART) performed relatively poorly in terms of MAB and IRMSE for small sample sizes, with BART performing better than GBM for small sample sizes, but with similar but slightly better performance for GBM for large sample sizes. Among the GPS-based methods, GBM and BART generally yielded the lowest MAB and IRMSE across the sample size settings, except for the doubly-robust gamma GPS estimator. 
Among the moment balancing approaches, entropy balancing with decorrelation induced only for first and second order moments and not interactions performed the best in terms of MAB and IRMSE, though for small sample sizes, entropy balancing frequently failed to arrive at a solution. For the heterogeneous setting, entropy balancing with decorrelation induced only for first order moments performed the best, likely due to numerical instability of adding more moment constraints. The CBPS estimator yielded worse performance than did entropy balancing overall, though it did not face the convergence problems of entropy balancing for small sample sizes.
Entropy balancing with higher order moment constraints failed to arrive at a solution more frequently and performed poorly even for large sample sizes, likely due to the strictness of the exact moment constraints. 

\begin{table}[ht]
    \centering
    \begin{adjustbox}{max width=0.95\textwidth}
    \begin{tabular}{lrrrrrrrrrrrr}
    \toprule
     & \multicolumn{2}{c}{$n=100$} & \multicolumn{2}{c}{$n=200$} & \multicolumn{2}{c}{$n=400$} & \multicolumn{2}{c}{$n=800$} & \multicolumn{2}{c}{$n=1600$} & \multicolumn{2}{c}{$n=3200$} \\ \cmidrule(lr){2-3}\cmidrule(lr){4-5}\cmidrule(lr){6-7}\cmidrule(lr){8-9}\cmidrule(lr){10-11}\cmidrule(lr){12-13}
     Method & MAB & IRMSE & MAB & IRMSE & MAB & IRMSE & MAB & IRMSE & MAB & IRMSE & MAB & \multicolumn{1}{r}{IRMSE} \\ 
    \midrule
    Unweighted  & 11.461 & 17.488 & 11.283 & 14.753 & 11.252 & 13.182 & 11.246 & 12.297 & 11.254 & 11.799 & 11.237 & 11.491 \\
    GPS (normal)  & 10.145 & 27.180 & 15.033 & 29.187 & 18.197 & 28.056 & 20.786 & 27.650 & 21.832 & 25.703 & 22.611 & 24.432 \\
    GPS (gamma)  & 9.340 & 23.994 & 9.892 & 19.269 & 9.993 & 15.831 & 10.112 & 13.978 & 10.084 & 12.693 & 10.006 & 11.162 \\
    GPS (normal,DR)  & 8.315 & 30.551 & 11.081 & 33.342 & 13.216 & 30.293 & 15.320 & 27.636 & 15.324 & 21.199 & 15.769 & 18.312 \\
    GPS (gamma,DR)  & 7.045 & 26.437 & 6.866 & 16.463 & 6.905 & 12.044 & 6.920 & 10.168 & 6.768 & 8.985 & 6.745 & 7.612 \\
    CBPS  & 9.871 & 25.398 & 12.463 & 23.375 & 14.055 & 22.016 & 14.064 & 18.802 & 14.970 & 17.546 & 15.712 & 16.847 \\
    GBM  & 8.190 & 23.428 & 7.551 & 17.298 & 7.105 & 12.498 & 7.379 & 10.432 & 7.524 & 9.225 & 7.746 & 8.589 \\
    BART  & 6.550 & 17.206 & 7.275 & 14.423 & 8.023 & 12.326 & 8.638 & 11.292 & 9.033 & 10.659 & 9.446 & 10.366 \\
    Entropy (1)  & ------ & ------ & 9.409 & 17.880 & 9.166 & 12.946 & 9.161 & 11.321 & 9.023 & 10.258 & 8.952 & 9.537 \\
    Entropy (2)  & ------ & ------ & 8.909 & 25.080 & 9.343 & 15.520 & 8.927 & 12.360 & 8.503 & 10.456 & 8.231 & 9.245 \\
    Entropy (2,int)  & ------ & ------ & ------ & ------ & ------ & ------ & ------ & ------ & 8.998 & 253.786 & 10.974 & 16.780 \\
    DCOW  & 4.684 & 12.204 & 3.866 & 8.383 & 3.495 & 6.245 & 3.252 & 4.947 & 2.992 & 4.075 & 2.750 & 3.416 \\
    DCOW (dm)  & 4.522 & 16.465 & 3.907 & 8.988 & 3.544 & 6.364 & 3.276 & 5.018 & 3.001 & 4.147 & 2.775 & 3.497 \\
    DCOW (DR)  & \textbf{3.904} & \textbf{9.284} & \textbf{3.335} & \textbf{6.455} & \textbf{2.663} & \textbf{4.586} & \textbf{2.196} & \textbf{3.388} & \textbf{1.919} & \textbf{2.661} & \textbf{1.753} & \textbf{2.189} \\
    DCOW (dm,DR)  & 3.905 & 11.567 & 3.459 & 6.919 & 2.828 & 4.806 & 2.356 & 3.568 & 2.057 & 2.817 & 1.873 & 2.320 \\
    \bottomrule 
    \end{tabular}
    \end{adjustbox}
    \caption{Mean absolute bias (MAB) and integrated root mean squared error (IRMSE) for the constant treatment effect setting.}
    \label{tab:nmes_const_tx_simulation_10000reps}
\end{table}

The results for the heterogeneous effect setting with 50 additional noise variables are displayed in Section \ref{supsec:main_sim} of the Supplementary Material. Again, DCOWs performed the best among all methods in terms of both MAB and IRMSE for all sample sizes except $n=100$, where the doubly-robust gamma GPS estimate was slightly better in terms of IRMSE.

Additional simulation results under the same setting of \cite{vegetabile2020nonparametric} can be found in Section \ref{supsec:veg_sim} of the Supplementary Material, wherein our weights also yielded the best performance across all sample size settings.


\section{Analysis of mechanical power data}
\label{sec:data}

We use the Medical Information Mart for Intensive Care III (MIMIC-III) database \citep{johnson2016mimic} to study the impact of a large degree of mechanical power of ventilation on mortality among critically ill patients in an ICU using electronic health record (EHR) data. 
Our study and the construction of the cohort from the MIMIC-III database is based on the original study of \citet{neto2018mechanical} and the code provided by the authors located at \url{https://github.com/alistairewj/mechanical-power}. Since there are widely-used formal guidelines that influence ventilation management among patients with respiratory distress \citep{papazian2019formal}, many observable factors in the EHR data are highly related to the mechanical power of ventilation. Many of these factors are also closely related to patient mortality. The guidelines involve consideration of both factors individually and many interactions among these factors. In addition, as the exposure, the mechanical power of ventilation, is itself a complex summary of multiple manipulable elements of a ventilator, the guidelines are not directly related to the exposure, but have a strong indirect effect on the power of ventilation. Thus, the dependence between observable factors and the exposure level is highly complex.

Patients included in the study were at least 16 years of age and received invasive mechanical ventilation for at least 48 hours. This restriction was utilized in the original analysis of \citet{neto2018mechanical} and focused on the population that requires extended use of ventilation. Patients who die within this 48-hour period are substantially sicker, and it is unclear whether manipulation of ventilation settings for such a population may induce meaningful changes to outcomes.
The study contains 5014 patients, and the treatment variable of interest is the amount of energy generated by the mechanical ventilator measured by the mean of the largest and smallest mechanical power of ventilation in Joules per minute in the second 24-hour period in the ICU, as in \citet{neto2018mechanical}. The outcome is an indicator of in-hospital mortality. As in the original analysis of \citet{neto2018mechanical}, we limit our analysis to patients receiving less than or equal to 50 Joules per minute, resulting in a sample size of 4933. We include $73$ pre-treatment covariates (some of which are discrete), resulting in a total dimension of $97$ of the vector $\bX$ of potential confounders. All methods considered in the simulation section were then applied to construct weights aiming to control for dependence between the $97$ covariates and mechanical power. Both GPS approaches, GBM, and BART resulted in several extraordinarily large weights; these weights were truncated to mitigate extreme variation. We attempted to construct entropy balancing weights that either exactly or approximately (within a correlation tolerance of $0.1$) balance pairwise interactions, but no numerical solution was found. 

\begin{figure}[ht]
\centering
\includegraphics[width=0.8\textwidth]{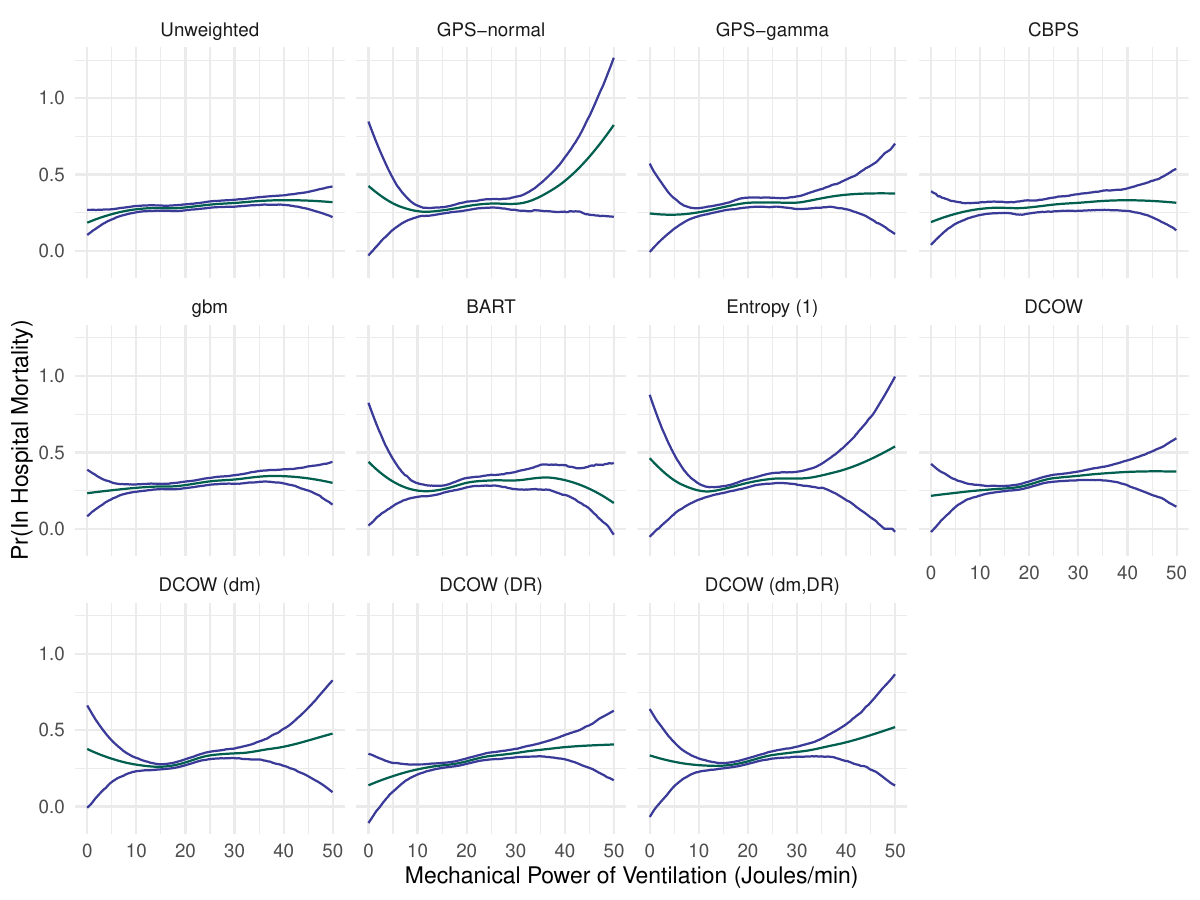}
\caption{\label{fig:bootstrap_mech_power} Shown are weighted local linear regression estimates of the ADRF of mechanical power on the probability of in-hospital mortality and pointwise 95\% confidence bands estimated via a nonparametric bootstrap. 38 replications of the bootstrap resulted in no numerical solution for entropy balancing weights and are left out of the confidence interval calculation.}
\end{figure}

For all methods, balance statistics, including our developed criterion \eqref{eqn:bal_criterion} and weighted correlations between first-order moments and pairwise interactions of covariates and mechanical power, are summarized in Table \ref{tab:mech_power_balance}. In terms of weighted marginal correlations, the version of our DCOWs that induces first-order marginal correlations between covariates and mechanical power to be zero performs the best, with the standard DCOWs and entropy balancing weights a close second. In terms of our proposed criterion \eqref{eqn:bal_criterion}, by definition, the DCOWs yield the smallest value. However, it is notable that only a small price is paid in terms of both \eqref{eqn:bal_criterion} and effective sample size (ESS) in order to exactly decorrelate marginal covariate moments and treatment. Among methods that do not target independence, entropy balancing has the smallest value of \eqref{eqn:bal_criterion}, indicating that decorrelating first-order moments in this particular dataset mitigates a vast majority of the dependence between covariates and mechanical power of ventilation. We also note that exactly decorrelating second-order moments using entropy balancing results in further instability, and thus worse mitigation of dependence. However, exact decorrelation of marginal covariate moments via entropy balancing results in much larger standard errors of the resulting ADRF, as evidenced in Figure \ref{fig:bootstrap_mech_power}. Interestingly, in terms of both marginal weighted correlations and our proposed independence metric, the flexible machine learning approaches (GBM and BART) perform significantly worse than a normal model for the conditional density. 

\begin{table}[ht]
\centering
\resizebox{0.8\textwidth}{!}{%
\begin{tabular}{rrrrrrrrrrr}
  \toprule
 & Unweighted & \specialcell[t]{GPS\\(normal)} & \specialcell[t]{GPS\\(gamma)} & CBPS & GBM & BART & \specialcell[t]{Entropy\\(1)} & \specialcell[t]{Entropy\\(2)} & DCOW & \specialcell[t]{DCOW\\(dm)} \\ 
\midrule
\eqref{eqn:bal_criterion} & 21.041 &  2.574 &  4.269 &  4.713 & 12.992 &  6.166 &  1.093 &  4.262 &  0.220 &  0.237 \\ 
  $\calD(\bw)$ & 21.041 &  2.349 &  4.014 &  4.328 & 12.960 &  5.710 &  1.005 &  4.083 &  0.144 &  0.151 \\ 
  $\calE(F^n_{A,\bw}, F^n_{A})$ & 0.000 & 0.027 & 0.003 & 0.010 & 0.003 & 0.043 & 0.030 & 0.001 & 0.051 & 0.059 \\ 
  $\calE(F^n_{\bX,\bw}, F^n_{\bX})$ & 0.000 & 0.198 & 0.251 & 0.374 & 0.029 & 0.414 & 0.058 & 0.178 & 0.026 & 0.027 \\ 
  ESS & 4933 & 2048 & 1832 & 1017 & 3892 &  782 & 2185 &  478 & 2232 & 2057 \\ 
  \midrule
    mean($|$Corr$|$) & 0.070 & 0.014 & 0.037 & 0.025 & 0.058 & 0.045 & 0.007 & 0.021 & 0.007 & 0.003 \\ 
  sd($|$Corr$|$) & 0.082 & 0.015 & 0.036 & 0.025 & 0.061 & 0.043 & 0.008 & 0.022 & 0.006 & 0.003 \\ 
  median($|$Corr$|$) & 0.041 & 0.010 & 0.026 & 0.018 & 0.036 & 0.035 & 0.005 & 0.016 & 0.005 & 0.002 \\ 
  95-pctl($|$Corr$|$) & 0.248 & 0.044 & 0.105 & 0.063 & 0.200 & 0.118 & 0.020 & 0.055 & 0.018 & 0.009 \\ 
  max($|$Corr$|$) & 0.788 & 0.200 & 0.854 & 0.470 & 0.568 & 0.939 & 0.123 & 0.507 & 0.084 & 0.045 \\ 
   \bottomrule
\end{tabular}
}
\caption{Summary statistics (mean, standard deviation, median, 95th percentile, and maximum) of the absolute weighted correlations of the first five powers of mechanical power of ventilation and all marginal moments of covariates, pairwise interactions of covariates, and up to 5th order polynomials of covariates. Summaries are over $8284\times 5$ weighted correlations of covariate moments and mechanical power moments.}
\label{tab:mech_power_balance}
\end{table}

\begin{figure}[ht]
\centering
\includegraphics[width=0.8\textwidth]{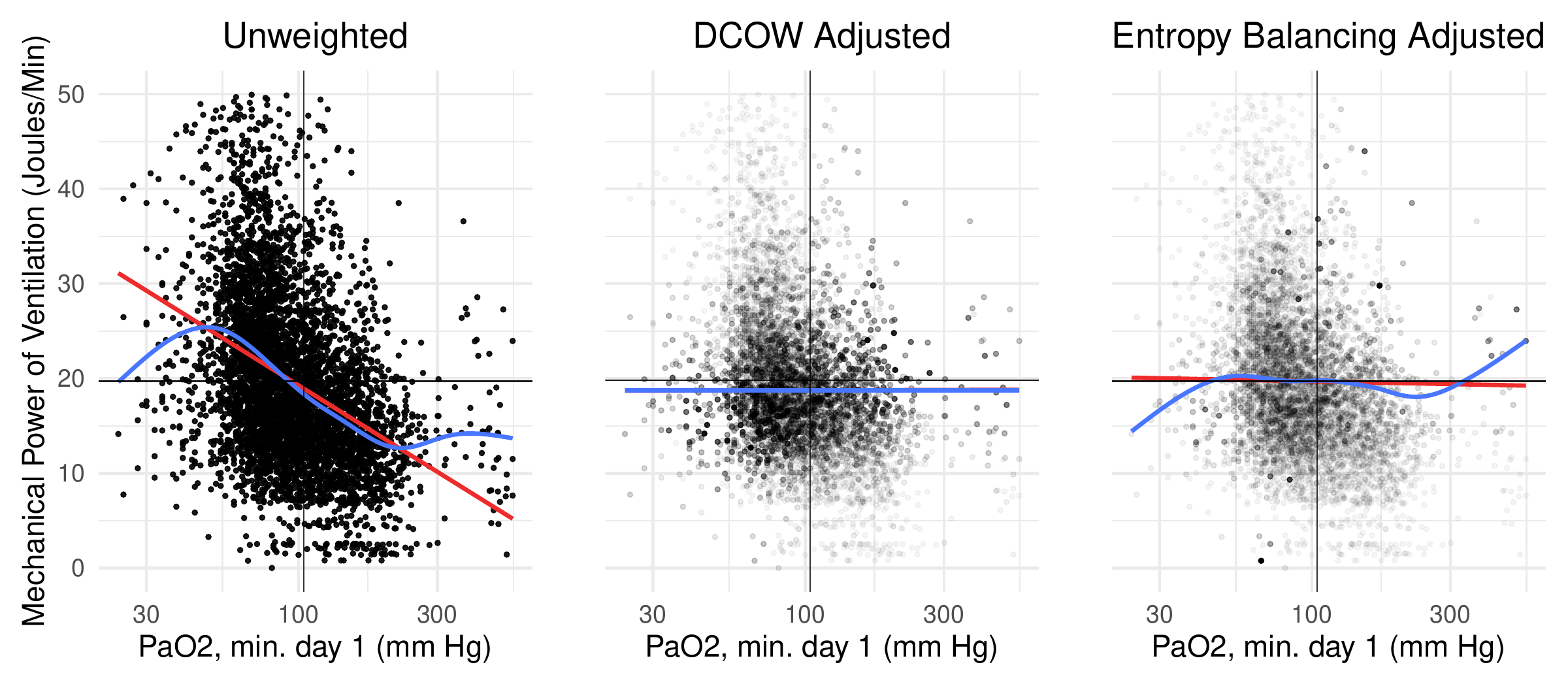}
\caption{\label{fig:mech_power_pao2} Shown are plots of the relationship between the minimum Pa$\text{O}_2$ on day 1 in the ICU (on a log base 10 scale) and the treatment, including an unadjusted plot (left) and plots adjusted by DCOWs and entropy balancing weights (right two plots). In the adjusted plots, the transparency of each point is proportional to its assigned weight, with lighter points indicating less weight. The blue line is a weighted nonparametric regression of the treatment on Pa$\text{O}_2$ (on the log base 10 scale) and the red line is a weighted linear regression.}
\end{figure}

For a concrete example of how DCOWs mitigate dependence, we illustrate the unadjusted marginal dependence of the Pa$\text{O}_2$/Fi$\text{O}_2$ ratio and the mechanical power of ventilation, where Pa$\text{O}_2$ is the partial pressure of oxygen in the arterial blood and Fi$\text{O}_2$ is the fraction of inspired oxygen. Pa$\text{O}_2$/Fi$\text{O}_2$ ratio is a pre-treatment covariate that characterizes acute hypoxemia, defines the presence and severity of acute respiratory distress syndrome (ADRS), and plays a critical role in ventilation guidelines \citep{papazian2019formal}. Both Pa$\text{O}_2$ and  Pa$\text{O}_2$/Fi$\text{O}_2$ ratio are also strongly associated with mortality and are thus critical confounders in this study. We display weighted marginal dependence of Pa$\text{O}_2$ and the mechanical power of ventilation with our proposed weights and with the entropy-balancing weights in Figure \ref{fig:mech_power_pao2}; we also show this relationship for the Pa$\text{O}_2$/Fi$\text{O}_2$ ratio in Section \ref{sec:data_supp} of the Supplementary Material. Compared with methods that aim to exactly decorrelate specified moments, our proposed weights can handle nonlinear dependence between covariates and treatment even in datasets with moderately high dimensions. While entropy balancing accounts for a significant proportion of dependence, there remains residual nonlinear dependence.

For each method, we construct pointwise confidence bands of the ADRF using a nonparametric bootstrap similar to \citet{wang1995bootstrap}. Weighted local linear regression estimates of the ADRF of mechanical power of ventilation on in-hospital mortality and 95\% confidence bands are displayed in Figure \ref{fig:bootstrap_mech_power}. Alternatively, it is possible to conduct inference using the results of Theorem \ref{thm:asymptotic_normality}; however, doing so requires careful selection of the bandwidth parameter $h$, such as by using undersmoothing, which involves choosing a smaller bandwidth than the optimal one. Rigorous assessment of the use of Theorem 3.7 for inference is left as future work. The entropy balancing approach with higher order moments was not included due to the small effective sample size and a large number of bootstrap replications for which no solution was found. Both the entropy balancing weights and the GPS estimated with a normal model exhibit reasonable control over the dependence between covariates and treatment but result in unacceptably high variances in the estimate of the ADRF. 

\section{Discussion}
\label{sec:discussion}

In this paper, we provide a detailed inspection of the role of weights in weighted nonparametric estimates of causal quantities involving continuous-valued treatments from observational data. This inspection shows clearly that the key source of bias depends on the degree to which the weights induce independence between the treatment and confounders. We then provide a measure that characterizes how well a set of weights mitigates the dependence between the treatment and confounders. This measure does not require any tuning parameters, making it straightforward to deploy in practice. Given some light smoothness conditions on the outcome data-generating model, this measure acts as an upper bound on the key source of systematic bias in a weighted nonparametric estimate of the ADRF. Our proposed weights, the DCOWS, which minimize our measure of dependence, provide an empirically robust means for estimating weights, as they directly target independence between the treatment and confounders. These weights are a natural complement to doubly-robust estimators, as they provide an anchor for the doubly-robust estimator: since they are guaranteed to be consistent on their own, the consistency of the doubly-robust estimator does not critically depend on the correctness of the outcome model. Thus, the outcome model can be safely used as a tool to reduce variability in the estimate of the ADRF.

In contrast to other weighting approaches aimed at removing bias due to confounding with continuous treatments, the DCOWs enjoy a number of benefits that make them particularly attractive for applied use: they do not require any modeling of the relationship between the treatment or outcome and the covariates, they do not require the choice of specific features of the covariate distribution to balance, they do not require parameter tuning or cross-fitting, they perform well empirically (in addition to their strong asymptotic properties), and they can be readily implemented without specialized software using existing quadratic programming solvers. We have released an open-source implementation of our method in the R statistical computing language that can be used off-the-shelf to estimate the weights, available at \url{https://github.com/jaredhuling/independenceWeights}. Although our method, like all methods that adjust for confounding by measured confounders, requires that a sufficient set of confounding variables has been collected by the researcher, its ease of use helps alleviate the analytical burden of estimating the causal effects of continuous treatments in the presence of confounding by measured variables.

\spacingset{1.5}

\allowdisplaybreaks

\newenvironment{definition}[1][Definition]{\begin{trivlist}
		\item[\hskip \labelsep {\bfseries #1}]}{\end{trivlist}}
\newenvironment{example}[1][Example]{\begin{trivlist}
		\item[\hskip \labelsep {\bfseries #1}]}{\end{trivlist}}
\newenvironment{rmq}[1][Remark]{\begin{trivlist}
		\item[\hskip \labelsep {\bfseries #1}]}{\end{trivlist}}

\appendix
\setcounter{secnumdepth}{0}


\section{Appendix: Regularity conditions}
\label{sec:appendix_a}

For $(\bX_1,A_1), \dots, (\bX_6,A_6)\distas{i.i.d.} F_{\bX,A}$, define the 6-th order kernel
$$k((\bX_1,A_1),\dots,(\bX_6,A_6)) = w^*(\bX_1,A_1) g_\bX(\bX_1,\bX_2,\bX_3,\bX_4)g_A(A_1,A_2,A_5,A_6)w^*(\bX_2,A_2)$$ with $w^*(\bx,a) = \frac{f_A(a)}{f_{A|\bX}(a|\bX=\bx)}$,
$g_\bX(\bX_1,\bX_2,\bX_3,\bX_4) = \| \bX_1-\bX_2\|_2 - \| \bX_1-\bX_3\|_2 - \| \bX_2-\bX_4\|_2+\| \bX_3-\bX_4\|_2$, and
$g_A(A_1,A_2,A_3,A_4) = | A_1-A_2| - | A_1-A_3| - | A_2-A_4|+| A_3-A_4|$. 
Further define the 4th order kernels
\begin{align*}
    k_\bX((\bX_1,A_1),\dots,(\bX_4,A_4)) = {} & w^*(\bX_1,A_1)\| \bX_1-\bX_3\|_2 + w^*(\bX_2,A_2)\| \bX_2-\bX_4\|_2 \\
    & -w^*(\bX_1,A_1)w^*(\bX_2,A_2)\| \bX_1-\bX_2\|_2 - \| \bX_3-\bX_4\|_2 \text{ and} \\
    k_A((\bX_1,A_1),\dots,(\bX_4,A_4)) = {} & w^*(\bX_1,A_1)| A_1-A_3| + w^*(\bX_2,A_2)| A_2-A_4| \\
    & - w^*(\bX_1,A_1)w^*(\bX_2,A_2)| A_1-A_2| - | A_3-A_4|.
\end{align*}

The following assumptions are required for several Lemmas and Theorems presented that rely on $V$-statistics. These conditions amount to finite moment conditions on squares of the Euclidean norms of $\bX$, $A$ and their products with the weights $w^*(\bX,A)$.
\begin{enumerate}[label=(A\arabic*)]
    \item $\bbE [k^2((\bX_1,A_1),\dots,(\bX_6,A_6))] < \infty$
    \item $\bbE [k_\bX^2((\bX_1,A_1),\dots,(\bX_4,A_4))] < \infty$ and $\bbE [k_A^2((\bX_1,A_1),\dots,(\bX_4,A_4))] < \infty$
\end{enumerate}


\begin{center}
{\large\bf Supplementary Material}
\end{center}
\vspace{-15pt}
\begin{description}
\item[Supplementary Material:] The Supplementary Material contains details for error decompositions of weighted nonparametric estimators, computational details of the proposed method, extended discussion of the existing literature, proofs of the theoretical results, and additional simulation studies. (pdf)
\end{description}
\vspace{-15pt}
\begin{center}
	{\large\bf Acknowledgments}
\end{center}
\vspace{-15pt}
We would like to thank the Associate Editor and anonymous referees for their constructive feedback and suggestions  Jue Hou for the helpful discussion. Chen's effort was partially supported by NSF grant DMS-2054346 and the University of Wisconsin School of Medicine and Public Health from the Wisconsin Partnership Program.

\def\spacingset#1{\renewcommand{\baselinestretch}%
{#1}\small\normalsize} \spacingset{0.5}

\bibliographystyle{Chicago}
\bibliography{manuscript}

\makeatletter\@input{xx_supp.tex}\makeatother
\end{document}


\def\spacingset#1{\renewcommand{\baselinestretch}%
{#1}\small\normalsize} \spacingset{1}

\renewcommand{\thesection}{\Alph{section}}
\renewcommand{\theequation}{S.\arabic{equation}}



\if1\blind
{
	\title{\bf Supplementary Material for ``Independence weights for causal inference with continuous treatments''}
	\author{Jared D. Huling,$^{1}$\thanks{corresponding authors: huling@umn.edu}
		Noah Greifer,$^{2}$\\
		Guanhua Chen$^{3}$\thanks{corresponding authors: gchen25@wisc.edu}\\
		\\
		$^{1}$Division of Biostatistics, University of Minnesota \\ [8pt]
		$^{2}$Institute for Quantitative Social Science, \\ Harvard University \\ [8pt]
		$^{3}$Department of Biostatistics and Medical Informatics, \\University of Wisconsin-Madison \\ [8pt]
	}
	\date{}
	\maketitle
} \fi

\if0\blind
{
	\bigskip
	\bigskip
	\bigskip
	\begin{center}
		{\Large \bf Supplementary Material for ``Independence weights for causal inference with continuous treatments'' }
	\end{center}
	\medskip
} \fi

\spacingset{1.75} 

\section{Properties of stabilized generalized propensity score weights}
\label{sec:appendix_aa}

The stabilized GPS weights $f_{A}(A)/f_{A|\bX}(A\:|\:\bX)$ have several key properties. Namely, when weighting by these weights they 1) result in independence of $\bX$ and $A$, 2) preserve the marginal distributions of $\bX$ and $A$, and 3) have mean 1.
For the following, we let $g(\bX)$ be \textit{any} function of $\bX$ only and let $h(A)$ be any function of $A$ only. It can be straightforwardly shown that 
\begin{align*}
\bbE_{\bX, A}\left(  g(\bX)h(A)\frac{f_A(A)}{f_{A|\bX}(A\:|\:\bX)}  \right) & =  \bbE_\bX(g(\bX))\bbE_A(h(A)), \\
\bbE_{\bX, A}\left(  g(\bX)\frac{f_A(A)}{f_{A|\bX}(A\:|\:\bX)}  \right) = {}  \bbE_{\bX}\left( g(\bX) \right), &\text{ and } \bbE_{\bX, A}\left(  h(A)\frac{f_A(A)}{f_{A|\bX}(A\:|\:\bX)}  \right) = {}  \bbE_{A}\left(h(A) \right).
\end{align*}
The first equality shows that when weighting by $\frac{f_A(A)}{f_{A|\bX}(A\:|\:\bX)}$, $\bX$ and $A$ are independent, as the above holds for any $g$ and $h$. Further, the second two equalities show that, on the weighted scale, the marginal distributions of $\bX$ and $A$ are preserved. Finally, the weights have mean 1, i.e. $\bbE_{\bX, A}\left(  \frac{f_A(A)}{f_{A|\bX}(A\:|\:\bX)}  \right) = 1$.



\section{Error decompositions}
\label{sec:appendix_b}

\subsection{Doubly robust estimation of the ADRF}

The error decompositions for a doubly-robust estimator (i.e., the augmented weighted Nadaraya-Watson estimator) of the causal distribution function curve defined in Section \ref{sec:drest} can be shown as follows:
\begin{align*}
    & \hat \mu_{NW}^{\bw,DR}(a_0) - \mu(a_0) \\ 
    = {} &  \int_\calX \hat{\mu}(\bx,a_0) \mathrm{d}F^n_\bX(\bx) - \int_\calX {\mu}(\bx,a_0) \mathrm{d}F_\bX(\bx) \\ 
    & + {\widehat{f}^n{}}^{-1}_{A,h}(a_0) \int_\calX\int_\calA (\mu(\bx, a) - \hat{\mu}(\bx,a_0)) K_h(a-a_0) \mathrm{d} F^n_{\bX,A,\bw}(\bx,a) 
    + \frac{1}{n}\sum_{i=1}^n \varepsilon_i w_i \widehat{f}^{-1}_{n,h}(a_0) {K_h(A_i - a_0)} \\
    = {} & \int_\calX (\hat{\mu}(\bx,a_0) - {\mu}(\bx,a_0)) \mathrm{d}F^n_\bX(\bx) + \int_\calX {\mu}(\bx,a_0) \mathrm{d}\left[ F^n_\bX - F_\bX \right](\bx) \\
    & + \frac{f_A(a_0)}{\widehat{f}^n_{A,h}(a_0)}\int_\calX\int_\calA({\mu}(\bx,a_0) - \hat{\mu}(\bx,a_0)) \mathrm{d} F^n_{\bX,A,\bw}(\bx,a) \\
    & + {\widehat{f}^n{}}^{-1}_{A,h}(a_0) \int_\calX\int_\calA \left\{(\mu(\bx, a) - \hat{\mu}(\bx,a_0))K_h(a-a_0) - (\mu(\bx, a_0) - \hat{\mu}(\bx,a_0)) f_A(a_0)\right\}  \mathrm{d} F^n_{\bX,A,\bw}(\bx,a) \\
    & + \frac{1}{n}\sum_{i=1}^n \varepsilon_i w_i \widehat{f}^{-1}_{n,h}(a_0) {K_h(A_i - a_0)} \\
    = {} & \int_\calX (\hat{\mu}(\bx,a_0) - {\mu}(\bx,a_0)) \mathrm{d}F^n_\bX(\bx) + \int_\calX {\mu}(\bx,a_0) \mathrm{d}\left[ F^n_\bX - F_\bX \right](\bx) \\
    & + \int_\calX\int_\calA({\mu}(\bx,a_0) - \hat{\mu}(\bx,a_0)) \mathrm{d} F^n_{\bX,A,\bw}(\bx,a) + \left(\frac{f_A(a_0)}{\widehat{f}^n_{A,h}(a_0)} - 1\right) \int_\calX\int_\calA({\mu}(\bx,a_0) - \hat{\mu}(\bx,a_0)) \mathrm{d} F^n_{\bX,A,\bw}(\bx,a) \\
    & + {\widehat{f}^n{}}^{-1}_{A,h}(a_0) \int_\calX\int_\calA \left\{(\mu(\bx, a) - \hat{\mu}(\bx,a_0))K_h(a-a_0) - (\mu(\bx, a_0) - \hat{\mu}(\bx,a_0)) f_A(a_0)\right\} \mathrm{d} F^n_{\bX,A,\bw}(\bx,a) \\
    & + \frac{1}{n}\sum_{i=1}^n \varepsilon_i w_i \widehat{f}^{-1}_{n,h}(a_0) {K_h(A_i - a_0)} \\
    = {} & \int_{\calX} \int_\calA (\mu(\bx, a_0) - \hat{\mu}(\bx,a_0)) \mathrm{d} \left[ F^n_{\bX,A,\bw} - F^n_{\bX}F^n_{A} \right](\bx,a)  \\
    & + \int_\calX {\mu}(\bx,a_0) \mathrm{d}\left[ F^n_\bX - F_X \right](\bx) + \left(\frac{f_A(a_0)}{\widehat{f}^n_{A,h}(a_0)} - 1\right) \int_\calX (\mu(\bx, a_0) - \hat{\mu}(\bx,a_0)) \mathrm{d}F^n_\bX(\bx) \\
    & + \left(\frac{f_A(a_0)}{\widehat{f}^n_{A,h}(a_0)} - 1\right) \int_\calX (\mu(\bx, a_0) - \hat{\mu}(\bx,a_0)) \mathrm{d} \left[ F^n_{\bX,A,\bw} - F^n_{\bX}F^n_{A} \right](\bx,a) \\
    & + {\widehat{f}^n{}}^{-1}_{A,h}(a_0) \int_\calX\int_\calA \left\{(\mu(\bx, a) - \hat{\mu}(\bx,a_0))K_h(a-a_0) - (\mu(\bx, a_0) - \hat{\mu}(\bx,a_0)) f_A(a_0)\right\} \mathrm{d} F^n_{\bX,A,\bw}(\bx,a) \\
    & + \frac{1}{n}\sum_{i=1}^n \varepsilon_i w_i \widehat{f}^{-1}_{n,h}(a_0) {K_h(A_i - a_0)},
\end{align*}
where the first equality holds by the definition of the doubly-robust estimator and the ADRF and the last holds due to Theorem 1.7.1 of \citet{durrett1996probability}.

The error decomposition for the non doubly-robust estimator is achieved by setting $\hat{\mu}(\bx,a_0)=0$ everywhere.

\subsection{Causal quantile estimation and a decomposition for weighted estimators of the causal distribution function curve}

Although we do not present a general theory,
we emphasize that the role of weights in their ability to induce independence between treatment and covariates is not unique to estimation of the ADRF and applies to a wide variety of estimands.
Consider estimation of the causal dose-response quantile function $q_{Y(a_0)}(\alpha) = \inf\{y: F_{Y(a_0)}(y) \leq \alpha\}$, where $F_{Y(a_0)}(y) = \bbP(Y(a_0) \leq y) = \bbE_{\bX}\left\{ \bbP(Y\leq y\vert \bX, A=a_0) \right\} = \bbE_{\bX}\left\{ F_{Y|\bX,A}(y\vert \bX,A=a_0) \right\}$.

The weighted nonparametric estimator of the CDF of $Y(a_0)$ is
\begin{align}
\widehat{F}_{Y(a_0)}^{\bw}(y)
= {} &  \frac{\sum_{i=1}^n I(Y_i \leq y) w_i K_h(A_i - a_0)}{ \sum_{i=1}^n K_h(A_i - a_0)} \label{eqn:weighted_nw_quantile_estimator}
\end{align}
and the resulting kernel quantile estimator of $q_{Y(a_0)}(\alpha) = \inf\{y: \alpha \leq F_{Y(a_0)}(y) \}$ at quantile $\alpha$ when $A=a_0$ is $\widehat{q}_{Y(a_0)}^{\bw}(\alpha) = \inf\{y: \alpha \leq \widehat{F}_{Y(a_0)}^{\bw}(y) \}$.

The error of \eqref{eqn:weighted_nw_quantile_estimator} at $A=a_0$ can be decomposed as
\begin{align}
    \widehat{F}_{Y(a_0)}^{\bw}(y) - {F}_{Y(a_0)}(y) = {} & \int_\calX F(y|\bx,a_0)\int_\calA \mathrm{d} \left[ F^n_{\bX,A,\bw} - F^n_{\bX}F^n_{A} \right](\bx,a) \nonumber \\
    & + R_{n}(a_0) + R_{n,h}(a_0) + R_{1n,h,\bw}(a_0) + R_{2n,h,\bw}(a_0) \nonumber  \\
    & + \frac{1}{n}\sum_{i=1}^n e_i(y) w_i \widehat{f}^{-1}_{n,h}(a_0) {K_h(A_i - a_0)}, \nonumber
\end{align}
where $e_i(y) = I(Y_i \leq y) - \bbP(Y\leq y|\bX_i,A_i)$ where by construction, $\bbE [e_i(y)] =0$, and $R_{n}(a_0)$, $R_{n,h}(a_0)$, $R_{1n,h,\bw}(a_0)$, and  $R_{2n,h,\bw}(a_0)$ behave similarly to the second through fifth terms in the decomposition \eqref{eqn:error_decomp_indep_first} with their consistency for zero only depending on either the sample size and/or bandwidth $h$ and not the weights. Thus, just as in estimation of the ADRF, the key source of systematic bias in causal quantile function estimation can be mitigated by the use of weights that make the covariates and treatment independent.

The error decomposition is shown via the following:
\begin{align*}
    & \widehat{F}_{Y(a_0)}^{\bw}(y) - {F}_{Y(a_0)}^{\bw}(y) \\
    = {} & {\widehat{f}^n{}}^{-1}_{A,h}(a_0) \int_\calX\int_\calA \bbP(Y\leq y|\bx,a)K_h(a-a_0) \mathrm{d} F^n_{\bX,A,\bw}(\bx, a) \\
    & - \int_\calX \bbP(Y\leq y|\bx,a_0) \mathrm{d} F_\bX(\bx) + {\widehat{f}^n{}}^{-1}_{A,h}(a_0) \frac{1}{n}\sum_{i=1}^nw_ie_i(y)K_h(A_i-a_0) \\
    = {} & \frac{f_A(a_0)}{\widehat{f}^n_{A,h}(a_0)}\int_\calX\int_\calA \bbP(Y\leq y|\bx,a_0) \mathrm{d} F^n_{\bX,A,\bw}(\bx, a) - \int_\calX \bbP(Y\leq y|\bx,a_0) \mathrm{d} F^n_\bX(\bx)\\
    & + {\widehat{f}^n{}}^{-1}_{A,h}(a_0) \int_\calX\int_\calA \left\{ \bbP(Y\leq y|\bx,a)K_h(a-a_0) - \bbP(Y\leq y|\bx,a_0)f_A(a_0) \right\}\mathrm{d} F^n_{\bX,A,\bw}(\bx, a) \\
    & + \int_\calX \bbP(Y\leq y|\bx,a_0) \mathrm{d} \left[ F^n_\bX - F_\bX \right](\bx) + {\widehat{f}^n{}}^{-1}_{A,h}(a_0) \frac{1}{n}\sum_{i=1}^nw_ie_i(y)K_h(A_i-a_0) \\
    = {} & \int_\calX\int_\calA \bbP(Y\leq y|\bx,a_0) \mathrm{d} F^n_{\bX,A,\bw}(\bx, a) - \int_\calX \bbP(Y\leq y|\bx,a_0) \mathrm{d} F^n_\bX(\bx) \\
    & + \left(\frac{f_A(a_0)}{\widehat{f}^n_{A,h}(a_0)} - 1\right) \int_\calX\int_\calA \bbP(Y\leq y|\bx,a_0) \mathrm{d} F^n_{\bX,A,\bw}(\bx, a) \\
    & + {\widehat{f}^n{}}^{-1}_{A,h}(a_0) \int_\calX\int_\calA \left\{ \bbP(Y\leq y|\bx,a)K_h(a-a_0) - \bbP(Y\leq y|\bx,a_0)f_A(a_0) \right\}\mathrm{d} F^n_{\bX,A,\bw}(\bx, a) \\
    & + \int_\calX \bbP(Y\leq y|\bx,a_0) \mathrm{d} \left[ F^n_\bX - F_\bX \right](\bx) + {\widehat{f}^n{}}^{-1}_{A,h}(a_0) \frac{1}{n}\sum_{i=1}^nw_ie_i(y)K_h(A_i-a_0) \\
    = {} & \int_\calX\int_\calA \bbP(Y\leq y|\bx,a_0) \mathrm{d}\left[ F^n_{\bX,A,\bw} - F^n_\bX F^n_A\right](\bx, a) \\
    & + \left(\frac{f_A(a_0)}{\widehat{f}^n_{A,h}(a_0)} - 1\right) \int_\calX\int_\calA \bbP(Y\leq y|\bx,a_0) \mathrm{d}\left[ F^n_{\bX,A,\bw} - F^n_\bX F^n_A\right](\bx, a) \\
    & + \left(\frac{f_A(a_0)}{\widehat{f}^n_{A,h}(a_0)} - 1\right) \int_\calX \bbP(Y\leq y|\bx,a_0) \mathrm{d}F^n_\bX(\bx) \\
    & + {\widehat{f}^n{}}^{-1}_{A,h}(a_0) \int_\calX\int_\calA \left\{ \bbP(Y\leq y|\bx,a)K_h(a-a_0) - \bbP(Y\leq y|\bx,a_0)f_A(a_0) \right\}\mathrm{d} F^n_{\bX,A,\bw}(\bx, a) \\
    & + \int_\calX \bbP(Y\leq y|\bx,a_0) \mathrm{d} \left[ F^n_\bX - F_\bX \right](\bx) + {\widehat{f}^n{}}^{-1}_{A,h}(a_0) \frac{1}{n}\sum_{i=1}^nw_ie_i(y)K_h(A_i-a_0), \\
\end{align*}
where the steps of the decomposition largely follow the same steps as that of the doubly-robust ADRF estimator.

\section{Computation}
\label{sec:computation}
We will show how the (P)DCOW optimization problem can be represented as a generic quadratic program 
$$
\min_{\bw}\: \bw\trans \bP \bw + \bq\trans \bw \
\text{ s.t. } \ \bA\trans \bw \le \bc.
$$
The (P)DCOW optimization criterion takes  $\bP = \bP_{DCOW} + \bP_{EB(A)} + \bP_{EB(\bX)} + \bP_{\lambda}$, $\bq\trans = \bq\trans_{EB(A)} + \bq\trans_{EB(\bX)}$, 
where $\bP_{DCOW} = \frac{1}{n^2} \bC \circ \bD$, $\bP_{EB(A)} = -\frac{1}{n^2} \bQ_A,  \bP_{EB(\bX)} = -\frac{1}{n^2} \bQ_X,  \bP_{\lambda} = \lambda\frac{1}{n^2}\bI_n$,
$\bq\trans_{EB(A)} = \frac{2}{n^2} \mathbf{1}\trans \bQ_A$, and 
$\bq\trans_{EB(\bX)} = \frac{2}{n^2} \mathbf{1}\trans \bQ_{\bX}$.
The terms $\bC = (C_{k\ell})$ and $\bD = (D_{k\ell})$ are as defined in \eqref{eqn:weighted_distance}, $\circ$ is the elementwise product, and $\bQ_A$ and $\bQ_{\bX}$ are the Euclidean distance matrices of standardized versions of $A$ and $\bX$, respectively.
From this formulation, we have
$\calV^2_{n,\bw}(\bX,A) = \bw\trans \bP_{DCOW} \bw$,
$\calE(F^n_{A,\bw}, F^n_{A}) = \bw\trans \bP_{EB(A)} \bw + \bq\trans_{EB(A)}\bw + c_A$,
$\calE(F^n_{\bX,\bw}, F^n_{\bX}) = \bw\trans \bP_{EB(X)} \bw + \bq\trans_{EB(X)}\bw + c_X$, and 
$\lambda\frac{1}{n^2}\sum_{i=1}^nw_i^2 = \bw\trans \bP_{\lambda} \bw$,
where $c_A$ and $c_X$ are constants,
so minimizing $\bw\trans \bP \bw + \bq\trans \bw$ amounts to minimizing $\calD(\bw) + \lambda\frac{1}{n^2}\sum_{i=1}^nw_i^2$, the solution to which are the PDCOWs (or the DCOWs when $\lambda = 0$). The constraints $\bA\trans$ and $\bc$ can be specified as
$\bA\trans = (-\bI_n, \mathbf{1}, -\mathbf{1})\trans$ and $\bc = ( \mathbf{0}\trans, n, -n)\trans$
so that $\bA\trans\bw \le \bc$ implies $w_i \ge 0$ for all $i$ and $\sum_{i} w_i = n$.
Although the optimization problem has the form of a linearly constrained quadratic program, the problem is not convex because $\bP$ is generally not positive semi-definite. Despite this, algorithms such as OSQP \citep{stellato2020osqp} or CCCP \citep{andersen2011interior} can reliably provide high quality solutions to the optimization problem.


\section{Extensions and implementation considerations}
\label{sec:extensions}

If in practice additional emphasis on correlations of particular moments of covariates and the treatment is of importance, the optimization criterion in \eqref{eqn:dcows} can be straightforwardly modified to optimize $\calD(\bw)$ subject to the constraints $$\sum_{i=1}^nw_i\left\{m_\bX(\bX_i) - \overline{m}_\bX(\bX_i)\right\}\left\{m_A(A_i) - \overline{m}_A(A_i)\right\}\trans = \bzero,$$ where ${m}_\bX$ is a vector of moments of $\bX$, $\overline{m}_\bX$ is the average of the moments over the sample, ${m}_A$ is a vector of moments of $A$, and similarly $\overline{m}_A$ is their sample average. These constraints allow for the user to construct weights that minimize dependence between $\bX$ and $A$ while emphasizing decorrelation of desired moments. If exactly zero correlation is infeasible, a constraint that the marginal correlations are less than a threshold can be used instead, similar to \citet{zubizarreta2015stable}. Additional emphasis on marginal correlation may improve finite sample performance if known moments are especially imbalanced or highly related to outcomes. These constraints can be straightforwardly incorporated into the quadratic programming problem described above. 

When the dimension $p$ of $\bX$ is high, the relative emphasis of the energy distance terms for $A$ and $\bX$ in $\calD(\bw)$ can be imbalanced, with less emphasis on the individual components of $\bX$ relative to $A$. To help mitigate this issue, one can instead use a version of $\calD(\bw)$ that aims to re-balance the terms to be more comparable on a per-unit basis like the following: $\calD_c(\bw) = \calV^2_{n,\bw}(\bX,A) + c_{\bX}\calE(F^n_{\bX,\bw}, F^n_{\bX}) + c_A\calE(F^n_{A,\bw}, F^n_{A})$, where $c_{\bX}+c_A=1$ and $c_{\bX}/c_A = \sqrt{p}$. This modification has no meaningful impact on the theoretical properties of $\calD_c(\bw)$, as our proof of Theorem \ref{thm:weighted_distance_criterion_duality} applies to this modified distance.



\section{Expanded discussion of the connection of proposed methods to existing literature}
\label{supsec:connection_other_methods}

Two main strands of research have been proposed to estimate the ADRF for a continuous treatment in the presence of confounding. One strand of research has focused on doubly-robust estimation using machine learning modeling of nuisance parameters, while the other has focused on the estimation of balancing weights, as we do here.

\citet{kennedy2017non} describes doubly-robust estimation of the causal ADRF using weighted local linear regression and shows asymptotic normality of the doubly-robust estimator under assumptions on the convergence-rates of the model for the conditional density of treatment and outcome and entropy restrictions of such models. In their work, the particular choices for the treatment and outcome models are not emphasized; machine-learning techniques were used for these models in their application. \citet{kennedy2017non} proves asymptotic normality of a doubly-robust local-linear estimator under the condition that either the GPS model or outcome model is consistent and the product of their convergence rates converges at rate $o_p(1/\sqrt{nh})$ or faster, where $h$ is the bandwidth of the local linear estimator. \citet{colangelo2020double} also focuses on doubly-robust estimation of a causal ADRF using sample splitting and weighted nonparametric regression. In their work, sample splitting is used in practice and for their theoretical results. They provide theory showing asymptotic normality under somewhat similar assumptions on the data-generating mechanism as our work, however their work requires root-mean-squared convergence of an explicit model for the conditional density of the treatment, whereas in our work we do not require convergence of our weights to the true inverse GPS. Their work emphasizes using machine learning-based estimators of the outcome model and conditional density model, which may require careful tuning of any hyperparameters involved. Sample splitting is not required for any of our theoretical results on DCOWs in estimation of the ADRF.

We now discuss works involving covariate-balancing type techniques; i.e. methods that construct weights directly through means of various optimization criteria. 
\citet{fong2018covariate} develop two weighting methods designed to reduce correlations between covariates and treatment. One version of their approach works by operating under a normal conditional density model for the treatment, supplementing it with moment-decorrelating conditions, and estimates the weights in a method of moments framework. Their second method, described as nonparametric, removes the assumption of a normal conditional density and works by maximizing an empirical likelihood subject to moment-decorrelating conditions. However, this still involves parametric choices on \textit{which} moments of covariates and treatment to decorrelate. As such, it is more accurately thought of as a parametric approach. Asymptotic results are not provided for the second approach. Both approaches only remove linear correlations between the treatment and user-specified basis functions of the covariates. \citet{yiu2018covariate} introduce so-called ``covariate association eliminating weights'', which are designed to minimize a (weighted) measure of dispersion under constraints that are designed to remove associations of covariates and treatment as measured through a parametric model. Thus, in general, this approach will not result in independence between covariates and treatment unless the model is either correctly specified or sufficiently flexible. It is unclear how to determine whether either of these scenarios holds in practice. Asymptotic results are not provided in \citet{yiu2018covariate}. Similar to \citet{yiu2018covariate}, both \citet{tubbicke2020entropy} and \citet{vegetabile2020nonparametric} propose to estimate weights that can minimize a (weighted) measure of dispersion subject to constraints that (weighted) marginal correlations of pre-specified moments of covariates and treatment are zero. Though these approaches give the user more flexibility in selecting exactly which moments are decorrelated, these choices still must be made by the analyst, and missing important moments can result in severe bias, as evidenced by our simulation studies. Asymptotic results are not provided in either work.
In contrast to all of the above methods, our approach works by constructing weights that minimize a measure of statistical \textit{dependence} subject to additional constraints on the weights to make them positive and sum to $n$. By doing so, our weights are nonparametric and, as we show, asymptotically completely mitigate statistical dependence between the covariates and treatment.

Similarly, the work of \citet{kallus2019kernel} focuses on constructing weights that minimize worst-case functional covariance between covariates and treatment. This viewpoint bears some relation to our work, as functional covariance corresponds with statistical independence testing \citep{gretton2007kernel, sejdinovic2013equivalence}. However, such kernel-based measures require careful tuning of the kernel and kernel hyperparameters for good empirical performance. As no theoretical results are provided in \citet{kallus2019kernel}, it is not clear what conditions on the choice of kernel, outcome model, and/or conditional density of the treatment are required for consistency or other stronger statistical guarantees. In contrast, as discussed in Remark \ref{remark2}, our approach works well even if the implied kernel of our distance measure does not correspond to the true outcome regression model. \citet{martinet2020balancing} operates in the same vein by providing a generalization of the work of \citet{kallus2019kernel} and shows risk consistency of their weighted estimate of a dose-response function under a strong assumption that the true GPS lives in a well-specified, finite-dimensional Reproducing Kernel Hilbert Space (RKHS). This assumption is in effect a parametric assumption about the data-generating process. In contrast, our methods emphasize mitigating statistical dependence instead of functional covariance over a particular function space. As we show in Theorem \ref{thm:weighted_distance_convergence}, our weighted joint CDF converges to the correct target distribution under only mild finite-moment assumptions on the covariates and treatment. We can further obtain consistency for the causal ADRF under a mild boundedness and continuity assumption on the outcome regression function as described in Section \ref{sec:methods}.




\section{Additional theoretical results}
\label{supsec:more_theory}

\subsection{Results for PDCOWs}

We present the following result, which shows how the proposed penalized distance acts as a bound on integration errors plus the variability of the weights. 
\begin{lem}\label{thm:lem_rmse}
Let $\calH$ be the native space induced by the radial kernel $\Phi(\cdot,\cdot) = -\| \cdot \|_2\times -| \cdot |$ on $\calX\times\calA$, and let $g(\cdot,\cdot)$ be any function where $g(\cdot,\cdot) \in \calH$. Then, for any weights $\bw$ satisfying $\sum_{i=1}^nw_i = n, w_i \ge 0$, and any $\bz=(z_1,\dots,z_n)$, we have
\begin{equation*}
\left[ \int_\calX g(\bx,a) \mathrm{d} \left[ F^n_{\bX, A, \bw} - F^n_\bX F^n_A  \right](\bx,a) \right]^2 + \frac{1}{n^2}\left( \sum_{i=1}^nw_iz_i \right)^2 \leq C_{gz} \left(\mathcal{D}_\gamma(\bw) + \lambda\frac{1}{n^2}\sum_{i=1}^nw_i^2\right),
\end{equation*}
where $C_{gz}=\max\{c_\gamma\|g\|_\calH, \max_iz_i^2/\lambda\} \geq 0$ is a constant depending on only $g$, $\gamma$, $\lambda$, and $\bz$.
\end{lem}
This shows that the penalized version of our proposed criterion acts as a bound on the integration error plus any term which involves the squares of the weights.

The following result shows the rate of convergence of $\calD(\bw^d)$ and $\calD(\bw^{pd})$ to zero and also shows that the squares of the PDCOWs are controlled asymptotically. 
\begin{lem}\label{thm:lem_criterion_rate}
Assume the moment conditions (A1) and (A2) listed in the Appendix hold, then $\calD(\bw^d)=O_p(1/n)$. If in addition we have $\bbE {w^*}^2(\bX,A) < \infty$, then $\calD(\bw^{pd})=O_p(1/n)$ and $\frac{1}{n^2}\sum_{i=1}^n{{w_i^{pd}{}}^2 }=O_p(1/n)$.
\end{lem}
\noindent Lemma \ref{thm:lem_criterion_rate} implies the rate of convergence of $\int|\varphi^n_{\bX,A,\bw}(\bt,s) - \varphi^n_{\bX}(\bt)\varphi^n_{A}(s)|^2 \omega(\bt, s) \mathrm{d}\bt \mathrm{d}s $ to zero for both $\bw^d$ and $\bw^{pd}$, indicating they result in independence of the treatment and covariates at a parametric rate. 

\subsection{Results for modified PDCOWs}

In the following, we show that the key properties of the PDCOWs are unchanged with the modified weights presented in equation \eqref{eqn:pdcows_mod} of the main text that are utilized in Theorem \ref{thm:asymptotic_normality}.

\begin{lem}\label{thm:weighted_distance_convergence_mod}
Let $\widetilde{\bw}^{pd}_n$ be the distance covariance optimal weights defined in \eqref{eqn:pdcows_mod} of the main text. Then if $\bbE\lVert \bX \rVert_2 < \infty$ and $\bbE\lvert A \rvert < \infty$, $\bbE {w^*}^2(\bX,A) < \infty$ holds, and $1/f_A(a_0)$ is uniformly bounded, then  
\begin{equation}\label{eqn:ecdf_convergence_pd_mod}
    \lim_{n \to \infty} F^n_{\bX, A, \widetilde{\bw}^{pd}}(\bx, a) = F_{\bX}(\bx)F_{A}(a)
\end{equation}
almost surely for every continuity point $(\bx, a)\in \bbR^{p+1}$.
\end{lem}

\begin{lem}\label{thm:lem_criterion_rate_mod}
Assume the moment conditions (A1) and (A2) listed in Appendix \ref{sec:appendix_a} hold, that $\bbE\lVert \bX \rVert_2 < \infty$ and $\bbE\lvert A \rvert < \infty$ hold, that $\bbE {w^*}^2(\bX,A) < \infty$ hold, and that $1/f_A(a_0)$ is uniformly bounded, then $\calD(\widetilde{\bw}^{pd})=O_p(1/n)$ and $\frac{1}{n^2}\sum_{i=1}^n{{\widetilde{w}_i^{pd}{}}^2 }=O_p(1/n)$.
\end{lem}

In particular, Lemma \eqref{thm:weighted_distance_convergence_mod} shows the modified weights still asymptotically result in independence of $\bX$ and $A$ and Lemma \eqref{thm:lem_criterion_rate_mod} shows that the modified weights result in $\calD$ converging to zero at the appropriate rate and the squares of the modified weights are controlled.

\subsection{Consistency of augmented estimator with a possibly misspecified outcome model}

Here we show a consistency result for the augmented estimator $\widehat{\mu}_{NW}^{\bw,DR}(a_0)$ with (P)DCOWs analogous to that shown for the weighting estimator $\widehat{\mu}_{NW}^{\bw}(a_0)$. Here, we allow the outcome model in the augmented estimator to be possibly misspecified in that it converges to a function that might not be the correct outcome model.

\begin{thm}\label{thm:thm3b}
Assume that the kernel $K(\cdot)$ is symmetric, second order, i.e. it meets the conditions that  $\int uK(u)\mathrm{d}u=0$, $\int K(u)\mathrm{d}u=1$, and $0 < \int u^2K(u)\mathrm{d}u < \infty$, and is bounded differentiable. Further, assume that the moment conditions required in Theorem \ref{thm:weighted_distance_convergence} hold and that $\mu(\bx,a_0)$ and is bounded and continuous on $\calX\times\calA$ and has second order derivatives, $f_A(a_0)$ is bounded and has second order derivatives, $1/f_A(a_0)$ is uniformly bounded. We further assume that the estimated outcome model $\hat{\mu}(\bx, a_0)$ is assumed to converge uniformly over $\bx \in \calX$ almost surely to $\widetilde{\mu}(\bx, a_0)$, which may or may not be equal to ${\mu}(\bx, a_0)$ and is such that $\bbE[\widetilde{\mu}(\bX, a_0)] = \widetilde{\mu}(a_0)$ for some $-\infty < \widetilde{\mu}(a_0) < \infty$, which may not be equal to $\mu(a_0)$. 
When $h \to 0$, $nh \to \infty$, then for $\bw=\bw^d \text{ and } \bw=\bw^{pd}$
\begin{equation}
\lim_{ n \to\infty} \widehat{\mu}_{NW}^{\bw,DR}(a_0) = \mu(a_0) 
\end{equation}
in probability for all continuity points $a_0 \in\calA$. 
\end{thm}

Thus, using (P)DCOWs perhaps unsurprisingly results in consistent estimates of the ADRF in an augmented estimator without the assumption of the outcome model being correctly specified.


\section{Technical Proofs}
\label{sec:proofs}

\begin{proof}[Proof of Theorem \ref{thm:weighted_distance_duality}]
From Theorem 1 of \citet{szekely2007measuring}, $\frac{1}{n^2}\sum_{k,\ell=1}^n C_{k\ell}D_{kl} \geq 0$. Thus, trivially $\calV^2_{n,\bw}(\bX,A) \geq 0$ since it is a quadratic form in $\bw$.

We aim to show that 
\begin{align}  
\calV^2_{n,\bw}(\bX,A)= {} & \int_{\bbR^{p+1}}\vert \varphi^n_{\bX, A, \bw}(\bt, s) - \varphi^n_{\bX,\bw}(\bt)\varphi^n_{A,\bw}(s) \nonumber \\ 
& \;\;\quad\quad + (\varphi^n_{\bX,\bw}(\bt) -\varphi^n_{\bX}(\bt))(\varphi^n_{A,\bw}(s) - \varphi^n_{A}(s)) \vert^2 \omega(\bt, s) \mathrm{d}\bt \; \mathrm{d}s.\nonumber
\end{align}
We begin by noting that the right hand side of the equation can be expressed as
\begin{align}  
{} & \int_{\bbR^{p+1}}\vert \varphi^n_{\bX, A, \bw}(\bt, s) - \varphi^n_{\bX,\bw}(\bt)\varphi^n_{A,\bw}(s) \nonumber \\ 
& \;\;\quad\quad + (\varphi^n_{\bX,\bw}(\bt) -\varphi^n_{\bX}(\bt))(\varphi^n_{A,\bw}(s) - \varphi^n_{A}(s)) \vert^2 \omega(\bt, s) \mathrm{d}\bt \; \mathrm{d}s \nonumber \\
= {} & \int_{\bbR^{p+1}}\vert \varphi^n_{\bX, A, \bw}(\bt, s) -  \varphi^n_{\bX,\bw}(\bt)\varphi^n_{A}(s) \nonumber \\
& \;\;\quad\quad - \varphi^n_{\bX}(\bt)\varphi^n_{A,\bw}(s)
+ \varphi^n_{\bX}(\bt)\varphi^n_{A}(s) \vert^2 \omega(\bt, s) \mathrm{d}\bt \; \mathrm{d}s.  \label{eqn:term_integral}
\end{align}
Throughout, we use $\mathrm{d}\omega$ as shorthand for $\omega(\bt, s) \mathrm{d}\bt \; \mathrm{d}s$, $\omega(\bt) \mathrm{d}\bt$, or $\omega(s) \mathrm{d}s$ whenever it is not ambiguous. For column vectors $\bt$ and $\bv$, we also use the notation $\langle \bt, \bv \rangle$ to denote the product $\bt\trans\bv$.
We prove the remainder of the theorem by expanding all the squared terms inside the integral in \eqref{eqn:term_integral}. The expansion is equal to 
\begin{equation}\label{eqn:expansion}
\widetilde{S}_1-2\widetilde{S}_2-2\widetilde{S}_3+2\widetilde{S}_4+\widetilde{S}_5+\widetilde{S}_6+2\widetilde{S}_7-2\widetilde{S}_8-2\widetilde{S}_9+\widetilde{S}_{10},
\end{equation}
where
\begin{alignat*}{3}
    \widetilde{S}_1 = {} & \varphi^n_{\bX, A, \bw}(\bt, s)\overline{\varphi^n_{\bX, A, \bw}(\bt, s)},
    \quad\quad && \widetilde{S}_2 &&  = {} \varphi^n_{\bX, A, \bw}(\bt, s)\overline{\varphi^n_{\bX,\bw}(\bt)\varphi^n_{A}(s)}, \\
    \widetilde{S}_3 = {} & \varphi^n_{\bX, A, \bw}(\bt, s)\overline{\varphi^n_{\bX}(\bt)\varphi^n_{A,\bw}(s)}, 
    \quad\quad && \widetilde{S}_4 &&  = {}  \varphi^n_{\bX,\bw}(\bt)\overline{\varphi^n_{\bX}(\bt)}\varphi^n_{A,\bw}(s)\overline{\varphi^n_{A}(s)}, \\
    \widetilde{S}_5 = {} & \varphi^n_{\bX,\bw}(\bt)\overline{\varphi^n_{\bX,\bw}(\bt)}\varphi^n_{A}(s)\overline{\varphi^n_{A}(s)},
    \quad\quad && \widetilde{S}_6 && = {}  \varphi^n_{\bX}(\bt)\overline{\varphi^n_{\bX}(\bt)}\varphi^n_{A,\bw}(s)\overline{\varphi^n_{A,\bw}(s)}, \\
    \widetilde{S}_7 = {} & \varphi^n_{\bX, A, \bw}(\bt, s)\overline{\varphi^n_{\bX}(\bt)\varphi^n_{A}(s)}, 
     \quad\quad && \widetilde{S}_8 && = {}  \varphi^n_{\bX,\bw}(\bt)\overline{\varphi^n_{\bX}(\bt)}\varphi^n_{A}(s)\overline{\varphi^n_{A}(s)}, \\
    \widetilde{S}_9 = {} & \varphi^n_{\bX}(\bt)\overline{\varphi^n_{\bX}(\bt)}\varphi^n_{A, \bw}(s)\overline{\varphi^n_{A}(s)}, \text{ and} 
    \quad\quad && \widetilde{S}_{10}  && = {} 
    \varphi^n_{\bX}(\bt) \overline{\varphi^n_{\bX}(\bt)}\varphi^n_{A}(s)\overline{\varphi^n_{A}(s)}.
\end{alignat*}

Defining $S_j = \int_{\bbR^{p+1}} \widetilde{S}_j \omega(\bt, s) \mathrm{d}\bt \; \mathrm{d}s$ for $j=1,\dots,10$ we note that the sum of ten integrals in the statement of Theorem \ref{thm:weighted_distance_duality} is equal to
\begin{equation*}
{S}_1-2{S}_2-2{S}_3+2{S}_4+{S}_5+{S}_6+2{S}_7-2{S}_8-2{S}_9+{S}_{10}.
\end{equation*}
We can re-express the first term as the following
\begin{align*}
    \widetilde{S}_1 = {} & \frac{1}{n^2}\sum_{k,\ell}w_kw_\ell\exp\{i\langle \bt,\bX_k\rangle + isA_k -i\langle\bt,\bX_\ell\rangle -isA_\ell\} \\
    = {} & \frac{1}{n^2}\sum_{k,\ell}w_kw_\ell\left\{\cos(\langle\bt,\bX_k-\bX_\ell\rangle) +  i\sin(\langle\bt,\bX_k-\bX_\ell\rangle)\right\} \\
    &\quad\quad\quad\times \left\{ \cos(s\{A_k-A_\ell\}) + i\sin(s\{A_k-A_\ell\})  \right\} \\
    = {} & \frac{1}{n^2}\sum_{k,\ell}w_kw_\ell\cos(\langle\bt,\bX_k-\bX_\ell\rangle)\cos(s\{A_k-A_\ell\}) + V_1,
\end{align*}
where $V_1$ are terms that integrate to 0 due to the anti-symmetry of the sine function when the integral \eqref{eqn:term_integral} is evaluated and the second equality holds by Euler's formula. Similarly, we have
\begin{align*}
    \widetilde{S}_2 = {} &                                               \frac{1}{n^3}\sum_{k,\ell=1}^nw_kw_\ell\sum_{m=1}^n\exp\{ i\langle \bt,\bX_k       \rangle + isA_k -i\langle \bt,\bX_\ell \rangle -isA_m \} \\
    = {} & \frac{1}{n^3}\sum_{k,\ell=1}^nw_kw_\ell\sum_{m=1}^n\cos(\langle \bt,\bX_k-\bX_\ell\rangle)\cos(s\{A_k-A_m\}) + V_2, \\
    \widetilde{S}_3 = {} & \frac{1}{n^3}\sum_{k,\ell=1}^nw_kw_\ell\sum_{m=1}^n\exp\{ i\langle \bt,\bX_k \rangle + isA_k -i\langle \bt,\bX_m \rangle -isA_\ell \} \\
    = {} & \frac{1}{n^3}\sum_{k,\ell=1}^nw_kw_\ell\sum_{m=1}^n\cos(\langle \bt,\bX_k-\bX_m\rangle)\cos(s\{A_k-A_\ell\}) + V_3,
\end{align*}
\begin{align*}
    \widetilde{S}_4 = {} & \frac{1}{n^4}\sum_{k,\ell=1}^nw_kw_\ell\sum_{j,m=1}^n\exp\{ i\langle \bt,\bX_k \rangle + isA_\ell -i\langle \bt,\bX_j \rangle -isA_m \} \\
    = {} & \frac{1}{n^4}\sum_{k,\ell=1}^nw_kw_\ell\sum_{j,m=1}^n\cos(\langle \bt,\bX_k-\bX_j\rangle)\cos(s\{A_\ell-A_m\}) + V_4, \\
    \widetilde{S}_5 = {} &
    \frac{1}{n^4}\sum_{k,\ell=1}^nw_kw_\ell\sum_{j,m=1}^n\exp\{ i\langle \bt,\bX_k \rangle + isA_j -i\langle \bt,\bX_\ell \rangle -isA_m \} \\
    = {} & \frac{1}{n^4}\sum_{k,\ell=1}^nw_kw_\ell\sum_{j,m=1}^n\cos(\langle \bt,\bX_k-\bX_\ell\rangle)\cos(s\{A_j-A_m\}) + V_5, 
\end{align*}
\begin{align*}
    \widetilde{S}_6 = {} &
    \frac{1}{n^4}\sum_{k,\ell=1}^nw_kw_\ell\sum_{j,m=1}^n\exp\{ i\langle \bt,\bX_j \rangle + isA_k -i\langle \bt,\bX_m \rangle -isA_\ell \} \\
    = {} & \frac{1}{n^4}\sum_{k,\ell=1}^nw_kw_\ell\sum_{j,m=1}^n\cos(\langle \bt,\bX_j-\bX_m\rangle)\cos(s\{A_k-A_\ell\}) + V_6, \\
    \widetilde{S}_7 = {} & \frac{1}{n^3}\sum_{k=1}^nw_k\exp\{i\langle \bt,\bX_k \rangle +isA_k\}\sum_{j,m=1}^n\exp\{ -i\langle \bt,\bX_j \rangle -isA_m \} \\
    = {} & \frac{1}{n^3}\sum_{k=1}^nw_k\sum_{j,m=1}^n\cos(\langle \bt,\bX_k-\bX_j\rangle)\cos(s\{A_k-A_m\}) + V_7, 
\end{align*}
\begin{align*}
    \widetilde{S}_8 = {} & \frac{1}{n^4}\sum_{k=1}^nw_k\exp\{i\langle \bt,\bX_k \rangle\}\sum_{j,\ell,m=1}^n\exp\{ -i\langle \bt,\bX_j \rangle+isA_\ell -isA_m \} \\
    = {} & \frac{1}{n^4}\sum_{k=1}^nw_k\sum_{j,\ell,m=1}^n\cos(\langle \bt,\bX_k-\bX_j\rangle)\cos(s\{A_\ell-A_m\}) + V_8, 
\end{align*}
\begin{align*}
    \widetilde{S}_9 = {} & \frac{1}{n^4}\sum_{k=1}^nw_k\exp\{isA_k\}\sum_{j,\ell,m=1}^n\exp\{ i\langle \bt,\bX_j \rangle -i\langle \bt,\bX_\ell \rangle -isA_m \} \\
    = {} & \frac{1}{n^4}\sum_{k=1}^nw_k\sum_{j,\ell,m=1}^n\cos(\langle \bt,\bX_j-\bX_\ell\rangle)\cos(s\{A_k-A_m\}) + V_9,  \text{ and }\\
    \widetilde{S}_{10} = {} &
    \frac{1}{n^4}\sum_{j,k,\ell,m=1}^n\exp\{   i\langle \bt,\bX_j \rangle -i\langle \bt,\bX_k \rangle + isA_\ell -isA_m  \} \\
    = {} & \frac{1}{n^4}\sum_{j,k,\ell,m=1}^n\cos(\langle \bt,\bX_j-\bX_k \rangle)\cos(s(A_\ell-A_m)) + V_{10},
\end{align*}
where similarly to $V_1$, $V_j$ for $j=2,\dots,10$ integrate to 0 when the integral \eqref{eqn:term_integral} is evaluated. 
To complete the proof, we will now show that $$\calV^2_{n,\bw}(\bX,A) = {S}_1-2{S}_2-2{S}_3+2{S}_4+{S}_5+{S}_6+2{S}_7-2{S}_8-2{S}_9+{S}_{10}. $$
Using the assumption that $\sum_{i=1}^nw_i=n$, we can expand $\frac{1}{n^2}\sum_{k,\ell=1}^nw_kw_\ell C_{k\ell}D_{kl}$ as
\begin{align*}
    & \frac{1}{n^2}\sum_{k,\ell=1}^nw_kw_\ell C_{k\ell}D_{kl} \\
   = {}  & \frac{1}{n^2}\sum_{k,\ell=1}^nw_kw_\ell\lVert \bX_k -\bX_\ell \rVert_2\lvert A_k-A_\ell \rvert - \frac{1}{n^3}\sum_{k,\ell=1}^nw_kw_\ell\sum_{m=1}^n\lVert \bX_k -\bX_\ell \rVert_2\lvert A_k-A_m \rvert \\
    & - \frac{1}{n^3}\sum_{k,\ell=1}^nw_kw_\ell\sum_{m=1}^n\lVert \bX_k -\bX_\ell \rVert_2\lvert A_\ell-A_m \rvert + \frac{1}{n^2}\sum_{k,\ell=1}^nw_kw_\ell\lVert \bX_k -\bX_\ell \rVert_2\frac{1}{n^2}\sum_{k,\ell=1}^n \lvert A_k-A_\ell \rvert \\
    & -\frac{1}{n^3}\sum_{k,\ell=1}^nw_kw_\ell\sum_{m=1}^n \lVert \bX_k -\bX_m \rVert_2\lvert A_k-A_\ell \rvert + \frac{1}{n^4}\sum_{k,\ell=1}^nw_kw_\ell\sum_{j,m=1}^n \lVert \bX_k -\bX_j \rVert_2\lvert A_k-A_m \rvert \\
    & + \frac{1}{n^4}\sum_{k,\ell=1}^nw_kw_\ell\sum_{j,m=1}^n \lVert \bX_k -\bX_j \rVert_2\lvert A_\ell-A_m \rvert 
    -\frac{1}{n^2}\sum_{k}^nw_k\sum_{m=1}^n\lVert \bX_k -\bX_m \rVert_2\frac{1}{n^2}\sum_{k,\ell=1}^n \lvert A_k-A_\ell \rvert \\
    & -\frac{1}{n^3}\sum_{k,\ell=1}^nw_kw_\ell\sum_{m=1}^n \lVert \bX_\ell -\bX_m \rVert_2\lvert A_k-A_\ell\rvert
    + \frac{1}{n^4}\sum_{k,\ell=1}^nw_kw_\ell\sum_{j,m=1}^n \lVert \bX_\ell -\bX_j \rVert_2\lvert A_k-A_m \rvert \\
    & + \frac{1}{n^4}\sum_{k,\ell=1}^nw_kw_\ell\sum_{j,m=1}^n \lVert \bX_\ell -\bX_j \rVert_2\lvert A_\ell-A_m \rvert 
    -\frac{1}{n^2}\sum_{k}^nw_k\sum_{m=1}^n\lVert \bX_k -\bX_m \rVert_2\frac{1}{n^2}\sum_{k,\ell=1}^n \lvert A_k-A_\ell \rvert \\
    & + \frac{1}{n^2}\sum_{k,\ell=1}^n \lVert \bX_k-\bX_\ell \rVert_2 \frac{1}{n^2}\sum_{k,\ell}^nw_kw_\ell\lvert A_k - A_\ell \rvert 
    - \frac{1}{n^2}\sum_{k,\ell=1}^n \lVert \bX_k-\bX_\ell \rVert_2 \frac{1}{n^2}\sum_{k}^nw_k\sum_{m=1}^n\lvert A_k - A_m \rvert \\
    & - \frac{1}{n^2}\sum_{k,\ell=1}^n \lVert \bX_k-\bX_\ell \rVert_2 \frac{1}{n^2}\sum_{k}^nw_k\sum_{m=1}^n\lvert A_k - A_m \rvert 
    + \frac{1}{n^2}\sum_{k,\ell=1}^n \lVert \bX_k-\bX_\ell \rVert_2 \frac{1}{n^2}\sum_{k,\ell}^n\lvert A_k - A_\ell \rvert \\
    = {} & \frac{1}{n^2}\sum_{k,\ell=1}^nw_kw_\ell\lVert \bX_k -\bX_\ell \rVert_2\lvert A_k-A_\ell \rvert
    -\frac{2}{n^3}\sum_{k,\ell=1}^nw_kw_\ell\sum_{m=1}^n\lVert \bX_k -\bX_\ell \rVert_2\lvert A_k-A_m \rvert \\
    & -\frac{2}{n^3}\sum_{k,\ell=1}^nw_kw_\ell\sum_{m=1}^n\lVert \bX_k -\bX_m \rVert_2\lvert A_k-A_\ell \rvert  +\frac{2}{n^3}\sum_{k=1}^nw_k\sum_{j,m=1}^n\lVert \bX_k -\bX_j \rVert_2\lvert A_k-A_m \rvert \\
    & + \frac{2}{n^4}\sum_{k,\ell=1}^nw_kw_\ell\sum_{j,m=1}^n\lVert \bX_k -\bX_j \rVert_2\lvert A_\ell-A_m \rvert  +\frac{1}{n^4}\sum_{k,\ell=1}^nw_kw_\ell\sum_{j,m=1}^n \lVert \bX_k -\bX_\ell \rVert_2\lvert A_j-A_m \rvert \\
    & + \frac{1}{n^4}\sum_{k,\ell=1}^nw_kw_\ell\sum_{j,m=1}^n \lVert \bX_j -\bX_m \rVert_2\lvert A_k-A_\ell \rvert 
    - \frac{2}{n^4}\sum_{k}^nw_k\sum_{j,\ell,m=1}^n \lVert \bX_k -\bX_j \rVert_2\lvert A_\ell-A_m \rvert \\
    & -\frac{2}{n^4}\sum_{k}^nw_k\sum_{j,\ell,m=1}^n \lVert \bX_j -\bX_\ell \rVert_2\lvert A_k-A_m \rvert 
    + \frac{1}{n^4}\sum_{j,k,\ell,m=1}^n\lVert \bX_j -\bX_k \rVert_2\lvert A_\ell-A_m \rvert. \numberthis \label{eqn:ten_sums}
\end{align*}
As in the proof of Theorem 1 of \citet{szekely2007measuring}, an application of Lemma 1 of \citet{szekely2007measuring} implies 
\begin{equation}\label{eqn:lem1}
    \lVert \bX_j -\bX_k \rVert_2\lvert A_\ell-A_m \rvert = \int_{\bbR^{p+1}}\left\{1 - \cos(\langle\bt,\bX_j -\bX_k\rangle) \right\}\left\{ 1 - \cos(s\{A_\ell-A_m\}) \right\} \omega(\bt, s) \mathrm{d}\bt \; \mathrm{d}s.
\end{equation}
In the following we use $\mathrm{d} \omega$ as short-hand for $\omega(\bt, s) \mathrm{d}\bt \; \mathrm{d}s$ or $\omega(s)\mathrm{d}s$ or $\omega(\bt) \mathrm{d}\bt$ when unambiguous.
An application of \eqref{eqn:lem1} to every term after the last equality in \eqref{eqn:ten_sums} and cancelling out terms after multiplying every term like $(1-\cos u)(1-\cos v) = 1 - \cos u - \cos v + \cos u\cos v$ yields
\begin{align*}
    & \frac{1}{n^2}\sum_{k,\ell=1}^nw_kw_\ell C_{k\ell}D_{kl} \\
    = {} & \frac{1}{n^2}\sum_{k,\ell=1}^nw_kw_\ell \int_{\bbR^{p+1}} \cos\langle\bt, \bX_k -\bX_\ell \rangle \cos(s\{A_k-A_\ell\}) \mathrm{d} \omega \hfill &  (=S_1) \\
    & -\frac{2}{n^3}\sum_{k,\ell=1}^nw_kw_\ell\sum_{m=1}^n\int_{\bbR^{p+1}} \cos\langle\bt, \bX_k -\bX_\ell \rangle \cos(s\{ A_k-A_m\}) \mathrm{d} \omega \hfill &  (=-2S_2)\\
    & -\frac{2}{n^3}\sum_{k,\ell=1}^nw_kw_\ell\sum_{m=1}^n \int_{\bbR^{p+1}} \cos\langle\bt, \bX_k -\bX_m \rangle \cos(s\{  A_k-A_\ell\}) \mathrm{d} \omega \hfill &  (=-2S_3) \\
    & + \frac{2}{n^3}\sum_{k=1}^nw_k\sum_{j,m=1}^n \int_{\bbR^{p+1}} \cos\langle\bt,  \bX_k -\bX_j \rangle \cos(s\{ A_k-A_m \}) \mathrm{d} \omega \hfill &  (=2S_7) \\
    & + \frac{2}{n^4}\sum_{k,\ell=1}^nw_kw_\ell\sum_{j,m=1}^n \int_{\bbR^{p+1}} \cos\langle\bt, \bX_k -\bX_j \rangle \cos(s\{  A_\ell-A_m\}) \mathrm{d} \omega \hfill &  (=2S_4)\\
    & +\frac{1}{n^4}\sum_{k,\ell=1}^nw_kw_\ell\sum_{j,m=1}^n \int_{\bbR^{p+1}} \cos\langle\bt,  \bX_k -\bX_\ell \rangle \cos(s\{ A_j-A_m \}) \mathrm{d} \omega \hfill &  (=S_5)\\
    & + \frac{1}{n^4}\sum_{k,\ell=1}^nw_kw_\ell\sum_{j,m=1}^n \int_{\bbR^{p+1}} \cos\langle\bt, \bX_j -\bX_m \rangle \cos(s\{ A_k-A_\ell \}) \mathrm{d} \omega \hfill &  (=S_6)\\
    & - \frac{2}{n^4}\sum_{k}^nw_k\sum_{j,\ell,m=1}^n \int_{\bbR^{p+1}} \cos\langle\bt, \bX_k -\bX_j \rangle \cos(s\{ A_\ell-A_m \}) \mathrm{d} \omega \hfill &  (=-2S_8)\\
    & -\frac{2}{n^4}\sum_{k}^nw_k\sum_{j,\ell,m=1}^n \int_{\bbR^{p+1}} \cos\langle\bt, \bX_j -\bX_\ell \rangle \cos(s\{ A_k-A_m \}) \mathrm{d} \omega \hfill &  (=-2S_9) \\
    & +\frac{1}{n^4}\sum_{j,k,\ell,m=1}^n \int_{\bbR^{p+1}} \cos\langle\bt,  \bX_j -\bX_k \rangle \cos(s\{  A_\ell-A_m \}) \mathrm{d} \omega \hfill &  (=S_{10})\\
    = {} & S_1 -2S_2-2S_3+2S_4+S_5+S_6+2S_7-2S_8-2S_9+S_{10}.
\end{align*}
Thus, the proof of the first result is complete. 
\end{proof}

\begin{proof}[Proof of Theorem \ref{thm:weighted_distance_criterion_duality}]
For this proof, we consider a more general version of our criterion \eqref{eqn:bal_criterion}. We consider a version which allows for potentially differing emphasis on the energy distance terms:
\begin{equation}\label{eqn:bal_criterion_gamma}
\calD_\gamma(\bw) = \calV^2_{n,\bw}(\bX,A) + \gamma\left\{\calE(F^n_{\bX,\bw}, F^n_{\bX}) + \calE(F^n_{A,\bw}, F^n_{A}) \right\},
\end{equation}
where $0 < \gamma \leq 1$ is any strictly positive number less than or equal to 1. The choice of $\gamma$ is not particularly important, however it should not be chosen to be arbitrarily large, as this may de-emphasize the dependence measure component of $\calD_\gamma(\bw)$ and over-emphasize the marginal distributions of $\bX$ and $A$. We recommend using either $\gamma = 1$ or $\gamma = 0.5$.

Clearly, if $\varphi^n_{\bX, A, \bw}(\bt, s) = \varphi^n_{\bX}(\bt)\varphi^n_A(s)$, $\varphi^n_{\bX,\bw}(\bt) = \varphi^n_{\bX}(\bt)$, and $\varphi^n_{A,\bw}(s) = \varphi^n_{A}(s)$ for all $(\bt, s)\in \bbR^{p+1}$, then $\calD_\gamma(\bw)$ = 0. We now prove the reverse direction of the if and only if statement also holds. 

Trivially, if $\calD_\gamma(\bw)=0$, then $\calE_{n}(F_{\bX,\bw}, F_{\bX}) + \calE_{n}(F_{A,\bw}, F_{A}) = 0$, which implies that $\varphi^n_{\bX,\bw}(\bt) = \varphi^n_{\bX}(\bt)$, and $\varphi^n_{A,\bw}(s) = \varphi^n_{A}(s)$ for all $(\bt, s)\in \bbR^{p+1}$ as per \citet{huling2020energy}.

By adding and subtracting $\varphi^n_{\bX,\bw}(\bt)\varphi^n_{A}(s)$, $\varphi^n_{\bX}(\bt)\varphi^n_{A,\bw}(s)$, and $\varphi^n_{\bX}(\bt)\varphi^n_{A}(s)$ and two applications of the triangle inequality, we have
\begin{align*}
    |\varphi^n_{\bX,A,\bw}(\bt,s) - \varphi^n_{\bX}(\bt)\varphi^n_{A}(s)| \leq {} &  |\varphi^n_{\bX,A,\bw}(\bt,s) - \varphi^n_{\bX,\bw}(\bt)\varphi^n_{A}(s) - \varphi^n_{\bX}(\bt)\varphi^n_{A,\bw}(s) + \varphi^n_{\bX}(\bt)\varphi^n_{A}(s)|\\ 
    & + |\varphi^n_A(s)|\cdot|\varphi^n_{\bx,\bw}(\bt) - \varphi_{\bx}(\bt)| + |\varphi^n_{\bx}(\bt)|\cdot|\varphi^n_{A,\bw}(s) - \varphi_{A}(s)| \\
    \leq {} & |\varphi^n_{\bX,A,\bw}(\bt,s) - \varphi^n_{\bX,\bw}(\bt)\varphi^n_{A}(s) - \varphi^n_{\bX}(\bt)\varphi^n_{A,\bw}(s) + \varphi^n_{\bX}(\bt)\varphi^n_{A}(s)| \\ 
    & + c\left\{|\varphi^n_{\bx,\bw}(\bt) - \varphi_{\bx}(\bt)| + |\varphi^n_{A,\bw}(s) - \varphi_{A}(s)|\right\},
\end{align*}
where $c = \sup_{\bt,s}(|\varphi^n_{\bx}(\bt)|, |\varphi^n_{A}(s)|)$. Due to the properties of characteristic functions, $c\leq 1$. Note that for scalars $z,a,b,d,\zeta$ with $\zeta \leq b$, it holds that if $z \leq a + bd$, then $\frac{\zeta}{b}z \leq a + \zeta d$.  Using this result and since $c\leq 1$, there exists a constant $0 < \eta=\sqrt{\gamma}\leq 1 $ such that
\begin{align*}
    \eta|\varphi^n_{\bX,A,\bw}(\bt,s) - \varphi^n_{\bX}(\bt)\varphi^n_{A}(s)| 
    \leq {} & |\varphi^n_{\bX,A, \bw}(\bt,s) - \varphi^n_{\bX,\bw}(\bt)\varphi^n_{A}(s) - \varphi^n_{\bX}(\bt)\varphi^n_{A,\bw}(s) + \varphi^n_{\bX}(\bt)\varphi^n_{A}(s)| \\ 
    & + \sqrt{\gamma}\left\{|\varphi^n_{\bx,\bw}(\bt) - \varphi^n_{\bx}(\bt)| + |\varphi^n_{A,\bw}(s) - \varphi^n_{A}(s)|\right\},
\end{align*}
thus
\begin{align}
    & \gamma\int|\varphi^n_{\bX,A,\bw}(\bt,s) - \varphi^n_{\bX}(\bt)\varphi^n_{A}(s)|^2 \mathrm{d}\omega \nonumber \\
    \leq {} & \int|\varphi^n_{\bX,A,\bw}(\bt,s) - \varphi^n_{\bX,\bw}(\bt)\varphi^n_{A}(s) -\varphi^n_{\bX}(\bt)\varphi^n_{A,\bw}(s) + \varphi^n_{\bX}(\bt)\varphi^n_{A}(s)|^2\mathrm{d}\omega \nonumber \\ 
    & + \gamma\left\{\int|\varphi^n_{\bx,\bw}(\bt) - \varphi^n_{\bx}(\bt)|^2\mathrm{d}\omega + \int|\varphi^n_{A,\bw}(s) - \varphi^n_{A}(s)|^2\mathrm{d}\omega\right\} \nonumber \\
    & + 2\sqrt{\gamma}\int|\varphi^n_{\bX,A,\bw}(\bt,s) - \varphi^n_{\bX,\bw}(\bt)\varphi^n_{A}(s) -\varphi^n_{\bX}(\bt)\varphi^n_{A,\bw}(s) + \varphi^n_{\bX}(\bt)\varphi^n_{A}(s)| \nonumber \\
    & \;\;\;\;\;\;\;\;\;\;\;\;\;\times \left(|\varphi^n_{\bx,\bw}(\bt) - \varphi^n_{\bx}(\bt)| + |\varphi^n_{A,\bw}(s) - \varphi^n_{A}(s)| \right)\mathrm{d}\omega \nonumber \\
    & + 2\int |\varphi^n_{\bx,\bw}(\bt) - \varphi^n_{\bx}(\bt)| \cdot |\varphi^n_{A,\bw}(s) - \varphi^n_{A}(s)|\mathrm{d}\omega \nonumber \\
    \leq {} & \int|\varphi^n_{\bX,A,\bw}(\bt,s) - \varphi^n_{\bX,\bw}(\bt)\varphi^n_{A}(s) -\varphi^n_{\bX}(\bt)\varphi^n_{A,\bw}(s) + \varphi^n_{\bX}(\bt)\varphi^n_{A}(s)|^2\mathrm{d}\omega \nonumber \\ 
    & + \gamma\left\{\int|\varphi^n_{\bx,\bw}(\bt) - \varphi^n_{\bx}(\bt)|^2\mathrm{d}\omega + \int|\varphi^n_{A,\bw}(s) - \varphi^n_{A}(s)|^2\mathrm{d}\omega\right\} \nonumber \\
    & + \sqrt{\gamma}\left(\int|\varphi^n_{\bX,A,\bw}(\bt,s) - \varphi^n_{\bX,\bw}(\bt)\varphi^n_{A}(s) -\varphi^n_{\bX}(\bt)\varphi^n_{A,\bw}(s) + \varphi^n_{\bX}(\bt)\varphi^n_{A}(s)|^2\mathrm{d}\omega \right. \nonumber \\
    & \;\;\;\;\;\;\;\;\;\;\;\;\; \left. +  \int |\varphi^n_{\bx,\bw}(\bt) - \varphi^n_{\bx}(\bt)|^2\mathrm{d}\omega + \int |\varphi^n_{A,\bw}(s) - \varphi^n_{A}(s)|^2 \mathrm{d}\omega \right) \nonumber \\
    & + \left(\int |\varphi^n_{\bx,\bw}(\bt) - \varphi^n_{\bx}(\bt)|^2 \mathrm{d}\omega + \int|\varphi^n_{A,\bw}(s) - \varphi^n_{A}(s)|^2\mathrm{d}\omega\right) \nonumber \\
    & \leq (1 + \max\{\sqrt{\gamma}, 1/\gamma + 1/\sqrt{\gamma}\} ) \calD_\gamma(\bw) = c_\gamma \calD_\gamma(\bw), \label{eqn:criterion_bound_indep_dist}
\end{align}
where the second inequality is due to the inequality $2ab\leq a^2+b^2$ for scalars $a,b$.

Therefore $\calD_\gamma(\bw)=0 \implies \int|\varphi^n_{\bX,A,\bw}(\bt,s) - \varphi^n_{\bX}(\bt)\varphi^n_{A}(s)|^2 \mathrm{d}\omega = 0$ and the proof is complete. If we choose $\gamma=1$, then $\int|\varphi^n_{\bX,A,\bw}(\bt,s) - \varphi^n_{\bX}(\bt)\varphi^n_{A}(s)|^2 \mathrm{d}\omega \leq 3\calD_1(\bw)$. When $\gamma=0.5$, $c_\gamma\approx 4.4$. 

\end{proof}


\begin{proof}[Proof of Lemma \ref{thm:lem1}]
The key term can be bounded as 
\begin{align*}
& \left[ \int_\calX g(\bx,a) \mathrm{d} \left[ F^n_{\bX, A, \bw} - F^n_\bX F^n_A  \right](\bx,a) \right]^2 \\ &
\leq \|g\|^2_\calH \int_{\calX\times\calA}\int_{\calX\times\calA} \|\bx-\bx'\|_2\cdot |a-a'| \mathrm{d} \left[ F^n_{\bX, A, \bw} - F^n_\bX F^n_A  \right](\bx,a) \; \mathrm{d} \left[ F^n_{\bX, A, \bw} - F^n_\bX F^n_A  \right](\bx',a') \\
& = \|g\|^2_\calH \int|\varphi^n_{\bX,A,\bw}(\bt,s) - \varphi^n_{\bX}(\bt)\varphi^n_{A}(s)|^2 \mathrm{d}\omega \\
& \leq 
c_\gamma \|g\|^2_\calH D_\gamma(\bw),
\end{align*}
where the first inequality holds due to the arguments in the proof of Theorem 4 of \citet{mak2018support}, the equality holds due to the representations of Equations (2.9) and (3.3) and Theorem 24 of \citet{sejdinovic2013equivalence}, and the second inequality holds due to our Theorem \ref{thm:weighted_distance_criterion_duality}. When $\gamma=1$, then $c_\gamma =3$. Thus, the result is shown.
\end{proof}

\begin{proof}[Proof of Theorem \ref{thm:weighted_distance_convergence}]
Define the true density ratio weights as $w^*(\bx,a) = f(A=a)/f(A=a|\bX=\bx)$. Then define $\bw^*_n = (w^*(\bX_1,A_1), \dots, w^*(\bX_n,A_n))\trans$ as the vector of density ratio weights and the normalized version of these weights as $$\widetilde{\bw}^*_n = (w^*(\bX_1,A_1), \dots, w^*(\bX_n,A_n))\trans / \left(\frac{1}{n}\sum_{i=1}^nw^*(\bX_i,A_i)\right).$$ We use the normalized weights as they converge to the same quantity as the unnormalized true weights, but are directly comparable to our weights via our distance criterion, whose specific formulation requires that the weights sum to $n$.
By the strong law of large numbers, we have $\lim_{n\to\infty}F^n_{\bX,A,\widetilde{\bw}^*_n}(\bx,a) = F_{\bX}(\bx)F_A(a)$ a.e. for every continuity point $(\bx,a)$, as in \citet{tokdar2010importance} and \citet{huling2020energy}. Similarly, $\lim_{n\to\infty}F^n_{\bX,\widetilde{\bw}^*_n}(\bx) = F_{\bX}(\bx)$ a.e. for every continuity point $\bx$ and $\lim_{n\to\infty}F^n_{A,\widetilde{\bw}^*_n}(a) = F_{A}(a)$ a.e. for every continuity point $a$. As in the proof of Theorem 2 of \citet{mak2018support}, by the Portmanteau and dominated convergence theorems, we have
\begin{align}
    &\lim_{n\to\infty}\bbE\left[ \lvert \varphi^n_{\bX}(\bt)\varphi^n_A(s) - \varphi^n_{\bX,A,\widetilde{\bw}^*_n}(\bt,s) \rvert^2 \right] = 0 \text{ for all } \bt,s \\
    &\lim_{n\to\infty}\bbE\left[ \lvert \varphi^n_{\bX}(\bt) - \varphi^n_{\bX,\widetilde{\bw}^*_n}(\bt) \rvert^2 \right] = 0 \text{ for all } \bt,s, \\
    &\lim_{n\to\infty}\bbE\left[ \lvert \varphi^n_A(s) - \varphi^n_{A,\widetilde{\bw}^*_n}(s) \rvert^2 \right] = 0 \text{ for all } \bt,s, \text{ and} \\
    &\lim_{n\to\infty}\bbE\left[ \lvert \varphi^n_{\bX,A,\widetilde{\bw}^*_n}(\bt,s) - \varphi^n_{\bX,\bw}(\bt)\varphi^n_A(s) - \varphi^n_{\bX}(\bt)\varphi^n_{A,\bw}(s) +  \varphi^n_{\bX}(\bt)\varphi^n_A(s)   \rvert^2 \right] = 0 \text{ for all } \bt,s 
\end{align}
Following the arguments in the proofs of Theorem 3.1 of \citet{huling2020energy} and Theorem 2 of \citet{mak2018support}, the expected distance criterion $\bbE\left[ \calD_\gamma(\widetilde{\bw}^*_n) \right]$ converges to zero as $n\to\infty$. Since for every $n$, we have $\calD_\gamma(\bw^d_n) \leq \calD_\gamma(\widetilde{\bw}^*_n)$ by definition of $\bw^d_n$, we also have $\calD_\gamma(\bw^d_n) \leq \bbE\left[\calD_\gamma(\widetilde{\bw}^*_n)\right]$. Thus, 
\begin{align} 
    &\lim_{n\to\infty}\left\{\int|\varphi^n_{\bX,A,\bw}(\bt,s) - \varphi^n_{\bX,\bw}(\bt)\varphi^n_{A}(s) -\varphi^n_{\bX}(\bt)\varphi^n_{A,\bw}(s) + \varphi^n_{\bX}(\bt)\varphi^n_{A}(s)|^2\mathrm{d}\omega \right. \nonumber \\ 
    & \left. \;\;\;\quad + \gamma\left(\int|\varphi^n_{\bx,\bw}(\bt) - \varphi^n_{\bx}(\bt)|^2\mathrm{d}\omega + \int|\varphi^n_{A,\bw}(s) - \varphi^n_{A}(s)|^2\mathrm{d}\omega\right)\right\} \nonumber \\
    = {} & \lim_{n\to\infty}\calD_\gamma(\bw^d_n)  = 0.\label{eqn:limit_opt_dist}
\end{align}

We now apply the bound \eqref{eqn:criterion_bound_indep_dist} from the proof of Theorem \ref{thm:weighted_distance_criterion_duality} with weights $\bw=\bw^d$, where here we drop the subscript $n$ in $\bw^d_n$ for notational simplicity.
Thus, we have
\begin{equation} \label{eqn:target_bound5}
    \lim_{n\to\infty} \int|\varphi^n_{\bX,A,\bw^d}(\bt,s) - \varphi^n_{\bX}(\bt)\varphi^n_{A}(s)|^2 \mathrm{d}\omega = 0
\end{equation}
a.s. Then by adding and subtracting $\varphi^n_{\bX}\varphi^n_{A}$ inside the square of $|\varphi^n_{\bX,A,\bw^d}(\bt,s) - \varphi_{\bX}(\bt)\varphi_{A}(s)|^2$, and applying the Minkowski inequality we have
\begin{align}
     & \int|\varphi^n_{\bX,A,\bw^d}(\bt,s) - \varphi_{\bX}(\bt)\varphi_{A}(s)|^2 \mathrm{d}\omega \label{eqn:target_dist}\\
      &\leq \left( \left[\int|\varphi^n_{\bX,A,\bw^d}(\bt,s) - \varphi^n_{\bX}(\bt)\varphi^n_{A}(s)|^2 \mathrm{d}\omega \right]^\frac{1}{2} + 
     \left[\int|\varphi_{\bX}(\bt)\varphi_{A}(s) - \varphi^n_{\bX}(\bt)\varphi^n_{A}(s)|^2 \mathrm{d}\omega \right]^{
     \frac{1}{2}} \right)^{2} \label{eqn:target_dist_bounds}
\end{align}
The first term inside the square in \eqref{eqn:target_dist_bounds} converges to zero a.e. due to \eqref{eqn:target_bound5} and the second term converges to zero a.e. by adding and subtracting $\varphi^n_{\bX}\varphi_A$, an application of the triangle inequality, Slutsky's theorem, the integrability of $\varphi_\bX$ and $\varphi_A$, and since $\varphi^n_{\bX}$ and $\varphi^n_{A}$ converge almost everywhere to $\varphi_\bX$ and $\varphi_A$, respectively. Thus, we have that 
\begin{equation*}
    \lim_{n\to\infty}\int|\varphi^n_{\bX,A,\bw^d}(\bt,s) - \varphi_{\bX}(\bt)\varphi_{A}(s)|^2 \mathrm{d}\omega =0 \\
\end{equation*}
almost everywhere. We now follow the arguments of \citet{huling2020energy} to show that this implies a.e. convergence of  $\varphi^n_{\bX,A,\bw^d}(\bt,s)$ to $\varphi_{\bX}(\bt)\varphi_{A}(s)$ almost surely. If we choose any subsequence $\{n_k\}_{k=1}^\infty$ of $\mathbb{N}_+$, we have the same property that $\lim_{k\to\infty}  \int|\varphi^{n_k}_{\bX,A,\bw^d_{n_k}}(\bt,s) - \varphi_{\bX}(\bt)\varphi_{A}(s)|^2 \mathrm{d}\omega =0$. By the Riesz-Fischer Theorem, a sequence of functions $f_n$ which converge to $f$ in $L_2$ has a subsequence $f_{n_k}$ which converges almost surely to $f$, implying the existence of a subsubsequence $\{n'_k\}_{k=1}^\infty \subseteq\{n_k\}_{k=1}^\infty$ such that ${\varphi}^{n'_k}_{\bX,A,\bw^d_{n'_k}}(\bt,s)$ converges to ${\varphi}_{\bX}(\bt)\varphi_A(s)$ almost surely as $k\to\infty$. Since $(n_k)$ was chosen arbitrarily, $\lim_{n\to\infty}\varphi^n_{\bX,A,\bw^d_n}(\bt,s) = \varphi_{\bX}(\bt)\varphi_{A}(s)$ almost surely. Thus the first statement of the proof is complete. By similar arguments, it can be shown that $\lim_{n\to\infty}\varphi^n_{\bX,\bw^d_n}(\bt) = \varphi_{\bX}(\bt)$ and $\lim_{n\to\infty}\varphi^n_{A,\bw^d_n}(s) = \varphi_{A}(s)$ almost surely almost surely.

Now, if $\bbE {w^*}^2(\bX,A) < \infty$, we have due to the law of large numbers and Slutsky's theorem that $\lim_{n\to\infty} \frac{1}{n}\sum_{i=1}^n{\widetilde{w}^*{}}_i^2 = \bbE {w^*}^2(\bX,A)$ almost surely and thus $\lim_{n\to\infty}\lambda \frac{1}{n^2}\sum_{i=1}^n{\widetilde{w}^*{}}_i^2=0$ almost surely. Thus using the same arguments as to show \eqref{eqn:target_bound5}, we also have
\begin{equation} \label{eqn:target_bound5_pd}
    \lim_{n\to\infty} \int|\varphi^n_{\bX,A,\bw^{pd}}(\bt,s) - \varphi^n_{\bX}(\bt)\varphi^n_{A}(s)|^2 \mathrm{d}\omega = 0
\end{equation}
and similar following arguments further show that $\lim_{n\to\infty}\varphi^n_{\bX,A,\bw^{pd}_n}(\bt,s) = \varphi_{\bX}(\bt)\varphi_{A}(s)$ almost surely. 
\end{proof}


\begin{proof}[Proof of Theorem \ref{thm:thm3}]
We investigate the limit of $\widehat{\mu}_{NW}^{\bw}(a_0) - \mu(a_0)$ as $nh\to\infty$ and $h\to 0$ for both $\bw=\bw^d$ and $\bw=\bw^{pd}$. For reference, we include the error decomposition of a weighted nonparametric estimate of the ADRF. As stated in the main text, given \textit{any} $\bw$, the error of \eqref{eqn:weighted_nw_estimator} at $A=a_0$ can be decomposed as 
\begin{align}
\widehat{\mu}_{NW}^{\bw}(a_0) - \mu(a_0)  = {} & \int_{\calX} \int_\calA \mu(\bx, a_0) \mathrm{d} \left[ F^n_{\bX,A,\bw}(\bx,a) - F^n_{\bX}(\bx)F^n_{A}(a) \right]  \label{eqn:error_decomp_indep} \\
& + \int_{\calX} \mu(\bx,a_0)\mathrm{d} \left[ F^n_{\bX} - F_\bX \right](\bx)   \label{eqn:error_decomp_sampling} \\ 
    & + \left(\frac{f_A(a_0)}{\widehat{f}^n_{A,h}(a_0)} - 1\right) \int_{\calX} \mu(\bx, a_0)\mathrm{d}F^n_{\bX}(\bx)  \label{eqn:error_decomp_density} \\
    & + \left(\frac{f_A(a_0)}{\widehat{f}^n_{A,h}(a_0)} - 1\right) \int_{\calX} \int_\calA \mu(\bx, a_0) \mathrm{d} \left[ F^n_{\bX,A,\bw}(\bx,a) - F^n_{\bX}(\bx)F^n_{A}(a) \right]  \label{eqn:error_decomp_wts_density1} \\
    &+{\widehat{f}^n{}}^{-1}_{A,h}(a_0)\int_\calX\int_\calA \left[\mu(\bx,a)K_h(a-a_0) - \mu(\bx,a_0)f_A(a_0)\right]\mathrm{d}F^n_{\bX,A,\bw}(\bx,a)  \label{eqn:error_decomp_wts_density2} \\
    & + \frac{1}{n}\sum_{i=1}^n \varepsilon_i w_i {\widehat{f}^n{}}^{-1}_{A,h}(a_0) {K_h(A_i - a_0)}, \label{eqn:error_decomp_wts_error}
\end{align}
where $\widehat{f}^n_{A,h}(a_0) = \int_\calA K_h(a-a_0)\mathrm{d}F^n_{A}(a)$ is a kernel density estimate of $f_A(a_0)$.
We use the decomposition of \eqref{eqn:error_decomp_indep}-\eqref{eqn:error_decomp_wts_error}. 
The key term in the error decomposition is the systematic bias term \eqref{eqn:error_decomp_indep}. We re-express this term plus the second term \eqref{eqn:error_decomp_sampling} as
\begin{align}
    & \int_{\calX} \int_\calA \mu(\bx, a_0) \mathrm{d} \left[ F^n_{\bX,A,\bw}(\bx,a) - F_{\bX}(\bx)F_{A}(a) \right]\nonumber
\end{align}
By the Portmanteau Theorem (Theorem 2.1, \citealp{billingsley1993convergence}) and Theorem \ref{thm:weighted_distance_convergence}, it follows that
\begin{equation*}
    \lim_{n\to\infty} \int_{\calX} \int_\calA \mu(\bx, a_0) \mathrm{d} F^n_{\bX,A,\bw}(\bx,a) = \int_{\calX} \int_\calA \mu(\bx, a_0) \mathrm{d} F_{\bX}(\bx)\mathrm{d}F_{A}(a)
\end{equation*}
almost surely for both $\bw=\bw^d$ and $\bw=\bw^{pd}$. The remaining terms in the error decomposition converge to zero without the need of careful consideration of the exact limiting behavior of the weights.

The term \eqref{eqn:error_decomp_density} is also independent of the weights but depends on the kernel density estimate $\widehat{f}^n_{A,h}(a_0)$. We note that under our assumptions on the kernel, due to standard asymptotic results for kernel density estimation we have $\widehat{f}^n_{A,h}(a_0)-f_{A}(a_0) = O_p((nh)^{-1/2}) + O_p(h^2)$.  We re-express \eqref{eqn:error_decomp_density} as
\begin{align}
    \left(\frac{f_A(a_0)}{\widehat{f}^n_{A,h}(a_0)} - 1\right) \int_{\calX} \mu(\bx, a_0)\mathrm{d}F^n_{\bX}(\bx) = {} & {\widehat{f}^n{}}^{-1}_{A,h}(a_0)(f_A(a_0) - \widehat{f}^n_{A,h}(a_0))\int_{\calX} \mu(\bx, a_0)\mathrm{d}F_{\bX}(\bx) \label{eqn:density_term_1} \\
    & + {\widehat{f}^n{}}^{-1}_{A,h}(a_0)(f_A(a_0) - \widehat{f}^n_{A,h}(a_0))\int_{\calX} \mu(\bx, a_0)\mathrm{d}\left[F^n_{\bX} - F_{\bX}\right](\bx).\label{eqn:density_term_2}
\end{align}
The term \eqref{eqn:density_term_1} is $O_p((nh)^{-1/2}) + O_p(h^2)$ and the term \eqref{eqn:density_term_2} is $[O_p((nh)^{-1/2}) + O_p(h^2)]O_p(n^{-1/2})$. Similarly, we have that \eqref{eqn:error_decomp_wts_density1} can be re-expressed as 
\begin{align}
    &\left(\frac{f_A(a_0)}{\widehat{f}^n_{A,h}(a_0)} - 1\right) \int_{\calX} \int_\calA \mu(\bx, a_0) \mathrm{d} \left[ F^n_{\bX,A,\bw}(\bx,a) - F^n_{\bX}(\bx)F^n_{A}(a) \right] \nonumber \\ 
    = {} & {\widehat{f}^n{}}^{-1}_{A,h}(a_0)(f_A(a_0) - \widehat{f}^n_{A,h}(a_0)) \int_{\calX} \int_\calA \mu(\bx, a_0) \mathrm{d} \left[ F^n_{\bX,A,\bw}(\bx,a) - F_{\bX}(\bx)F_{A}(a) \right] \label{eqn:density_wts_term_1} \\
    & + {\widehat{f}^n{}}^{-1}_{A,h}(a_0)(f_A(a_0) - \widehat{f}^n_{A,h}(a_0)) \int_{\calX} \int_\calA \mu(\bx, a_0) \mathrm{d} \left[ F_{\bX}(\bx)\left\{ F_{A}(a) - F^n_{A}(a) \right\} \right] \label{eqn:density_wts_term_2} \\
    & + {\widehat{f}^n{}}^{-1}_{A,h}(a_0)(f_A(a_0) - \widehat{f}^n_{A,h}(a_0)) \int_{\calX} \int_\calA \mu(\bx, a_0) \mathrm{d} \left[ F^n_{A}(a) \left\{ F_{\bX}(\bx) - F^n_{\bX}(\bx)\right\} \right]. \label{eqn:density_wts_term_3}
\end{align}
By the Portmanteau Theorem, \eqref{eqn:density_wts_term_1} is $[O_p((nh)^{-1/2}) + O_p(h^2)]o_p(1)$ for both $\bw=\bw^d$ and $\bw=\bw^{pd}$. Terms \eqref{eqn:density_wts_term_2} and \eqref{eqn:density_wts_term_3} are both $[O_p((nh)^{-1/2}) + O_p(h^2)]O_p(n^{-1/2})$. The term \eqref{eqn:error_decomp_wts_density2} can be re-expressed as 
\begin{align}
    & {\widehat{f}^n{}}^{-1}_{A,h}(a_0)\int_\calX\int_\calA \left[\mu(\bx,a)K_h(a-a_0) - \mu(\bx,a_0)f_A(a_0)\right]\mathrm{d}F^n_{\bX,A,\bw}(\bx,a) \nonumber \\
    = {} & {\widehat{f}^n{}}^{-1}_{A,h}(a_0)\int_\calX\int_\calA \left[\mu(\bx,a)K_h(a-a_0) - \mu(\bx,a_0)f_A(a_0)\right] \mathrm{d}F_A(a)\mathrm{d}F_\bX(\bx) \label{eqn:density_wts2_term_1} \\
    & + {\widehat{f}^n{}}^{-1}_{A,h}(a_0)\int_\calX\int_\calA \left[\mu(\bx,a)K_h(a-a_0) - \mu(\bx,a_0)f_A(a_0)\right]\mathrm{d}\left[F^n_{\bX,A,\bw} - F_\bX F_A\right](\bx,a). \label{eqn:density_wts2_term_2}
\end{align}
The term inside the first integral of \eqref{eqn:density_wts2_term_1} can be written as
\begin{align}
    & \int_\calA \left[\mu(\bx,a)K_h(a-a_0) - \mu(\bx,a_0)f_A(a_0)\right] \mathrm{d}F_A(a) \label{eqn:term_kernel} \\
    = {} & \int_\calA \mu(\bx,a)K_h(a-a_0) \mathrm{d}F_A(a) \label{eqn:term_kernel1} \\
    & - \mu(\bx,a_0)f_A(a_0). \nonumber
\end{align}
Term \eqref{eqn:term_kernel1} can be expressed as
\begin{align}
    & \frac{1}{h}\int \mu(\bx,a)K((a-a_0)/h ) f_A(a)\mathrm{d}a \nonumber \\
    = {} & \int \mu(\bx,a_0 + uh)K( u ) f_A(a_0 +uh)\mathrm{d}u \nonumber \\
    = {} & \int K(u) \left\{\mu(\bx,a_0) + uh\mu'(\bx,a_0)+u^2h^2\mu''(\bx,a_0)/2\right\} \left(f_A(a_0) + uhf'_A(a_0)\right)\mathrm{d}u + o(h^2) \nonumber\\
    = {} & \Big(\underbrace{\int K(u)\mathrm{d}u}_{=1}\Big)\mu(\bx,a_0)f_A(a_0) + \Big(\underbrace{\int uK(u)\mathrm{d}u}_{=0}\Big)h\left\{\mu(\bx,a_0)f'_A(a_0) + \mu'(\bx,a_0)f_A(a_0)\right\} \nonumber\\
    & + \Big(\underbrace{\int u^2K(u)\mathrm{d}u}_{\equiv \kappa_2}\Big)h^2\left\{\mu'(\bx,a_0)f'_A(a_0) + \mu''(\bx,a_0)f_A(a_0)/2\right\} + o(h^2) \nonumber \\
    = {} & \mu(\bx,a_0)f_A(a_0) + h^2\kappa_2B(\bx, a_0)f_A(a_0) + o(h^2), \nonumber
\end{align}
where $B(\bx, a_0) = \mu''(\bx,a_0)/2 + \mu'(\bx,a_0)f'_A(a_0) / f_A(a_0)$ and the second equality holds by taking Taylor expansions of $\mu$ and $f_A$. Hence \eqref{eqn:term_kernel} is equal to $h^2\kappa_2B(\bx, a_0)f_A(a_0) + o(h^2)$. Thus, since the expectation is a linear operator, this implies that \eqref{eqn:density_wts2_term_1} is $O_p((nh)^{-1/2}) + O_p(h^2)$.
Similarly to \eqref{eqn:density_wts_term_1} and using similar arguments as above for \eqref{eqn:density_wts2_term_1}, \eqref{eqn:density_wts2_term_2} is $[O_p((nh)^{-1/2}) + O_p(h^2)]o_p(1)$ due to the Portmanteau Theorem for both $\bw=\bw^d$ and $\bw=\bw^{pd}$. The last term in the decomposition of $\widehat{\mu}_{NW}^{\bw}(a_0) - \mu(a_0)$ is $\frac{1}{n}\sum_{i=1}^n \varepsilon_i w_i {\widehat{f}^n{}}^{-1}_{A,h}(a_0) {K_h(A_i - a_0)}$, which has mean zero due to the definition of $\varepsilon_i$ and converges to 0 as $n\to\infty$ due to the law of large numbers regardless of the weights. Thus, all terms in the decomposition of $\widehat{\mu}_{NW}^{\bw}(a_0) - \mu(a_0)$ converge to zero in probability as $nh\to\infty$ and $h\to 0$ for both $\bw=\bw^d$ and $\bw=\bw^{pd}$.

Now we prove consistency of $\widehat{\mu}_{NWs}^{\bw}(a_0)$ for $\mu(a_0)$. We begin by noting that 
\begin{align*}
    \widehat{\mu}_{NW}^{\bw}(a_0) - \widehat{\mu}_{NWs}^{\bw}(a_0) = {} & \frac{\frac{1}{n} \sum_{i=1}^n Y_i w_i K_h(A_i - a_0) \left\{D_\bw(a_0) - D(a_0)\right\}}{D_\bw(a_0) D(a_0)},
\end{align*}
where 
\begin{align*}
    D_\bw(a_0) = \frac{1}{n} \sum_{i=1}^n w_iK_h(A_i - a_0) \text{ and }
    D(a_0) = \frac{1}{n} \sum_{i=1}^n K_h(A_i - a_0).
\end{align*}
Thus, $\widehat{\mu}_{NWs}^{\bw}(a_0)$ is consistent for $\mu(a_0)$ if $D_\bw(a_0) - D(a_0)$ is consistent for 0. We have already shown consistency of $D(a_0)$ for $f_A(a_0)$. We can express
\begin{align}
    D_\bw(a_0) - D(a_0) = {} &  \int_\calA K_h(a - a_0) \mathrm{d} \left[ F^n_{A,\bw} - F^n_{A} \right](a) \nonumber \\
    = {} &  \int_\calA K_h(a - a_0) \mathrm{d} \left[ F^n_{A} - F_{A} \right](a) \label{eqn:wtd_kernel_error_2}\\
    & + \int_\calA K_h(a - a_0) \mathrm{d} \left[ F^n_{A,\bw} - F_{A} \right](a) \label{eqn:wtd_kernel_error_1}.
\end{align}

The term \eqref{eqn:wtd_kernel_error_2} is $O_p(1/\sqrt{n})$ since the kernel has finite mean and variance and is thus $o_p(1)$. Similar to how \eqref{eqn:density_wts_term_1} was showed to converge to 0, by Theorem \ref{thm:weighted_distance_convergence} and the Portmanteau Theorem, \eqref{eqn:wtd_kernel_error_1} is $o_p(1)$. Thus,  $D_\bw(a_0) - D(a_0) = o_p(1)$ and $D_\bw(a_0)$ converges to $f_A(a_0)$ and thus by Slutksy's theorem, $\widehat{\mu}_{NW}^{\bw}(a_0) - \widehat{\mu}_{NWs}^{\bw}(a_0) = o_p(1)$. Thus, the consistency result is shown for $\widehat{\mu}_{NWs}^{\bw}(a_0)$.
\end{proof}

\begin{proof}[Proof of Lemma \ref{thm:lem_rmse}]
The proof follows from Lemma \ref{thm:lem1} and the Cauchy-Schwarz inequality.
\end{proof}

\begin{proof}[Proof of Lemma \ref{thm:lem_criterion_rate}]

We begin by noting that $\calV^2_{n,\bw^*_n}(\bX,A)$, $\calE(F^n_{\bX,\bw^*_n}$, and $F^n_{\bX}) + \calE(F^n_{A,\bw^*_n}, F^n_{A})$ can all be expressed as V-statistics, where $\bw^*_n$ and $\widetilde{\bw}^*_n$ are defined in the proof of Theorem \ref{thm:weighted_distance_convergence}. 

Similar to \citet{lyons2013distance} and \citet{jakobsen2017distance}, we can express $\calV^2_{n,\bw^*_n}(\bX,A)$ as a V-statistic with asymmetric kernel $k$ of degree 6, where $$k((\bX_1,A_1),\dots,(\bX_6,A_6)) = w^*(\bX_1,A_1) g_\bX(\bX_1,\bX_2,\bX_3,\bX_4)g_A(A_1,A_2,A_5,A_6)w^*(\bX_2,A_2)$$ with $w^*(\bx,a) = \frac{f_A(a)}{f_{A|\bX}(a|\bX=\bx)}$,
\begin{align*}
g_\bX(\bX_1,\bX_2,\bX_3,\bX_4) = {} & \| \bX_1-\bX_2\|_2 - \| \bX_1-\bX_3\|_2 - \| \bX_2-\bX_4\|_2+\| \bX_3-\bX_4\|_2, \text{ and} \\
g_A(A_1,A_2,A_3,A_4) = {} & | A_1-A_2| - | A_1-A_3| - | A_2-A_4|+| A_3-A_4|. 
\end{align*}
In other words, we have
\begin{align*}
    \calV^2_{n,\bw^*_n}(\bX,A) = \frac{1}{n^6}\sum_{i_1=1}^n\cdots\sum_{i_6=1}^nk((\bX_{i_1},A_{i_1}),\dots,(\bX_{i_6},A_{i_6})).
\end{align*}
Then, under our assumption that  $\bbE k^2<\infty$, using either the  results for V-statistics from Theorem 5.3.1 of \citet{korolyuk1989theory} or using the symmetrization argument of \citet{jakobsen2017distance} that allows for the invocation of more standard results for V-statistics in \citet{serfling1980approximation}, we have $\calV^2_{n,\bw^*_n}(\bX,A) = O_p(1/n)$, which implies that $\calV^2_{n,\widetilde{\bw}^*_n}(\bX,A) = O_p(1/n)$ due to Slutsky's theorem since $w^*_i$ and $\widetilde{w}^*_i$ only differ by a normalizing constant and have the same mean.

Similarly as in \citet{huling2020energy}, we can express both $\calE(F^n_{\bX,\bw^*_n}, F^n_{\bX})$ and $\calE(F^n_{A,\bw^*_n}, F^n_{A})$ as V-statistics with asymmetric kernels $k_\bX$ and $k_A$, respectively, both of order 4. In particular, the kernels are
\begin{align*}
    k_\bX((\bX_1,A_1),\dots,(\bX_4,A_4)) = {} & w^*(\bX_1,A_1)\| \bX_1-\bX_3\|_2 + w^*(\bX_2,A_2)\| \bX_2-\bX_4\|_2 \\
    & -w^*(\bX_1,A_1)w^*(\bX_2,A_2)\| \bX_1-\bX_2\|_2 - \| \bX_3-\bX_4\|_2 \text{ and} \\
    k_A((\bX_1,A_1),\dots,(\bX_4,A_4)) = {} & w^*(\bX_1,A_1)| A_1-A_3| + w^*(\bX_2,A_2)| A_2-A_4| \\
    & - w^*(\bX_1,A_1)w^*(\bX_2,A_2)| A_1-A_2| - | A_3-A_4|.
\end{align*}

As in the proof of Lemma A.2 in \citet{huling2020energy} and with similar arguments as above, under the assumed moment conditions that $\bbE k_A^2<\infty$ and $\bbE k_\bX^2<\infty$, we have that $\calE(F^n_{\bX,\bw^*_n}, F^n_{\bX})=O_p(1/n)$ and $\calE(F^n_{A,\bw^*_n}, F^n_{A})=O_p(1/n)$, which implies by similar arguments as the above that $\calE(F^n_{\bX,\widetilde{\bw}^*_n}, F^n_{\bX})=O_p(1/n)$ and $\calE(F^n_{A,\widetilde{\bw}^*_n}, F^n_{A})=O_p(1/n)$. Thus, since each term in $\calD(\widetilde{\bw}^*_n)$ is $O_p(1/n)$, we have that $\calD(\widetilde{\bw}^*_n) = O_p(1/n)$.

Then, since $\calD(\bw_n^{d}) \leq \calD(\widetilde{\bw}^*_n)$ for each $n$, we must have that $\calD(\bw^{d})=O_p(1/n)$ and thus we have proved the first statement of the Lemma. 

We now prove the final statements of the Lemma. Since $\bbE {w^*}^2(\bX,A) < \infty$, by the central limit theorem we have $\lambda\frac{1}{n^2}\sum_{i=1}^n{w^*}_i^2 = O_p(1/n)$, which implies that $\lambda\frac{1}{n^2}\sum_{i=1}^n{\widetilde{w}^*{}}_i^2 = O_p(1/n)$ since $w^*_i$ and $\widetilde{w}^*_i$ only differ by a normalizing constant and have the same mean. Then since $\calD(\bw_n^{pd})+\lambda\frac{1}{n^2}\sum_{i=1}^n{{w_i^{pd}{}}^2 } \leq \calD({\widetilde{\bw}^*{}}_n)+\lambda\frac{1}{n^2}\sum_{i=1}^n{\widetilde{w}^*{}}_i^2$ for each $n$, we must have that $\calD(\bw^{pd})+\lambda\frac{1}{n^2}\sum_{i=1}^n{{w_i^{pd}{}}^2 }=O_p(1/n)$. Since $\calD(\bw^{pd})=O_p(1/n)$, we also must have $\frac{1}{n^2}\sum_{i=1}^n{{w_i^{pd}{}}^2 }=O_p(1/n)$. Thus, the result is shown.
\end{proof}

\begin{proof}[Proof of Theorem \ref{thm:thm4}]
In the proof of Theorem \ref{thm:thm3}, the convergence rates of each term in the decomposition of the error term in  \eqref{eqn:error_decomp_indep}-\eqref{eqn:error_decomp_wts_error} except the terms \eqref{eqn:error_decomp_indep} plus \eqref{eqn:error_decomp_sampling}, which with the PDCOV weights is
\begin{align}
    & \int_{\calX} \int_\calA \mu(\bx, a_0) \mathrm{d} \left[ F^n_{\bX,A,\bw^{pd}}(\bx,a) - F_{\bX}(\bx)F_{A}(a) \right]\nonumber
\end{align}
and the final term \eqref{eqn:error_decomp_wts_error}, which with the PDCOV weights is $\frac{1}{n}\sum_{i=1}^n \varepsilon_i w^{pd}_i {\widehat{f}^n{}}^{-1}_{A,h}(a_0) {K_h(A_i - a_0)}.$
All other terms have convergence rates of at most $O_p(1/\sqrt{nh})$. Under the conditions of Lemmas \ref{thm:lem1} and \ref{thm:lem_criterion_rate}, we have
\begin{equation*}
    \left[\int_{\calX} \int_\calA \mu(\bx, a_0) \mathrm{d} \left[ F^n_{\bX,A,\bw^{pd}}(\bx,a) - F_{\bX}(\bx)F_{A}(a) \right] \right]^2 \leq 3||\mu||_\calH\calD_\gamma(\bw^{pd}) = O_p(1/n),
\end{equation*}
where the inequality holds due to Lemma \ref{thm:lem1} because $\mu(\bx, a_0) \in \calH_{\calX}$ implies $\mu(\bx, a_0) \in \calH$ viewing $\mu(\bx, a_0)$ as a constant in $a$ and the convergence rate of $\calD_\gamma(\bw^{pd})$ is due to Lemma \ref{thm:lem_criterion_rate}. Thus 
\begin{align*}
    & \int_{\calX} \int_\calA \mu(\bx, a_0) \mathrm{d} \left[ F^n_{\bX,A,\bw^{pd}}(\bx,a) - F_{\bX}(\bx)F_{A}(a) \right] = O_p(1/\sqrt{n}).
\end{align*}

We now investigate the convergence rate of 
$\frac{1}{n}\sum_{i=1}^n \varepsilon_i w^{pd}_i {\widehat{f}^n{}}^{-1}_{A,h}(a_0) {K_h(A_i - a_0)}.$ 
Define\\ $\overline{\sigma}^2 = \sup_{(\bx,a)\in \calX\times\calA}\text{Var}(Y(a) - \mu(\bx,a))$. 
Note that $\frac{1}{n}\sum_{i=1}^n\varepsilon_i{w}^{pd}_i = O_p(n^{-1/2})$ due to an application of Chebyshev's inequality as follows. For any $\delta >0$, conditional on $(\bX_i, A_i)_{i=1}^n$ we have with probability greater than $1-\delta^2$ that 
\begin{align*}
    \Big\lvert \frac{1}{n}\sum_{i=1}^n\varepsilon_i{w^{pd}_i{}}\Big\rvert \leq {} & \frac{1}{\delta}\frac{1}{n^{1/2}} \sqrt{\frac{1}{n} \sum_{i=1}^n\sigma^2(\bX_i,A_i){w^{pd}_i{}}^2} \\
    \leq {} & \frac{\overline{\sigma}}{\delta}\frac{1}{n^{1/2}} \sqrt{\frac{1}{n} \sum_{i=1}^n{w^{pd}_i{}}^2} = O_p(1/n^{1/2}),
\end{align*}
where $\overline{\sigma}^2 = \sup_{(\bx,a)\in \calX\times\calA}\text{Var}(Y(a) - \mu(\bx,a))$ and the second inequality holds since $\sigma^2(\bx,a)$ is assumed to be uniformly bounded over $(\bx,a)\in \calX\times\calA$ and due to Lemma \ref{thm:lem_criterion_rate}.

We now decompose the term of interest as
\begin{align}
\frac{1}{n}\sum_{i=1}^n \varepsilon_i w^{pd}_i {\widehat{f}^n{}}^{-1}_{A,h}(a_0) {K_h(A_i - a_0)} = {} & {\widehat{f}^n{}}^{-1}_{A,h}(a_0) f_A(a_0) \frac{1}{n}\sum_{i=1}^n \varepsilon_i w^{pd}_i \label{eqn:wtd_error_decomp1} \\ 
& + {\widehat{f}^n{}}^{-1}_{A,h}(a_0) \frac{1}{n}\sum_{i=1}^n\varepsilon_i w^{pd}_i(K_h(A_i - a_0) - f_A(a_0)) \label{eqn:wtd_error_decomp2},
\end{align}
where due to the above, we have $\frac{1}{n}\sum_{i=1}^n \varepsilon_i w^{pd}_i = O_p(1/\sqrt{n})$ as well as results for kernel density estimators and Slutsky's theorem the term \eqref{eqn:wtd_error_decomp1} is $O_p(1/\sqrt{n})$. Further, ${\widehat{f}^n{}}^{-1}_{A,h}(a_0)$ converge to a constant and by Slutsky's theorem the rate of convergence of \eqref{eqn:wtd_error_decomp2} would solely depend on the convergence rate of $\sum_{i=1}^n\varepsilon_i w^{pd}_i(K_h(A_i - a_0) - f_A(a_0))$. We show that such a term  equals to $O_p({h}^{2}) + O_p(1/\sqrt{nh})$ using the arguments as below:

\begin{align*}
   \left(\frac{1}{n}\sum_{i=1}^n \varepsilon_i w^{pd}_i(K_h(A_i - a_0) - f_A(a_0)) \right)^2 = {} & O_p(1) \left(\frac{1}{n}\sum_{i=1}^n  w^{pd}_i(K_h(A_i - a_0) - f_A(a_0)) \right)^2 \\ \leq {} & O_p(1)\left(\frac{1}{n}\sum_{i=1}^n {w^{pd}_i{}}^2 \right)\times 
   \left( \frac{1}{n}\sum_{i=1}^n (K_h(A_i - a_0) - f_A(a_0))^2 \right) \nonumber \\
   = {} & O_p(1) \times( O_p({h}^{4}) + O_p(1/nh)) \\ 
   = {} &  O_p({h}^{4}) + O_p(1/nh)
\end{align*}

The first inequality is due to Cauchy-Schwarz inequality. The first equality is due to the assumption of finite variance of the residuals and the second equality due to Lemma \ref{thm:lem_criterion_rate} and the convergence rate of the mean square error of a kernel density estimator is $O_p({h}^{4}) + O_p(1/nh)$.

\end{proof}

\begin{proof}[Proof of Lemma \ref{thm:weighted_distance_convergence_mod}]
The proof follows the same arguments as the proof of Theorem \ref{thm:weighted_distance_convergence} with minor modification. In particular, denoting the ``true'' normalized weights as $\widetilde{\bw}^*_n$, we have that due to the positivity condition and the assumption that $1/f_A(a_0)$ is uniformly bounded, we also have that $\widetilde{w}^*_i = \widetilde{w}^*(\bX_i, A_i)$ are uniformly bounded. Thus, there is an $n^*$ large enough such that for all $n>n^*$, each element in $\widetilde{\bw}^*_n$ is less than $Bn^{1/3}$ for the constant $B$ defined in \eqref{eqn:pdcows_mod}.

Thus, for all $n>n^*$, we also have that $\calD_\gamma(\widetilde{\bw}^{pd}_n) + \frac{\lambda}{n}\sum_{i=1}^n{\widetilde{w}_i^{pd}{}}^2 \leq \calD_\gamma(\widetilde{\bw}^*_n) + \frac{\lambda}{n}\sum_{i=1}^n{\widetilde{w}^*{}}_i^2$ by definition of $\widetilde{\bw}^{pd}_n$. By the arguments in the proof of Theorem \ref{thm:weighted_distance_convergence}, $\calD_\gamma(\widetilde{\bw}^*_n) + \frac{\lambda}{n}\sum_{i=1}^n{\widetilde{w}^*{}}_i^2$ converges a.s. to 0. Thus, the remainder of the proof follows the arguments in the proof of Theorem \ref{thm:weighted_distance_convergence}.
\end{proof}

\begin{proof}[Proof of Lemma \ref{thm:lem_criterion_rate_mod}]
The proof follows almost exactly the arguments in the proof of Lemma \ref{thm:lem_criterion_rate}. The crux of the arguments rely on the inequality $\calD(\widetilde{\bw}_n^{pd})+\lambda\frac{1}{n^2}\sum_{i=1}^n{{\widetilde{w}_i^{pd}{}}^2 } \leq \calD({\widetilde{\bw}^*{}}_n)+\lambda\frac{1}{n^2}\sum_{i=1}^n{\widetilde{w}^*{}}_i^2$. We note that similar to the proof of Lemma \ref{thm:weighted_distance_convergence_mod}, there is an $n^*$ large enough such that for all $n>n^*$, each element in $\widetilde{\bw}^*_n$ is less than $Bn^{1/3}$ for the constant $B$ defined in \eqref{eqn:pdcows_mod}. Thus, as long as $n>n^*$, the inequality $\calD(\widetilde{\bw}_n^{pd})+\lambda\frac{1}{n^2}\sum_{i=1}^n{{\widetilde{w}_i^{pd}{}}^2 } \leq \calD({\widetilde{\bw}^*{}}_n)+\lambda\frac{1}{n^2}\sum_{i=1}^n{\widetilde{w}^*{}}_i^2$ holds. Thus, the remaining arguments of the proof of Lemma \ref{thm:lem_criterion_rate} carry through and the result holds.
\end{proof}

\begin{proof}[Proof of Theorem \ref{thm:asymptotic_normality}]
We begin by decomposing the error of a weighted doubly-robust estimator as 
\begin{align}
    & \hat \mu_{NW}^{\bw,DR}(a_0) - \mu(a_0) \nonumber \\ 
    = {} & \int_{\calX} \int_\calA (\mu(\bx, a_0) - \hat{\mu}(\bx,a_0)) \mathrm{d} \left[ F^n_{\bX,A,\bw} - F^n_{\bX}F^n_{A} \right](\bx,a) \label{dr_error_decomp_term1} \\
    & + \int_\calX {\mu}(\bx,a_0) \mathrm{d}\left[ F^n_\bX - F_\bX \right](\bx) \label{dr_error_decomp_term2} \\ 
    & + \left(\frac{f_A(a_0)}{\widehat{f}^n_{A,h}(a_0)} - 1\right) \int_\calX (\mu(\bx, a_0) - \hat{\mu}(\bx,a_0)) \mathrm{d}F^n_\bX(\bx) \label{dr_error_decomp_term3} \\
    & + \left(\frac{f_A(a_0)}{\widehat{f}^n_{A,h}(a_0)} - 1\right) \int_\calX (\mu(\bx, a_0) - \hat{\mu}(\bx,a_0)) \mathrm{d} \left[ F^n_{\bX,A,\bw} - F^n_{\bX}F^n_{A} \right](\bx,a) \label{dr_error_decomp_term4} \\
    & + {\widehat{f}^n{}}^{-1}_{A,h}(a_0) \int_\calX\int_\calA \left\{(\mu(\bx, a) - \hat{\mu}(\bx,a_0))K_h(a-a_0) \right. \nonumber \\
    & \;\;\quad\quad\quad\quad\quad\quad\quad\quad \left. - (\mu(\bx, a_0) - \hat{\mu}(\bx,a_0)) f_A(a_0)\right\} \mathrm{d} F^n_{\bX,A,\bw}(\bx,a). \label{dr_error_decomp_term5} \\
    & + \frac{1}{n}\sum_{i=1}^n \varepsilon_i w_i \widehat{f}^{-1}_{n,h}(a_0) {K_h(A_i - a_0)} \label{dr_error_decomp_term6}
\end{align}

We now consider the above decomposition with $\bw = \bwtpd$.
We first explore term \ref{dr_error_decomp_term1}. 
Under the conditions of Lemma \ref{thm:lem_criterion_rate_mod} and assuming that $\mu(\cdot,a_0)-\hat{\mu}(\cdot,a_0) \in \calH_{\calX}$, we have
\begin{equation*}
    \left[\int_{\calX} \int_\calA (\mu(\bx, a_0) - \hat{\mu}(\bx, a_0)) \mathrm{d} \left[ F^n_{\bX,A,\bwtpd}(\bx,a) - F^n_{\bX}(\bx)F^n_{A}(a) \right] \right]^2 \leq 3||\mu-\hat{\mu}||_\calH\calD_\gamma(\bwtpd) = O_p(1/n),
\end{equation*}
where the inequality holds due to Lemma \ref{thm:lem1} and the convergence rate of $\calD_\gamma(\bwtpd)$ is due to Lemma \ref{thm:lem_criterion_rate_mod} and the assumption that $||\mu-\hat{\mu}||_\calH = O_p(1)$. Thus 
\begin{align*}
    & \int_{\calX} \int_\calA (\mu(\bx, a_0) - \hat{\mu}(\bx, a_0)) \mathrm{d} \left[ F^n_{\bX,A,\bwtpd}(\bx,a) - F^n_{\bX}(\bx)F^n_{A}(a) \right] = O_p(n^{-1/2}).
\end{align*}

The term \eqref{dr_error_decomp_term2} is $O_p(n^{-1/2})$ due to the central limit theorem and our assumptions on the mean outcome regression function $\mu$.

We now explore term \eqref{dr_error_decomp_term3}. 
The term \eqref{dr_error_decomp_term3} is also independent of the weights but depends on the kernel density estimate $\widehat{f}^n_{A,h}(a_0)$. Under our assumptions on the kernel, due to standard asymptotic results for kernel density estimation we have $\widehat{f}^n_{A,h}(a_0)-f_{A}(a_0) = O_p((nh)^{-1/2}) + O_p(h^2)$.  We re-express \eqref{dr_error_decomp_term3} as
\begin{align}
    = {} & {\widehat{f}^n{}}^{-1}_{A,h}(a_0)(f_A(a_0) - \widehat{f}^n_{A,h}(a_0))\int_{\calX} (\mu(\bx, a_0) - \hat{\mu}(\bx, a_0)) \mathrm{d}F_{\bX}(\bx) \nonumber \\
    & + {\widehat{f}^n{}}^{-1}_{A,h}(a_0)(f_A(a_0) - \widehat{f}^n_{A,h}(a_0))\int_{\calX} (\mu(\bx, a_0) - \hat{\mu}(\bx, a_0))\mathrm{d}\left[F^n_{\bX} - F_{\bX}\right](\bx)\nonumber \\
    = {} & {f_A}^{-1}(a_0)(f_A(a_0) - \widehat{f}^n_{A,h}(a_0))\int_{\calX} (\mu(\bx, a_0) - \hat{\mu}(\bx, a_0)) \mathrm{d}F_{\bX}(\bx) \label{eqn:density_term_1_dr} \\
    & + {\widehat{f}^n{}}^{-1}_{A,h}(a_0)(f_A(a_0) - \widehat{f}^n_{A,h}(a_0))\int_{\calX} (\mu(\bx, a_0) - \hat{\mu}(\bx, a_0))\mathrm{d}\left[F^n_{\bX} - F_{\bX}\right](\bx).\label{eqn:density_term_2_dr} \\
    &+ O_p((nh)^{-1})
\end{align}

The term \eqref{eqn:density_term_1_dr} is $O_p((nh)^{-1/2})(r + o_p(1)) + O_p(h^2)(r + o_p(1))$ and the term \eqref{eqn:density_term_2_dr} is $[O_p((nh)^{-1/2}) + O_p(h^2)]O_p(n^{-1/2})(r + o_p(1))$, where $r=0$ under our assumption on the convergence of $\int_{\calX} (\mu(\bx, a_0) - \hat{\mu}(\bx, a_0))^2 \mathrm{d}F_{\bX}(\bx)$.

Similarly, we have that \eqref{dr_error_decomp_term4} can be re-expressed as 
\begin{align}
    = {} & {\widehat{f}^n{}}^{-1}_{A,h}(a_0)(f_A(a_0) - \widehat{f}^n_{A,h}(a_0)) \int_{\calX} \int_\calA (\mu(\bx, a_0) - \hat{\mu}(\bx, a_0)) \mathrm{d} \left[ F^n_{\bX,A,\bw}(\bx,a) - F_{\bX}(\bx)F_{A}(a) \right] \label{eqn:density_wts_term_1_dr} \\
    & + {\widehat{f}^n{}}^{-1}_{A,h}(a_0)(f_A(a_0) - \widehat{f}^n_{A,h}(a_0)) \int_{\calX} \int_\calA (\mu(\bx, a_0) - \hat{\mu}(\bx, a_0)) \mathrm{d} \left[ F_{\bX}(\bx)\left\{ F_{A}(a) - F^n_{A}(a) \right\} \right] \label{eqn:density_wts_term_2_dr} \\
    & + {\widehat{f}^n{}}^{-1}_{A,h}(a_0)(f_A(a_0) - \widehat{f}^n_{A,h}(a_0)) \int_{\calX} \int_\calA (\mu(\bx, a_0) - \hat{\mu}(\bx, a_0)) \mathrm{d} \left[ F^n_{A}(a) \left\{ F_{\bX}(\bx) - F^n_{\bX}(\bx)\right\} \right]. \label{eqn:density_wts_term_3_dr}
\end{align}
By the Portmanteau Theorem, \eqref{eqn:density_wts_term_1_dr} is $[O_p((nh)^{-1/2}) + O_p(h^2)]o_p(1)$ for $\bw=\bwtpd$. Terms \eqref{eqn:density_wts_term_2_dr} and \eqref{eqn:density_wts_term_3_dr} are both $[O_p((nh)^{-1/2}) + O_p(h^2)]O_p(n^{-1/2})o_p(r + 1)$ with $r=0$. Thus, \eqref{dr_error_decomp_term4} is $o_p((nh)^{-1/2})$.

We now explore term \eqref{dr_error_decomp_term5}. The term \eqref{dr_error_decomp_term5} can be re-expressed as 
\begin{align}
    & {\widehat{f}^n{}}^{-1}_{A,h}(a_0)\int_\calX\int_\calA \left[(\mu(\bx,a) - \hat{\mu}(\bx,a_0))K_h(a-a_0) - (\mu(\bx,a_0) - \hat{\mu}(\bx,a_0))f_A(a_0)\right]\mathrm{d}F^n_{\bX,A,\bw}(\bx,a) \nonumber \\
    = {} & {\widehat{f}^n{}}^{-1}_{A,h}(a_0)\int_\calX\int_\calA \left[(\mu(\bx,a) - \hat{\mu}(\bx,a_0))K_h(a-a_0) - (\mu(\bx,a_0) - \hat{\mu}(\bx,a_0))f_A(a_0)\right] \mathrm{d}F_A(a)\mathrm{d}F_\bX(\bx) \nonumber \\
    & + {\widehat{f}^n{}}^{-1}_{A,h}(a_0)\int_\calX\int_\calA \left[(\mu(\bx,a) - \hat{\mu}(\bx,a_0))K_h(a-a_0) - (\mu(\bx,a_0) - \hat{\mu}(\bx,a_0))f_A(a_0)\right] \\
    & \quad\quad\quad\quad\times \mathrm{d}\left[F^n_{\bX,A,\bw} - F_\bX F_A\right](\bx,a). \nonumber \\
    = {} & {f_{A}}^{-1}(a_0)\int_\calX\int_\calA \left[(\mu(\bx,a) - \hat{\mu}(\bx,a_0))K_h(a-a_0) - (\mu(\bx,a_0) - \hat{\mu}(\bx,a_0))f_A(a_0)\right]\nonumber \\
    & \quad\quad\quad\quad\times \mathrm{d}F_A(a)\mathrm{d}F_\bX(\bx) \label{eqn:dr_density_wts2_term_1} \\
    & + {f_{A}}^{-1}(a_0)\int_\calX\int_\calA \left[(\mu(\bx,a) - \hat{\mu}(\bx,a_0))K_h(a-a_0) - (\mu(\bx,a_0) - \hat{\mu}(\bx,a_0))f_A(a_0)\right] \nonumber \\ 
    & \quad\quad\quad\quad\times \mathrm{d}\left[F^n_{\bX,A,\bw} - F_\bX F_A\right](\bx,a). \label{eqn:dr_density_wts2_term_2} \\
    & + o_p((nh)^{-1/2}). \nonumber
\end{align}
The term inside the first integral of \eqref{eqn:dr_density_wts2_term_1} can be written as
\begin{align}
    & \int_\calA \left[(\mu(\bx,a) - \hat{\mu}(\bx,a_0))K_h(a-a_0) - (\mu(\bx,a_0)-\hat{\mu}(\bx,a_0))f_A(a_0)\right] \mathrm{d}F_A(a) \label{eqn:dr_term_kernel} \\
    = {} & \int_\calA (\mu(\bx,a)-\hat{\mu}(\bx,a_0))K_h(a-a_0) \mathrm{d}F_A(a) \label{eqn:dr_term_kernel1} \\
    & - (\mu(\bx,a_0)-\hat{\mu}(\bx,a_0))f_A(a_0). \nonumber
\end{align}

The first term in \eqref{eqn:dr_term_kernel1} can be expressed as
\begin{align}
    & \frac{1}{h}\int \mu(\bx,a)K((a-a_0)/h ) f_A(a)\mathrm{d}a - \hat{\mu}(\bx,a_0)\frac{1}{h}\int K((a-a_0)/h ) f_A(a)\mathrm{d}a \nonumber \\
    = {} & \int \mu(\bx,a_0 + uh)K( u ) f_A(a_0 +uh)\mathrm{d}u - \hat{\mu}(\bx,a_0)\int K( u ) f_A(a_0 +uh)\mathrm{d}u \nonumber \\
    = {} & \int K(u) \left\{\mu(\bx,a_0) + uh\mu'(\bx,a_0)+u^2h^2\mu''(\bx,a_0)/2\right\} \left(f_A(a_0) + uhf'_A(a_0)\right)\mathrm{d}u \nonumber\\
    &- \hat{\mu}(\bx,a_0)\int K( u ) \left(f_A(a_0) + uhf'_A(a_0)\right)\mathrm{d}u  + o(h^2) \nonumber \\
    = {} & \Big(\underbrace{\int K(u)\mathrm{d}u}_{=1}\Big)\mu(\bx,a_0)f_A(a_0) + \Big(\underbrace{\int uK(u)\mathrm{d}u}_{=0}\Big)h\left\{\mu(\bx,a_0)f'_A(a_0) + \mu'(\bx,a_0)f_A(a_0)\right\} \nonumber\\
    & + \Big(\underbrace{\int u^2K(u)\mathrm{d}u}_{\equiv \kappa_2}\Big)h^2\left\{\mu'(\bx,a_0)f'_A(a_0) + \mu''(\bx,a_0)f_A(a_0)/2\right\}  - \hat{\mu}(\bx,a_0)f_A(a_0) + o(h^2) \nonumber \\
    = {} & (\mu(\bx,a_0)-\hat{\mu}(\bx,a_0))f_A(a_0) + h^2\kappa_2B(\bx,a_0)f_A(a_0) + o(h^2), \nonumber
\end{align}
where $B(\bx,a_0) = \mu''(\bx,a_0)/2 + \mu'(\bx,a_0)f'_A(a_0) / f_A(a_0)$ and the second equality holds by taking Taylor expansions of $\mu$ and $f_A$. Hence \eqref{eqn:dr_term_kernel} is equal to $h^2\kappa_2B(\bx,a_0)f_A(a_0) + o(h^2)$. Thus, since the expectation is a linear operator, this implies that \eqref{eqn:dr_density_wts2_term_1} is $O_p((nh)^{-1/2}) + O_p(h^2)$.
Similarly to \eqref{eqn:density_wts_term_1} and using similar arguments as above for \eqref{eqn:density_wts2_term_1}, \eqref{eqn:dr_density_wts2_term_2} is $[O_p((nh)^{-1/2}) + O_p(h^2)]o_p(1)$ due to the Portmanteau Theorem for both $\bw=\bwtpd$. 
Therefore, term \eqref{eqn:dr_density_wts2_term_1} and hence term \ref{dr_error_decomp_term5} are $h^2\kappa_2\int_\calX B(\bx,a_0)\mathrm{d}F_\bX(\bx) + o_p((nh)^{-1/2})$.

We now explore term \eqref{dr_error_decomp_term6}. As shown, this term is
\begin{align}
 & {f_A}^{-1}(a_0) \frac{1}{n}\sum_{i=1}^n\varepsilon_i w^{pd}_i(K_h(A_i - a_0) - f_A(a_0)) \label{eqn:wtd_error_decomp2_dr} + O_p(n^{-1/2}).
\end{align}

Thus, the two leading terms that determine the asymptotic distribution are \eqref{dr_error_decomp_term3} and \eqref{dr_error_decomp_term6}. 
Define $B(a_0) = \int_\calX B(\bx,a_0)\mathrm{d}F_\bX(\bx)$.  We thus have 
\begin{align}
& \hat \mu_{NW}^{\bwtpd,DR}(a_0) - \mu(a_0) - h^2\kappa_2B(a_0) \nonumber \\ 
= {} &  {f_A}^{-1}(a_0)(f_A(a_0) - \widehat{f}^n_{A,h}(a_0))\int_{\calX} (\mu(\bx, a_0) - \hat{\mu}(\bx, a_0)) \mathrm{d}F_{\bX}(\bx) \\
& + {f_A}^{-1}(a_0) \frac{1}{n}\sum_{i=1}^n\varepsilon_i \widetilde{w}^{pd}_i(K_h(A_i - a_0) - f_A(a_0)) + o_p((nh)^{-1/2}) \nonumber \\
= {} & \frac{1}{n}\sum_{i=1}^n\left[K_h(A_i-a_0) - f_A(a_0) \right] \frac{\int_{\calX} (\mu(\bx, a_0) - \hat{\mu}(\bx, a_0)) \mathrm{d}F_{\bX}(\bx)}{f_A(a_0)} \label{dr_decomp_non_wtd_term} \\
& + {f_A}^{-1}(a_0)\frac{1}{n}\sum_{i=1}^n\varepsilon_i\widetilde{w}^{pd}_iK_h(A_i-a_0) \label{dr_decomp_kernel_wtd_resid} \\
& - \frac{1}{n}\sum_{i=1}^n\varepsilon_i\widetilde{w}^{pd}_i + o_p((nh)^{-1/2}). \label{dr_decomp_pure_wtd_resid}
\end{align}

Note that $\frac{1}{n}\sum_{i=1}^n\varepsilon_i\widetilde{w}^{pd}_i = O_p(n^{-1/2})$ due to an application of Chebyshev's inequality as follows. For any $\delta >0$, conditional on $(\bX_i, A_i)_{i=1}^n$ we have with probability greater than $1-\delta^2$ that 
\begin{align*}
    \Big\lvert \frac{1}{n}\sum_{i=1}^n\varepsilon_i\widetilde{w{}}^{pd}_i\Big\rvert \leq {} & \frac{1}{\delta}\frac{1}{n^{1/2}} \sqrt{\frac{1}{n} \sum_{i=1}^n\sigma^2(\bX_i,A_i){\widetilde{w}^{pd}{}}^2_i} \\
    \leq {} & \frac{\overline{\sigma}}{\delta}\frac{1}{n^{1/2}} \sqrt{\frac{1}{n} \sum_{i=1}^n{\widetilde{w}^{pd}{}}^2_i} = O_p(1/n^{1/2}),
\end{align*}
where $\overline{\sigma}^2 = \sup_{(\bx,a)\in \calX\times\calA}\text{Var}(Y(a) - \mu(\bx,a))$.

We now explore the term \eqref{dr_decomp_non_wtd_term}.
\begin{align*}
    & \frac{1}{n}\sum_{i=1}^n\left[K_h(A_i-a_0) - f_A(a_0) \right] \frac{\int_{\calX} (\mu(\bx, a_0) - \hat{\mu}(\bx, a_0)) \mathrm{d}F_{\bX}(\bx)}{f_A(a_0)} \\
    = {} & O_p((nh)^{-1/2})\frac{\int_{\calX} (\mu(\bx, a_0) - \hat{\mu}(\bx, a_0)) \mathrm{d}F_{\bX}(\bx)}{f_A(a_0)} \\
    = {} & O_p((nh)^{-1/2})o_p(1) = o_p((nh)^{-1/2}),
\end{align*}
where the first equality holds by standard properties of nonparametric density estimation and the second holds due to the assumption that $\int_{\calX} (\mu(\bx, a_0) - \hat{\mu}(\bx, a_0))^2 \mathrm{d}F_{\bX}(\bx) = o_p(1)$. 

Thus, the asymptotic distribution completely depends on the term \eqref{dr_decomp_kernel_wtd_resid}, which we now explore by applying the conditional central limit theorem (CLT) of \citet{wong2017kernel}, which we re-state below following this proof as Theorem \ref{thm:partial_clt}.

In the notation of the Supplementary Material of \citet{wong2017kernel}, we define \\ 
$Z_n = (nh)^{1/2}\left\{ \frac{1}{n}\sum_{i=1}^n \varepsilon_i{\widetilde{w}^{pd}{}}_i\frac{K_h(A_i-a_0)}{f_A(a_0)} \right\}  = (nh)^{1/2}\times$ term \eqref{dr_decomp_kernel_wtd_resid},
$$C_i = \frac{\varepsilon_i{\widetilde{w}^{pd}{}}_i\frac{K_h(A_i-a_0)}{f_A(a_0)}}{\left( \sigma^2 \sum_{j=1}^n {\widetilde{w}^{pd}{}}^2_j\frac{K^2_h(A_j-a_0)}{f^2_A(a_0)} \right)^{1/2}} = \frac{\varepsilon_i{\widetilde{w}^{pd}{}}_iK_h(A_i-a_0)}{\left( \sigma^2 \sum_{j=1}^n {\widetilde{w}^{pd}{}}^2_jK^2_h(A_j-a_0) \right)^{1/2}},$$
and $$D_j = \frac{\mu(\bX_j, a_0) - \mu(a_0)}{\left[n \text{Var}\{\mu(\bX,a_0)\}\right]^{1/2}},$$
where we use the notation $D_j$ in the place of \citet{wong2017kernel}'s notation $A_j$, since we already use $A$ to denote the treatment value.
We note that $$(nh)^{1/2}\left\{ \frac{1}{n}\sum_{i=1}^n \varepsilon_i{\widetilde{w}^{pd}{}}_i\frac{K_h(A_i-a_0)}{f_A(a_0)} + \frac{1}{n}\sum_{i=1}^n\left[{\mu}(\bX_i,a_0) - {\mu}(a_0) \right] \right\}$$ and $Z_n$ have the same limiting distribution since $$\int_\calX {\mu}(\bx,a_0) \mathrm{d}\left[ F^n_\bX - F_X \right](\bx) = \frac{1}{n}\sum_{i=1}^n\left[{\mu}(\bX_i,a_0) - {\mu}(a_0) \right] = O_p(n^{-1/2}).$$ 
Thus, \begin{equation*}
    \frac{\sqrt{nh}f_A(a_0)}{\sigma \sqrt{\frac{1}{n}\sum_{i=1}^n{\widetilde{w}^{pd}{}}^2_iK^2_h(A_i-a_0)}}\left(\hat \mu_{NW}^{\bw,DR}(a_0) - \mu(a_0) - h^2\kappa_2B(a_0)\right) 
\end{equation*}
and 
\begin{equation*}
    \frac{\sqrt{nh}f_A(a_0)}{\sigma \sqrt{\frac{1}{n}\sum_{i=1}^n{\widetilde{w}^{pd}{}}^2_iK^2_h(A_i-a_0)}}\left(\hat \mu_{NW}^{\bw,DR}(a_0) - \frac{1}{n}\sum_{i=1}^n\mu(\bX_i, a_0) - h^2\kappa_2B(a_0)\right) 
\end{equation*}
have the same limiting distribution.

We now investigate the Lyapunov-like condition of the conditional CLT (Theorem \ref{thm:partial_clt}) using $\delta=1$. Letting $\calD_n = \{D_1, \dots, D_n, \calB_1, \dots, \calB_n\}$, where $\calB_j = \{A_j, \bX_j\}$, 
we have
\begin{align*}
    \bbE\left[  \sum_{i=1}^n|C_i|^3 \Big\vert \calD_n\right] & \leq \frac{\max_i\bbE|\varepsilon_i|^3\sum_{i=1}^n{\widetilde{w}^{pd}{}}^3_iK_h^3(A_i-a_0)}{\sigma^3\left(\sum_{j=1}^n{\widetilde{w}^{pd}{}}^2_jK_h^2(A_j-a_0)\right)^{3/2}} \\
    & \leq \frac{\max_i\bbE|\varepsilon_i|^3\max_i K_h(A_i-a_0)\max_i {\widetilde{w}^{pd}{}}_i}{\sigma^3\left(\sum_{j=1}^n{\widetilde{w}^{pd}{}}^2_jK_h^2(A_j-a_0)\right)^{1/2}} = \frac{o_p(n^{1/2})}{O_p(n^{1/2})} = o_p(1).\\
\end{align*}

Letting $g^2(\calD_n) =  \frac{\sigma^2}{f^2_A(a_0)n}\sum_{i=1}^n{\widetilde{w}^{pd}{}}^2_iK^2_h(A_i-a_0)$, we have that $\bbE[g^2(\calD_n)]\leq M$ for some constant $M$ and $g^2(\calD_n)\leq M + o_p(1)$ because $\frac{\sigma^2}{f^2_A(a_0)n}\sum_{i=1}^n{\widetilde{w}^{pd}{}}^2_iK^2_h(A_i-a_0) \leq \frac{\sigma^2}{f^2_A(a_0)}\max_iK^2_h(A_i-a_0)\frac{1}{n}\sum_{i=1}^n{\widetilde{w}^{pd}{}}^2_i = O_p(1)$ by Lemma \ref{thm:lem_criterion_rate_mod}. Thus all of the criteria to apply Theorem \ref{thm:partial_clt} are met.

In the notation of \citet{wong2017kernel}, we then define 
$Z_n^* = g(\calD_n)\sum_{i=1}^n\text{var}^{1/2}(C_i\vert \calD_n)G_i = \sigma n^{-1/2}\sum_{i=1}^nK_h(A_i-a_0){\widetilde{w}^{pd}{}}_iG_i$, where $G_1, \dots, G_n$ are i.i.d. standard normal random variables. By Theorem S1 of \citet{wong2017kernel} and letting $\phi_n(t), \phi^*_n(t)$ be the characteristic functions of $Z_n$ and $Z_n^*$, respectively, we have $|\phi_n(t) - \phi^*_n(t)| \xrightarrow{n\rightarrow\infty} 0$ for all $t\in\bbR$ where $\phi^*_n$ is twice differentiable.

Thus, we have as $n\rightarrow\infty$, $h\rightarrow 0$, $nh\rightarrow\infty$ that
\begin{equation*}
    \frac{\sqrt{nh}f_A(a_0)}{\sigma \sqrt{\frac{1}{n}\sum_{i=1}^n{\widetilde{w}^{pd}{}}^2_iK^2_h(A_i-a_0)}}\left(\hat \mu_{NW}^{\bw,DR}(a_0) - \mu(a_0) - h^2\kappa_2B(a_0)\right) \xrightarrow{d} \calN(0,1)
\end{equation*}
$\bw = \widetilde{\bw}^{pd}_n$ and the result is proven.
\end{proof}

Here we present/re-state the statement of the partially conditional central limit theorem (CLT) of \citet{wong2017kernel}. 

\begin{manualtheorem}{S1 of \citet{wong2017kernel}}[Partially conditional CLT]\label{thm:partial_clt}
Let $(D_1, \calB_1), \dots, (D_n, \calB_n)$ be independent and identically distributed where $D_1,\dots, D_n$ are random variables and $\calB_1, \dots, \calB_n$ are sets of random variables. Let $\{C_1, \dots, C_n\}$ be another set of random variables. Write $\calD_n = \{D_1, \dots, D_n, \calB_1, \dots, \calB_n\}$. Assume these random variables satisfy 
\begin{align*}
    & \bbE[D_j] = 0, \bbE[C_j|\calD_n] = 0 \text{ for } j=1, \dots, n, \\
    & \sum_{j=1}^n\mbox{Var}(D_j) = 1, \sum_{j=1}^n\mbox{Var}(C_j|\calD_n) = 1,
\end{align*}
and there exists $\delta>0$ such that $\sum_{j=1}^n\bbE[|C_j|^{2+\delta}|\calD_n] \rightarrow 0$ in probability. Additionally, $C_1, \dots, C_n$ are conditionally independent given $\calD_n$. Let $g$ be a non-random function mapping from the support of $\calD_n$ to the positive real numbers $\bbR^+$ such that there exists a constant $M>0$ such that $\bbE[g^2(\calD_n)] \leq M$ and $g^2(\calD_n) \leq M + o_p(1)$. For any positive real number $\tau$, consider the two random variables 
\begin{align*}
    Z_n = \tau\sum_{j=1}^n D_j + g(\calD_n)\sum_{j=1}^nC_j \text{ and } Z_n^* = \tau F + g(\calD_n) \sum_{j=1}^n\left\{\mbox{Var}(C_j|\calD_n)\right\}^{1/2}G_j,
\end{align*}
where $F, G_1, \dots, G_n$ are i.i.d. standard normal random variables independent of $C_1, \dots, C_n$ and $\calD_n$. Let $\phi_n$ and $\phi^*_n$ be the characteristic functions of $Z_n$ and $Z_n^*$, respectively. Then $|\phi_n(t) - \phi^*_n(t)|\rightarrow 0$ for every $t\in\bbR$. Further, $\bbE[{Z_n^*{}}^2] = \tau^2 + \bbE[g^2(\calD_n)] \leq \tau^2 + M$ and $\phi^*_n$ is twice differentiable.

\end{manualtheorem}

\begin{proof}[Proof of Theorem \ref{thm:thm3b}]
We investigate the limit of $\widehat{\mu}_{NW}^{\bw,DR}(a_0) - \mu(a_0)$ as $nh\to\infty$ and $h\to 0$ for both $\bw=\bw^d$; we omit the proof for $\bw=\bw^{pd}$ as it trivially follows in the same manner with the same arguments as in the proof of Theorem \ref{thm:thm3}. For reference, we include the error decomposition of a weighted nonparametric estimate of the ADRF. Recall that $\hat{\mu}(\bx, a_0)$ is assumed to converge uniformly almost surely to $\widetilde{\mu}(\bx, a_0)$, which may or may not be equal to ${\mu}(\bx, a_0)$.
As stated in the main text, given \textit{any} $\bw$, the error of \eqref{eqn:weighted_nw_estimator} at $A=a_0$ can be decomposed as 
\begin{align}
&\widehat{\mu}_{NW}^{\bw,DR}(a_0) - \mu(a_0) \nonumber \\ 
= {} & \int_{\calX} \int_\calA \left(\mu(\bx, a_0) - \hat{\mu}(\bx, a_0)\right) \mathrm{d} \left[ F^n_{\bX,A,\bw}(\bx,a) - F^n_{\bX}(\bx)F^n_{A}(a) \right]  \label{eqn:error_decomp_indep_aug} \\
& + \int_{\calX} \mu(\bx,a_0)\mathrm{d} \left[ F^n_{\bX} - F_\bX \right](\bx)   \label{eqn:error_decomp_sampling_aug} \\ 
    & + \left(\frac{f_A(a_0)}{\widehat{f}^n_{A,h}(a_0)} - 1\right) \int_{\calX} \left(\mu(\bx, a_0) - \hat{\mu}(\bx, a_0)\right)\mathrm{d}F^n_{\bX}(\bx)  \label{eqn:error_decomp_density_aug} \\
    & + \left(\frac{f_A(a_0)}{\widehat{f}^n_{A,h}(a_0)} - 1\right) \int_{\calX} \int_\calA \left(\mu(\bx, a_0) - \hat{\mu}(\bx, a_0)\right) \mathrm{d} \left[ F^n_{\bX,A,\bw}(\bx,a) - F^n_{\bX}(\bx)F^n_{A}(a) \right]  \label{eqn:error_decomp_wts_density1_aug} \\
    &+{\widehat{f}^n{}}^{-1}_{A,h}(a_0)\int_\calX\int_\calA \left[\left(\mu(\bx, a_0) - \hat{\mu}(\bx, a_0)\right)K_h(a-a_0) - \left(\mu(\bx, a_0) - \hat{\mu}(\bx, a_0)\right)f_A(a_0)\right]\mathrm{d}F^n_{\bX,A,\bw}(\bx,a)  \label{eqn:error_decomp_wts_density2_aug} \\
    & + \frac{1}{n}\sum_{i=1}^n \varepsilon_i w_i {\widehat{f}^n{}}^{-1}_{A,h}(a_0) {K_h(A_i - a_0)}, \label{eqn:error_decomp_wts_error_aug}
\end{align}
where $\widehat{f}^n_{A,h}(a_0) = \int_\calA K_h(a-a_0)\mathrm{d}F^n_{A}(a)$ is a kernel density estimate of $f_A(a_0)$.

Using $\bw=\bw^d$, 
the convergence of terms \eqref{eqn:error_decomp_sampling_aug} and \eqref{eqn:error_decomp_wts_error_aug} directly follows from previous work, as these terms are identical to their corresponding terms in the error decomposition of $\widehat{\mu}_{NW}^{\bw}(a_0)$. The term \eqref{eqn:error_decomp_indep_aug} can be decomposed as

\begin{align}
    & \int_{\calX} \int_\calA \left(\mu(\bx, a_0) - \widetilde{\mu}(\bx, a_0)\right) \mathrm{d} \left[ F^n_{\bX,A,\bw}(\bx,a) - F^n_{\bX}(\bx)F^n_{A}(a) \right]  \nonumber \\ 
    & + \int_{\calX} \int_\calA \left(\widetilde{\mu}(\bx, a_0) - \hat{\mu}(\bx, a_0)\right) \mathrm{d} \left[ F^n_{\bX,A,\bw}(\bx,a) - F^n_{\bX}(\bx)F^n_{A}(a) \right] \nonumber \\
    = {} &  \int_{\calX} \int_\calA \left(\mu(\bx, a_0) - \widetilde{\mu}(\bx, a_0)\right) \mathrm{d} \left[ F^n_{\bX,A,\bw}(\bx,a) - F_{\bX}(\bx)F_{A}(a) \right] \label{eqn:aug_decomp_term_1a} \\
    & + \int_{\calX} \left(\mu(\bx, a_0) - \widetilde{\mu}(\bx, a_0)\right) \mathrm{d} \left[ F_{\bX}(\bx)- F^n_{\bX}(\bx) \right] \label{eqn:aug_decomp_term_1b} \\
    & + \int_{\calX} \int_\calA \left(\widetilde{\mu}(\bx, a_0) - \hat{\mu}(\bx, a_0)\right) \mathrm{d} \left[ F^n_{\bX,A,\bw}(\bx,a) - F_{\bX}(\bx)F_{A}(a) \right] \label{eqn:aug_decomp_term_1c} \\
    & + \int_{\calX} \int_\calA \left(\widetilde{\mu}(\bx, a_0) - \hat{\mu}(\bx, a_0)\right) \mathrm{d} \left[ F^n_{\bX}(\bx) - F_{\bX}(\bx) \right]. \label{eqn:aug_decomp_term_1d}
\end{align}

Similar to the proof of consistency for $\widehat{\mu}_{NW}^{\bw}(a_0)$, term \eqref{eqn:aug_decomp_term_1a} is $o_p(1)$ by the Portmanteau Theorem. Term \eqref{eqn:aug_decomp_term_1b} is $o_p(1)$ by the weak law of large numbers (WLLN). What remains to show for the convergence of term \eqref{eqn:error_decomp_indep_aug} is that \eqref{eqn:aug_decomp_term_1c} and \eqref{eqn:aug_decomp_term_1d} are $o_p1(1)$.

By the assumption of uniform convergence, we have that 
\begin{align*}
    & \int_{\calX} \int_\calA \left(\widetilde{\mu}(\bx, a_0) - \hat{\mu}(\bx, a_0)\right) \mathrm{d} \left[ F^n_{\bX,A,\bw}(\bx,a) - F_{\bX}(\bx)F_{A}(a) \right] \\
    & \leq \int_{\calX} \int_\calA | \widetilde{\mu}(\bx, a_0) - \hat{\mu}(\bx, a_0) |  \mathrm{d} \left[ F^n_{\bX,A,\bw}(\bx,a) - F_{\bX}(\bx)F_{A}(a) \right] \\
    & \leq \sup_{\bx\in\calX}| \widetilde{\mu}(\bx, a_0) - \hat{\mu}(\bx, a_0) | \int_{\calX} \int_\calA  \mathrm{d} \left[ F^n_{\bX,A,\bw}(\bx,a) - F_{\bX}(\bx)F_{A}(a) \right] \\
    & = o_p(1)o_p(1) = o_p(1)
\end{align*}
by the uniform almost sure convergence of $\hat{\mu}(\bx, a_0)$ to $\widetilde{\mu}(\bx, a_0)$ and by our Theorem \ref{thm:weighted_distance_convergence}. Similar can be shown for term \eqref{eqn:aug_decomp_term_1d} but using the Glivenko-Cantelli Theorem.

The convergence of \eqref{eqn:error_decomp_density_aug}, \eqref{eqn:error_decomp_wts_density1_aug}, and \eqref{eqn:error_decomp_wts_density2_aug} (to $0$) can be shown using a similar inequality as above that makes use of the uniform almost sure convergence of $\hat{\mu}(\bx, a_0)$ to $\widetilde{\mu}(\bx, a_0)$ and by using the same steps as used in the proof of consistency of $\widehat{\mu}_{NW}^{\bw}(a_0)$. We show how these arguments are extended by showing the convergence of \eqref{eqn:error_decomp_wts_density2_aug} for 0 in probability below.

The term \eqref{eqn:error_decomp_wts_density2_aug} can be re-expressed as 
\begin{align}
    & {\widehat{f}^n{}}^{-1}_{A,h}(a_0)\int_\calX\int_\calA \left[\left(\mu(\bx, a_0) - \hat{\mu}(\bx, a_0)\right)K_h(a-a_0) - \left(\mu(\bx, a_0) - \hat{\mu}(\bx, a_0)\right)f_A(a_0)\right]\mathrm{d}F^n_{\bX,A,\bw}(\bx,a)   \nonumber \\
    = {} & {\widehat{f}^n{}}^{-1}_{A,h}(a_0)\int_\calX\int_\calA \left[\left(\mu(\bx, a_0) - \hat{\mu}(\bx, a_0)\right)K_h(a-a_0) - \left(\mu(\bx, a_0) - \hat{\mu}(\bx, a_0)\right)f_A(a_0)\right] \mathrm{d}F_A(a)\mathrm{d}F_\bX(\bx) \label{eqn:density_wts2_term_1_aug} \\
    & + {\widehat{f}^n{}}^{-1}_{A,h}(a_0)\int_\calX\int_\calA \left[\left(\mu(\bx, a_0) - \hat{\mu}(\bx, a_0)\right)K_h(a-a_0) \right. \nonumber \\
    & \quad\quad\quad\quad\quad\quad\quad\quad\; \left. - \left(\mu(\bx, a_0) - \hat{\mu}(\bx, a_0)\right)f_A(a_0)\right]\mathrm{d}\left[F^n_{\bX,A,\bw} - F_\bX F_A\right](\bx,a). \label{eqn:density_wts2_term_2_aug}
\end{align}
The term inside the first integral of \eqref{eqn:density_wts2_term_1_aug} can be written as
\begin{align}
    & \int_\calA \left[\left(\mu(\bx, a_0) - \hat{\mu}(\bx, a_0)\right)K_h(a-a_0) - \left(\mu(\bx, a_0) - \hat{\mu}(\bx, a_0)\right)f_A(a_0)\right] \mathrm{d}F_A(a) \label{eqn:term_kernel_aug} \\
    = {} & \int_\calA \left(\mu(\bx, a_0) - \widetilde{\mu}(\bx, a_0)\right)K_h(a-a_0) \mathrm{d}F_A(a) \label{eqn:term_kernel1_aug} \\
    & - \left(\mu(\bx, a_0) - \widetilde{\mu}(\bx, a_0)\right)f_A(a_0) \label{eqn:term_kernel1b_aug} \\
    & + \int_\calA \left(\widetilde{\mu}(\bx, a_0) - \hat{\mu}(\bx, a_0)\right)K_h(a-a_0) \mathrm{d}F_A(a) \nonumber \\
    & - \left(\widetilde{\mu}(\bx, a_0) - \hat{\mu}(\bx, a_0)\right)f_A(a_0). \nonumber
\end{align}

Term \eqref{eqn:term_kernel1_aug} can be expressed as
\begin{align}
    & \frac{1}{h}\int \left(\mu(\bx, a_0) - \hat{\mu}(\bx, a_0)\right)K((a-a_0)/h ) f_A(a)\mathrm{d}a \nonumber \\
    = {} & \int \left(\mu(\bx, a_0+uh) - \hat{\mu}(\bx, a_0+uh)\right)K( u ) f_A(a_0 +uh)\mathrm{d}u \nonumber \\
    = {} & \int K(u) \left\{\left(\mu(\bx, a_0) - \hat{\mu}(\bx, a_0)\right) + uh\frac{\partial}{\partial a_0} \left(\mu(\bx, a_0) - \hat{\mu}(\bx, a_0)\right) +\frac{1}{2}u^2h^2\frac{\partial^2}{\partial a^2_0}\left(\mu(\bx, a_0) - \hat{\mu}(\bx, a_0)\right)\right\} \\ 
    & \quad\quad\times\left(f_A(a_0) + uh\frac{\partial}{\partial a_0}f_A(a_0)\right)\mathrm{d}u + o(h^2) \nonumber\\
    = {} & \Big(\underbrace{\int K(u)\mathrm{d}u}_{=1}\Big)\left(\mu(\bx, a_0) - \hat{\mu}(\bx, a_0)\right)f_A(a_0) \nonumber \\
    & + \Big(\underbrace{\int uK(u)\mathrm{d}u}_{=0}\Big)h\left\{\left(\mu(\bx, a_0) - \hat{\mu}(\bx, a_0)\right)\frac{\partial}{\partial a_0}f_A(a_0) + \frac{\partial}{\partial a_0}\left(\mu(\bx, a_0) - \hat{\mu}(\bx, a_0)\right)f_A(a_0)\right\} \nonumber\\
    & + \Big(\underbrace{\int u^2K(u)\mathrm{d}u}_{\equiv \kappa_2}\Big)h^2\left\{\frac{\partial}{\partial a_0}\left(\mu(\bx, a_0) - \hat{\mu}(\bx, a_0)\right)\frac{\partial}{\partial a_0}f_A(a_0) \right. \nonumber \\
    & \quad\quad\quad\quad\quad\quad\quad\quad\quad\quad \left. + \frac{\partial^2}{\partial a^2_0}\left(\mu(\bx, a_0) - \hat{\mu}(\bx, a_0)\right)f_A(a_0)/2\right\} + o(h^2) \nonumber \\
    = {} & \left(\mu(\bx, a_0) - \hat{\mu}(\bx, a_0)\right)f_A(a_0) + h^2\kappa_2\widetilde{B}(\bx, a_0)f_A(a_0) + o(h^2), \nonumber
\end{align}
where $\widetilde{B}(\bx, a_0) = \frac{\partial^2}{\partial a^2_0}\left(\mu(\bx, a_0) - \hat{\mu}(\bx, a_0)\right)f_A(a_0)/2 + \frac{\partial}{\partial a_0}\left(\mu(\bx, a_0) - \hat{\mu}(\bx, a_0)\right)\left\{\frac{\partial}{\partial a_0}f_A(a_0)\right\} / f_A(a_0)$ and the second equality holds by taking Taylor expansions of $\mu$ and $f_A$. Hence \eqref{eqn:term_kernel1_aug}$+$\eqref{eqn:term_kernel1b_aug} is equal to $h^2\kappa_2\widetilde{B}(\bx, a_0)f_A(a_0) + o(h^2)$. Thus, since the expectation is a linear operator, this implies that \eqref{eqn:density_wts2_term_1_aug} is

\begin{align*}
    & {\widehat{f}^n{}}^{-1}_{A,h}(a_0)\int_\calX\int_\calA \left[\left(\mu(\bx, a_0) - \hat{\mu}(\bx, a_0)\right)K_h(a-a_0) - \left(\mu(\bx, a_0) - \hat{\mu}(\bx, a_0)\right)f_A(a_0)\right] \mathrm{d}F_A(a)\mathrm{d}F_\bX(\bx) \\
    = {} & O_p((nh)^{-1/2}) + O_p(h^2) \\
    & + {\widehat{f}^n{}}^{-1}_{A,h}(a_0)\int_\calX \left\{ \int_\calA \left(\widetilde{\mu}(\bx, a_0) - \hat{\mu}(\bx, a_0)\right)K_h(a-a_0) \mathrm{d}F_A(a) \right. \\
    & \left. \quad\quad\quad\quad\quad\quad\quad \vphantom{\int_\calA} -\left(\widetilde{\mu}(\bx, a_0) - \hat{\mu}(\bx, a_0)\right)f_A(a_0) \right\}\mathrm{d}F_\bX(\bx) \\
    & \leq O_p((nh)^{-1/2}) + O_p(h^2) \\
    & + \sup_{\bx \in \calX} \vert \widetilde{\mu}(\bx, a_0) - \hat{\mu}(\bx, a_0) \vert \int_\calA (K_h(a-a_0) + f_A(a_0)) \mathrm{d}F_A(a) \\
    & = O_p((nh)^{-1/2}) + O_p(h^2) + o_p(1)\int_\calA (K_h(a-a_0) + f_A(a_0)) \mathrm{d}F_A(a) \\
    & = O_p((nh)^{-1/2}) + O_p(h^2) + o_p(1).
\end{align*}

Thus, following the same arguments as in the proof of Theorem \ref{thm:thm3}, the term \eqref{eqn:error_decomp_wts_density2_aug} is $O_p((nh)^{-1/2}) + O_p(h^2) + o_p(1) + [O_p((nh)^{-1/2}) + O_p(h^2)]o_p(1)$. Thus, the consistency result holds.
\end{proof}


\section{Main text simulation study}
\label{supsec:main_sim}

\subsection{Additional details of main text simulation study}

In the NMES dataset, two of the covariates are continuous ($X_1$: age started to smoke and $X_2$: age last smoked) and the remaining 7 are discrete ($X_{3k}$: gender (male, female); $X_{4k}$: race (African American, other); $X_{5k}$: seat belt use, (low, medium, high); $X_{6k}$: education ($<$ high school, high school, $1-3$ years of college, $4+$ years of college); $X_{7k}$: marital status (never married, widowed/divorced, married); $X_{8k}$: census region (Northeast, Midwest, South, West); $X_{9k}$: poverty status (poor, near poor, low income, middle income, high income), with anywhere from $\ell_k=$ 2 to 5 levels.

The overall covariate vector thus has dimension 18 after converting categorical variables to dummy-coded sets and removing the dummy variables for the reference categories. In our simulation, we leave the treatment level (pack years) and covariates intact and simulate outcomes for each unit from the following model:
\begin{align*}
    Y = {} &  \gamma_1b_1X_1 + \gamma_2b_2\widetilde{X}_1^2 + \gamma_3b_3\widetilde{X}_1^3 + \gamma_4b_4X_2 + \gamma_5b_5\widetilde{X}_2^2 + \gamma_6b_6\widetilde{X}_2^3 \\
    & + \sum_{j=3}^8\sum_{k=1}^{\ell_j}\left\{\alpha_{jk}b_{1jk} + \eta_{1jk}b_{2jk}X_1 + \eta_{2jk}b_{3jk}\widetilde{X}_1^2 +\eta_{3jk}b_{4jk}X_2 + \eta_{4jk}b_{5jk}\widetilde{X}_2^2 \right\}I(X_j = k) \\
    & - \overline{m}(\bX) + f(A) (1 + \delta(\bX))) + \varepsilon,
\end{align*}
where $\widetilde{X}_{j} = X_j-\overline{X}_j$, $\varepsilon \sim N(0,4)$ (i.i.d.), $\gamma_1, \gamma_3, \eta_{1jk}, \eta_{3jk}\sim$ Unif$(-0.5, 0.5)$, $\gamma_2, \gamma_4, \eta_{2jk}, \eta_{4jk}\sim$ Unif$(-0.1, 0.1)$, $\gamma_3, \gamma_6\sim$ Unif$(-0.01, 0.01)$, $\alpha_{jk}\sim$ Unif$(-10, 10)$, $b_j, b_{ijk}\sim$ Bernoulli$(0.5)$, $\overline{m}(\bX)$ is the mean of the main effect terms preceding it, and the treatment effect curve $f(A) = A/4 + \frac{2}{(A/100 + 1/2)^3} - (A-40)^2/100$. Thus, in the notation of Section \ref{sec:simulations} of the main text, $$\btheta_1 = (\gamma_1, \gamma_2, \gamma_3, \gamma_4, \gamma_5, \gamma_6, b_1, b_2, b_3, b_4, b_5, b_6, \eta_{1jk}, \eta_{2jk}, \eta_{3jk}, \eta_{4jk}, b_{1jk}, b_{2jk}, b_{3jk}, b_{4jk}, b_{5jk} \text{ for all } j,k)^{\trans} .$$ For the constant treatment effect setting, $\delta(\bX) = 0$, and for the heterogeneous treatment effect setting, $$
\delta(\bX) = \gamma_{I1}\widetilde{X}_1 + \gamma_{I2}\widetilde{X}_2 +  \sum_{j=3}^8\sum_{k=1}^{\ell_j}\left\{\eta_{I1jk}b_{I1jk}\widetilde{X}_1 + \eta_{I2jk}b_{I2jk}\widetilde{X}_2 \right\}\widetilde{X}_{j,k}
$$ 
where $\widetilde{X}_{j,k} = I(X_j = k) - \overline{I(X_j = k)}$, $\eta_{I1jk}, \eta_{I2jk}\sim$ Unif$(-0.5, 0.5)$, and $b_{I1jk}, b_{I3jk}\sim$ Bernoulli$(0.5)$. In the notation of the main text, 
$$\btheta_2 = (\gamma_{I1}, \gamma_{I2}, \gamma_{I1jk}, \gamma_{I2jk}, b_{I1jk}, b_{I2jk} \text{ for all } j,k)^{\trans}.$$

We generate 100 different draws of the coefficients ($\eta$, $\gamma$, and $b$ terms in $\btheta_1$ and $\btheta_2$) in the outcome model above, allowing for the simulation study to explore a wide variety of outcome models.

We also conduct a simulation study where 50 noise variables are added to the covariates in such a way that they are correlated both with treatment and response. To generate the additional variables, we first draw random variables from a normal distribution with mean $=(\overline{X}_1+\overline{X}_2)/2$ and variance $= (\text{var}(X_1)+\text{var}(X_2))/2$, where $X_1$ is the age started to smoke and $X_2$ is the age last smoked. We then use a Cholesky decomposition to induce an $AR(1)$ correlation structure with correlation parameter $\rho=-1/3$, where the correlation structure is with respect to the expanded vector $(Y,A,X_{nv1}, \dots, X_{nv50})$, where $X_{nvj}$ are the added variables and the $AR(1)$ correlation structure is modified in such a way that $A$ and $Y$ remain unchanged but the added variables have strong correlations jointly with $A$ and $Y$ and thus impact the finite sample error in estimating the ADRF if the added variables are not properly adjusted for.
This variance structure is used so that the variation of the noise variables is comparable to that of the continuous covariates in the NMES data.

\subsection{Additional simulation results from main text}

The MAB and IRMSE results for the heterogeneous treatment effect setting of the simulation study presented in the main text are displayed in Table \ref{tab:nmes_het_tx_simulation_10000reps}.

The results for the heterogeneous effect setting with 50 additional noise variables are displayed in Table \ref{tab:nmes_het_tx_nv50_simulation_10000reps}. Here, DCOWs perform the best among all methods in terms of both MAB and IRMSE for all sample sizes except $n=100$, where the doubly-robust gamma GPS estimate was slightly better in terms of IRMSE.

\begin{table}[ht]
    \centering
    \begin{adjustbox}{max width=0.95\textwidth}
    \begin{tabular}{lrrrrrrrrrrrr}
    \toprule
     & \multicolumn{2}{c}{$n=100$} & \multicolumn{2}{c}{$n=200$} & \multicolumn{2}{c}{$n=400$} & \multicolumn{2}{c}{$n=800$} & \multicolumn{2}{c}{$n=1600$} & \multicolumn{2}{c}{$n=3200$} \\ \cmidrule(lr){2-3}\cmidrule(lr){4-5}\cmidrule(lr){6-7}\cmidrule(lr){8-9}\cmidrule(lr){10-11}\cmidrule(lr){12-13}
     Method & MAB & IRMSE & MAB & IRMSE & MAB & IRMSE & MAB & IRMSE & MAB & IRMSE & MAB & \multicolumn{1}{r}{IRMSE} \\ 
    \midrule
    Unweighted  & 20.075 & 28.198 & 19.865 & 24.382 & 19.863 & 22.361 & 19.826 & 21.165 & 19.860 & 20.552 & 19.838 & 20.151 \\
    GPS (normal)  & 16.490 & 36.634 & 20.904 & 36.562 & 23.999 & 34.649 & 26.269 & 33.482 & 27.298 & 31.335 & 27.982 & 29.873 \\
    GPS (gamma)  & 15.597 & 33.153 & 15.491 & 26.710 & 15.263 & 22.321 & 14.942 & 19.740 & 14.698 & 18.013 & 14.414 & 16.031 \\
    GPS (normal,DR)  & 12.062 & 41.168 & 15.197 & 40.612 & 17.611 & 36.146 & 19.466 & 32.248 & 19.562 & 25.719 & 19.958 & 22.685 \\
    GPS (gamma,DR)  & 12.021 & 38.538 & 11.194 & 25.270 & 10.856 & 18.518 & 10.553 & 15.730 & 10.341 & 14.197 & 10.235 & 11.970 \\
    CBPS  & 17.990 & 36.480 & 20.534 & 33.170 & 22.074 & 31.068 & 20.669 & 26.016 & 20.196 & 23.072 & 20.213 & 21.503 \\
    GBM  & 15.174 & 34.035 & 13.646 & 25.480 & 12.849 & 19.368 & 12.832 & 16.467 & 12.908 & 14.925 & 13.089 & 14.087 \\
    BART  & 13.372 & 26.339 & 13.141 & 21.772 & 13.753 & 18.962 & 14.286 & 17.460 & 14.712 & 16.654 & 15.118 & 16.245 \\
    Entropy (1)  & ------ & ------ & 12.066 & 24.966 & 11.030 & 16.764 & 10.545 & 13.917 & 10.229 & 12.200 & 10.061 & 11.053 \\
    Entropy (2)  & ------ & ------ & 15.737 & 36.405 & 15.872 & 24.570 & 14.401 & 19.456 & 13.181 & 16.191 & 12.242 & 13.928 \\
    Entropy (2,int)  & ------ & ------ & ------ & ------ & ------ & ------ & ------ & ------ & 14.777 & 363.289 & 16.761 & 24.311 \\
    DCOW  & 6.718 & 18.517 & 6.083 & 13.254 & 5.818 & 10.552 & 5.569 & 8.852 & 5.400 & 7.703 & 5.329 & 6.830 \\
    DCOW (dm)  & 7.999 & 27.130 & 6.366 & 14.836 & 5.943 & 11.074 & 5.699 & 9.226 & 5.536 & 8.038 & 5.500 & 7.129 \\
    DCOW (DR)  & \textbf{5.778} & \textbf{16.726} & \textbf{5.220} & \textbf{12.206} & \textbf{4.891} & \textbf{9.780} & \textbf{4.735} & \textbf{8.311} & \textbf{4.829} & \textbf{7.422} & \textbf{5.055} & \textbf{6.753} \\
    DCOW (dm,DR)  & 6.447 & 21.900 & 5.569 & 13.414 & 5.158 & 10.321 & 5.007 & 8.760 & 5.088 & 7.823 & 5.276 & 7.074 \\
    \bottomrule 
    \end{tabular}
    \end{adjustbox}
    \caption{Mean absolute bias (MAB) and integrated root mean squared error (IRMSE) for the heterogeneous treatment effect setting.}
    \label{tab:nmes_het_tx_simulation_10000reps}
\end{table}

\begin{table}[ht]
    \centering
    \begin{adjustbox}{max width=\textwidth}
    \begin{tabular}{lrrrrrrrrrrrr}
    \toprule
     & \multicolumn{2}{c}{$n=100$} & \multicolumn{2}{c}{$n=200$} & \multicolumn{2}{c}{$n=400$} & \multicolumn{2}{c}{$n=800$} & \multicolumn{2}{c}{$n=1600$} & \multicolumn{2}{c}{$n=3200$} \\ \cmidrule(lr){2-3}\cmidrule(lr){4-5}\cmidrule(lr){6-7}\cmidrule(lr){8-9}\cmidrule(lr){10-11}\cmidrule(lr){12-13}
     Method & MAB & IRMSE & MAB & IRMSE & MAB & IRMSE & MAB & IRMSE & MAB & IRMSE & MAB & \multicolumn{1}{r}{IRMSE} \\ 
    \midrule
    Unweighted  & 20.075 & 28.198 & 19.865 & 24.382 & 19.863 & 22.361 & 19.826 & 21.165 & 19.860 & 20.552 & 19.838 & 20.151 \\
    GPS (normal)  & 19.733 & 29.907 & 16.453 & 31.114 & 22.881 & 39.654 & 30.521 & 46.508 & 37.341 & 50.345 & 42.592 & 51.453 \\
    GPS (gamma)  & 16.508 & 43.818 & 18.124 & 35.312 & 18.156 & 28.509 & 18.197 & 26.357 & 18.704 & 25.993 & 19.429 & 24.759 \\
    GPS (normal,DR)  & 13.993 & \textbf{20.345} & 14.330 & 21.161 & 17.472 & 33.781 & 22.528 & 42.084 & 26.955 & 41.828 & 30.583 & 39.910 \\
    GPS (gamma,DR)  & 13.655 & 74.504 & 15.441 & 46.099 & 15.694 & 26.679 & 15.769 & 22.863 & 16.455 & 22.531 & 17.155 & 20.910 \\
    CBPS  & 17.821 & 32.365 & 17.366 & 29.713 & 20.020 & 30.437 & 23.032 & 30.789 & 25.835 & 31.782 & 27.623 & 32.080 \\
    GBM  & 16.372 & 30.524 & 14.500 & 24.168 & 14.961 & 21.193 & 15.211 & 19.662 & 15.348 & 18.582 & 15.277 & 17.573 \\
    BART  & 14.451 & 25.927 & 13.831 & 23.202 & 15.749 & 22.567 & 17.324 & 22.518 & 17.971 & 21.877 & 17.587 & 20.575 \\
    Entropy (1)  & ------ & ------ & ------ & ------ & ------ & ------ & 14.089 & 18.085 & 13.380 & 15.623 & 13.123 & 14.327 \\
    Entropy (2)  & ------ & ------ & ------ & ------ & ------ & ------ & ------ & ------ & 18.963 & 28.307 & 17.360 & 19.206 \\
    Entropy (2,int)  & ------ & ------ & ------ & ------ & ------ & ------ & ------ & ------ & ------ & ------ & ------ & ------ \\
    DCOW  & \textbf{12.589} & 24.710 & \textbf{11.362} & 18.520 & \textbf{11.310} & 15.067 & 11.435 & 13.282 & 11.275 & 12.169 & 10.825 & 11.239 \\
    DCOW (dm)  & 20.054 & 28.215 & 14.120 & 31.635 & 13.428 & 20.513 & 12.623 & 15.718 & 11.289 & 12.633 & 10.117 & 10.691 \\
    DCOW (DR)  & 13.481 & 21.034 & 12.426 & \textbf{16.655} & 11.545 & \textbf{14.065} & \textbf{10.784} & \textbf{12.206} & \textbf{10.201} & \textbf{10.966} & \textbf{9.495} & \textbf{9.875} \\
    DCOW (dm,DR)  & 14.084 & 20.742 & 13.024 & 20.867 & 11.838 & 16.022 & 11.035 & 13.202 & 10.377 & 11.400 & 9.650 & 10.106 \\
    \bottomrule 
    \end{tabular}
    \end{adjustbox}
    \caption{Mean absolute bias (MAB) and integrated root mean squared error (IRMSE) for the heterogeneous treatment effect setting with 50 noise variables added.}
    \label{tab:nmes_het_tx_nv50_simulation_10000reps}
\end{table}

\section{ Vegetabile's Simulation }
\label{supsec:veg_sim}

We also consider the simulation settings in a recent paper \cite{vegetabile2020nonparametric} for evaluating the weighted estimator for ADRF by different weights. There are five covariates: $X_1 \sim N(-0.5,1)$,$X_2 \sim N(1,1)$,$X_3 \sim N(0,1)$,$X_4 \sim N(1,1)$, and $X_5 \sim \text{Bernoulli}(0.3)$. The observed dose $A \sim \chi^{2}(df=3, \tau_a)$, where the non-centrality parameter $\tau_{a} = 5|X_1| + 6|X_2| + |X_4| + 3|X_5|$. The choice of $\tau_{a}$ yields high concentration of treatment at zero with a
longer right tail of higher valued treatments, which is common in many applications (e.g., the substance abuse treatment program discussed in \cite{vegetabile2020nonparametric}).

The outcome is generated as follows,
$$Y = \frac{1}{50} \left[ (-0.15A^2 + A(X_1^2 + X_2^2) - 15) + ((X_1 + 3) + 2(X_2 - 25)^2 + X_3) - C + \epsilon \right]$$
$$C = E((X_1 + 3) + 2(X_2 - 25)^2 + X_3); \epsilon \sim N(0,1)$$
which implies that the marginal dose-response curve is $E(Y(a)) = -0.03a^2 + 0.065a - 0.3$. There is strong confounding effect in this setting and the relationship between $a $ and $Y(a)$ is nonlinear, which makes the ADRF estimation challenging. Furthermore, \cite{vegetabile2020nonparametric} assume the covariates $\bf{X}$ are not directly observable and only transformed covariates $\bf{Z}$ are. In particular, $Z_1 = \exp(X_1/2), Z_2 = \frac{X_2}{1 + exp(X_1)} + 10, Z_3 = \frac{X_1 X_3}{25} + 0.6$, $Z_4 = (X_4 - 1)^2, Z_5 = X_5$. These transformations induce correlations among observed covariates and introduce strong nonlinear relationship between $A$ and $Z$, which further complicates the estimation.

We generate the training and testing data using the same data generation mechanism as described above. We consider various sample size for training data and The testing data size is $10000$. We compare the same methods as we do in the previous section and report the (empirical) MAB and IRMSE using the testing data. Following the setup in \cite{vegetabile2020nonparametric}, we only evaluate the estimation performance in the high-density region of $A$: $A \in [1.5,45]$. 

Our weights yield the best performance, with the doubly robust estimator using our weights performing slightly worse than the non-doubly robust version likely due to extreme misspecification of the outcome regression model. Entropy balancing approaches also performed well, which was observed in the \cite{vegetabile2020nonparametric}. In this example, entropy balancing with second order moments and entropy balancing with interactions and second order moments performed better than just first order moments. Entropy balancing with only first order moments decorrelated performs no better than the naive unweighted estimator. Among machine learning-based GPS estimation approaches, BART performed the best across all sample sizes, however the performance of gbm is only slightly worse.  In this case, parametric models for the GPS performed significantly worse than a naive unweighted estimator, with the doubly robust estimators helping only marginally.

\begin{table}[ht]
    \centering
    \begin{adjustbox}{max width=0.75\textwidth}
    \begin{tabular}{lrrrrrrrr}
    \toprule
     & \multicolumn{2}{c}{$n=250$} & \multicolumn{2}{c}{$n=500$} & \multicolumn{2}{c}{$n=1000$} & \multicolumn{2}{c}{$n=2000$}\\ \cmidrule(lr){2-3}\cmidrule(lr){4-5}\cmidrule(lr){6-7}\cmidrule(lr){8-9}
     Method & MAB & IRMSE & MAB & IRMSE & MAB & IRMSE & MAB & IRMSE  \\ 
    \midrule
    Unweighted  & 0.338 & 0.429 & 0.343 & 0.395 & 0.338 & 0.369 & 0.338 & 0.356 \\
    GPS (normal)  & 0.526 & 0.766 & 0.619 & 0.793 & 0.688 & 0.811 & 0.763 & 0.858 \\
    GPS (gamma)  & 0.512 & 0.723 & 0.597 & 0.767 & 0.663 & 0.794 & 0.736 & 0.847 \\
    GPS (normal,DR)  & 0.443 & 1.024 & 0.531 & 1.195 & 0.539 & 0.860 & 0.609 & 0.851 \\
    GPS (gamma,DR)  & 0.441 & 1.001 & 0.503 & 0.950 & 0.525 & 0.817 & 0.588 & 0.793 \\
    CBPS  & 0.441 & 0.703 & 0.441 & 0.640 & 0.429 & 0.619 & 0.452 & 0.671 \\
    GBM  & 0.162 & 0.348 & 0.137 & 0.266 & 0.123 & 0.212 & 0.121 & 0.185 \\
    BART  & 0.138 & 0.307 & 0.121 & 0.242 & 0.104 & 0.192 & 0.100 & 0.167 \\
    Entropy (1)  & 0.333 & 0.447 & 0.352 & 0.420 & 0.357 & 0.399 & 0.364 & 0.390 \\
    Entropy (2)  & 0.175 & 0.376 & 0.171 & 0.287 & 0.169 & 0.239 & 0.169 & 0.211 \\
    Entropy (2,int)  & 0.196 & 0.428 & 0.172 & 0.304 & 0.165 & 0.245 & 0.164 & 0.214 \\
    DCOW  & \textbf{0.133} & 0.228 & \textbf{0.113} & 0.168 & \textbf{0.092} & \textbf{0.122} & 0.075 & \textbf{0.093} \\
    DCOW (dm)  & 0.139 & 0.244 & 0.115 & 0.175 & 0.092 & 0.124 & \textbf{0.074} & 0.094 \\
    DCOW (DR)  & 0.138 & \textbf{0.217} & 0.118 & \textbf{0.166} & 0.098 & 0.125 & 0.080 & 0.098 \\
    DCOW (dm,DR)  & 0.140 & 0.225 & 0.121 & 0.171 & 0.100 & 0.128 & 0.082 & 0.100 \\
    \bottomrule 
    \end{tabular}
    \end{adjustbox}
    \caption{Mean absolute bias (MAB) and integrated root mean squared error (IRMSE) for the simulation setup from \citet{vegetabile2020nonparametric}.}
    \label{tab:vegetabile_simulation_1000reps}
\end{table}

\section{ Simulation Investigating the Convergence of PDCOWs }
\label{supsec:wt_convergence_sim}

\begin{figure}[H]
	\centering
	\includegraphics[width=0.75\textwidth]{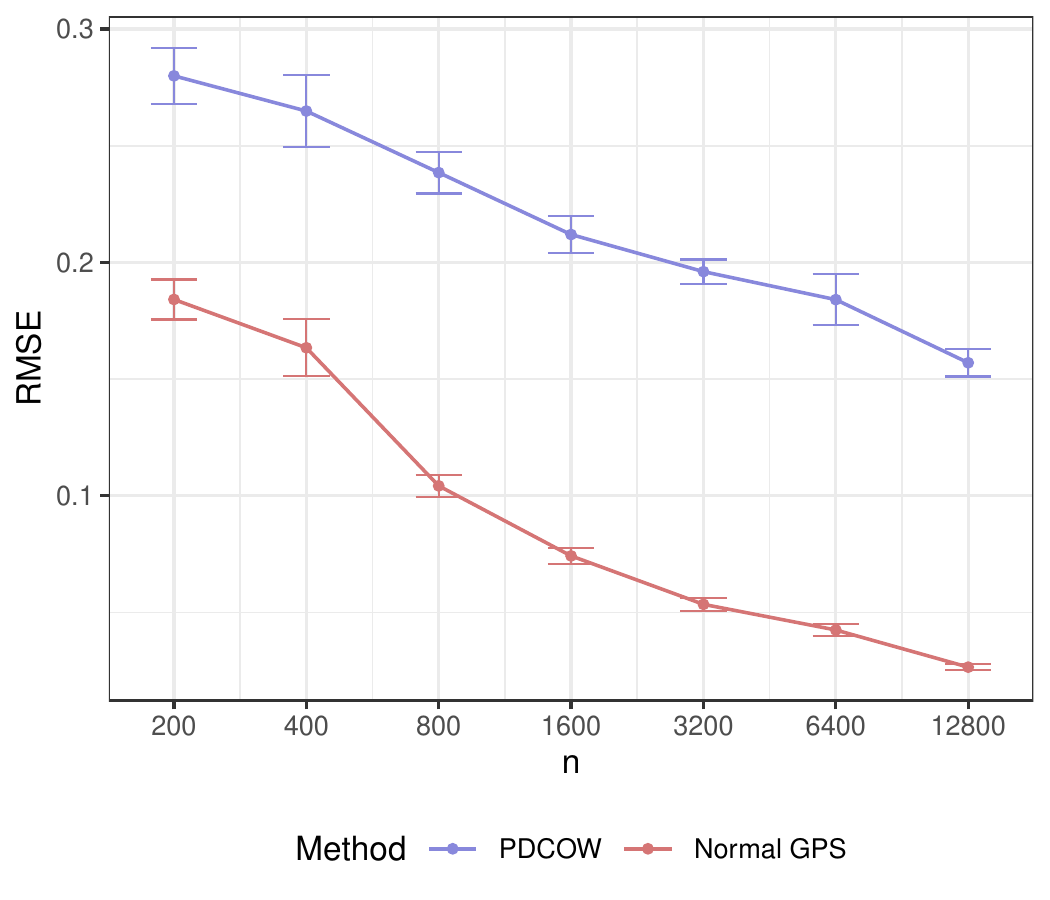}
	\caption{\label{fig:weight_est_sim} Shown are the average RMSEs of estimates of the individual weights compared with the ``true'' weights $w^*(X_i, A_i)$ across 100 replications of a simulation experiment. PDCOWs are our proposed weights and ``Normal GPS'' are weights computed using estimated stabilized inverse GPS weights under the correctly specified conditional density model. Error bars represent plus and minus 1 standard deviation across the replications.}
\end{figure}

In this section we aim to investigate whether the PDCOWs $\widetilde{w}^*_i$ converge to the ``true'' stabilized inverse GPS weights $w^*(\bX_i, A_i)$. In particular, we aim to understand whether the root mean squared error (RMSE) 
\begin{equation}\label{eqn:wt_rmse}
\sqrt{\frac{1}{n}\sum_{i=1} \left(w_i - w^*(\bX_i, A_i)\right)^2}
\end{equation}
of the PDCOWs (i.e. for $w_i = \widetilde{w}^*_i$) may plausibly converge to zero in some sense. For this simulation, we generate a continuous-valued treatment that is conditionally normal given a three-dimensional covariate vector $\bX = (X_1, X_2, X_3)^{\trans}$. 
In particular, we generate $A|\bX\sim N(\mu(\bX), 1)$, where $\mu(\bX) = X_1 + 1.2X_2 + 0.6 X_3$ and $X_j$ are i.i.d. normal random variables with mean 0 and standard deviation 0.2. Thus, the treatment $A$ is marginally normal with mean zero and standard deviation $\sqrt{1+0.2^2\times(1 + 1.2^2+0.6^2)} \approx 1.0545$.
We vary the sample size as $n=100 \times 2 ^ j$ for $j=1, \dots, 7$. 
For each simulated dataset we compute the PDCOWs with $\lambda = 10$ and additionally compute stabilized inverse GPS weights estimated by fitting a \textit{correctly-specified} normal conditional density model for $A|\bX$ (denoted as ``Normal GPS''); for each method we compute the RMSE \eqref{eqn:wt_rmse}. For each sample size, we replicate the simulation experiment 100 times. The average and standard deviations of the RMSEs for each method and sample size are displayed in Figure \ref{fig:weight_est_sim}.
From Figure \ref{fig:weight_est_sim}, not only does the RMSE using the PDCOWs consistently decrease as the sample size increases, but it decreases at what appears to be {a similar rate} as the correctly specified parametric model; thus, our PDCOWs appear in this setting thus appear to converge to $w^*(X_i, A_i)$ in mean-square at a or near a parametric rate. 
Unsurprisingly, using the correctly specified model results in better estimation of the individual weights and thus our performance in estimating weights isn't ``optimal'' since the goal of our PDCOWs is not optimal recovery of the true $w^*(X_i, A_i)$, it is instead to minimize dependence, which has a direct connection to bias as we showed.

Whether our PDCOWs will converge to the stabilized inverse GPS weights in RMSE is still an open question; this brief simulation suggests the PDCOWs might converge to the true GPS weights in certain scenarios.


\section{Additional mechanical power data results}
\label{sec:data_supp}

\begin{figure}[H]
\centering
\includegraphics[width=1\textwidth]{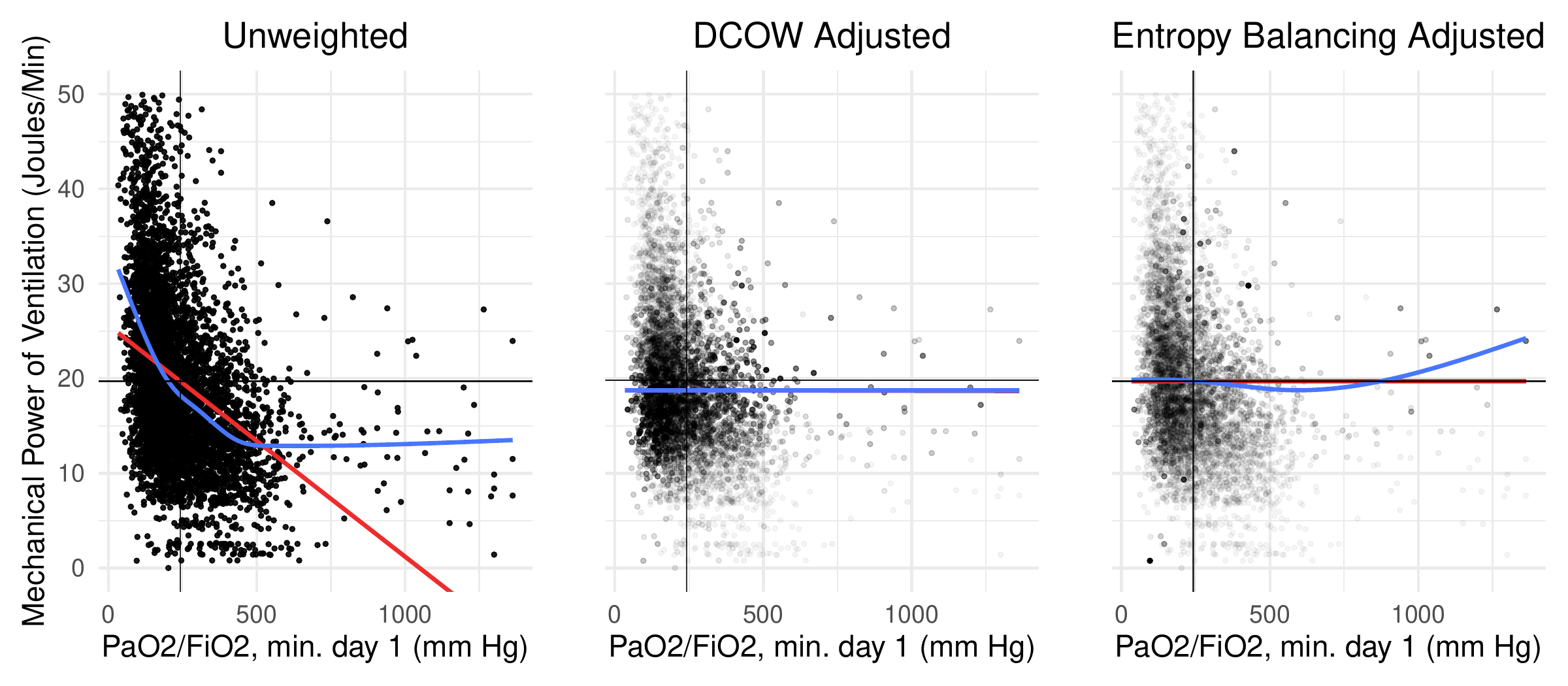}
\caption{\label{fig:mech_power_pao2_ratio} Shown are plots of the relationship between the minimum Pa$\text{O}_2$/Fi$\text{O}_2$ ratio on day 1 in the ICU and the treatment, including an unadjusted plot (left) and plots adjusted by DCOWs and entropy balancing weights (right two plots). In the adjusted plots, the transparency of each point is proportional to its assigned weight, with lighter points indicating less weight. The blue line is a weighted nonparametric regression of the treatment on Pa$\text{O}_2$/Fi$\text{O}_2$ and the red line is a weighted linear regression.}
\end{figure}

\bibliographystyle{Chicago}
\bibliography{manuscript}

\makeatletter\@input{xx.tex}\makeatother